\documentclass[a4paper, 12pt]{article}
\usepackage[symbol]{footmisc}
\usepackage{algorithm}
\usepackage{algpseudocode}
\usepackage{amsmath}

\usepackage{titlesec}
\usepackage{amssymb}
\usepackage{bm}
\usepackage{physics}
\usepackage{bigints}
\usepackage{bbm}
\usepackage{comment}

\usepackage{fullpage}

\newcommand{\lowerromannumeral}[1]{\romannumeral#1\relax} 
\newcommand{\upperRomannumeral}[1]{\uppercase\expandafter{\romannumeral#1}}

\usepackage{graphicx} 
\usepackage{multicol} 
\usepackage{lipsum}
\usepackage[style=authoryear,backend=bibtex,maxcitenames=2,maxbibnames=99,giveninits=true]{biblatex}
\bibliography{MixedLPCM}
\usepackage{xcolor}

\newtheorem{proposition}{Proposition}

\usepackage{mathtools}

\usepackage{booktabs}   
\definecolor{whitesmoke}{rgb}{0.96, 0.96, 0.96}
\definecolor{shadecolor}{named}{whitesmoke}

\definecolor{powderblue(web)}{rgb}{0.69, 0.88, 0.9}
\definecolor{antiquewhite}{rgb}{0.98, 0.92, 0.84}
\definecolor{navajowhite2}{RGB}{ 238,207,161}
\usepackage{caption}
\usepackage{gensymb} 
\usepackage[export]{adjustbox}

\usepackage{hyperref}

\begin{document}

\begin{center}

{\Large \bfseries Mixed Latent Position Cluster Models for Networks}

\vspace{5 mm}

{\large Chaoyi Lu$^{1,2}$, Riccardo Rastelli$^{1,*}$}\\
$^{1}$School of Mathematics and Statistics, University College Dublin, Ireland\\
$^{2}$Insight Research Ireland Centre for Data Analytics, University College Dublin, Ireland\\
$^{*}$\texttt{riccardo.rastelli@ucd.ie}

\vspace{5 mm}


\vspace{5mm}

\end{center}


\begin{abstract}

\noindent
Over the last two decades, the Latent Position Model (LPM) has become a prominent tool to obtain model-based visualizations of networks.
However, the geometric structure of the LPM is inherently symmetric, in the sense that outgoing and incoming edges are assumed to follow the same statistical distribution.
As a consequence, the canonical LPM framework is not ideal for the analysis of directed networks.
In addition, edges may be weighted to describe the duration or intensity of a connection.
This can lead to disassortative patterns and other motifs that cannot be easily captured by the underlying geometry.
To address these limitations, we develop a novel extension of the LPM, called the Mixed Latent Position Cluster Model (MLPCM), which can deal with asymmetry and non-Euclidean patterns, while providing new interpretations of the latent space.
We dissect the directed edges of the network by formally disentangling how a node behaves from how it is perceived by others.
This leads to a dual representation of a node's profile, identifying its ``overt'' and ``covert'' social positions.
In order to efficiently estimate the parameters of our model, we develop a variational Bayes approach to approximate the posterior distribution.
Unlike many existing variational frameworks, our algorithm does not require any additional numerical approximations.
Model selection is performed by introducing a novel partially integrated complete likelihood criteria, which builds upon the literature on penalized likelihood methods.
We demonstrate the accuracy of our proposed methodology using synthetic datasets, and we illustrate its practical utility with an application to a dataset of international arms transfers.

\end{abstract}



\section{Introduction}

The Latent Position Model (LPM) is a prominent statistical framework for network analysis.
Following the pioneering work of \textcite{hoff2002latent} this model has been widely applied in the social sciences and in a variety of other research areas.
From a methodological point of view, a number of extensions and variants of the LPM have also been proposed, including \textcite{sarkar2005dynamic,sewell2016latent,kim2018review, lu2025zero, jiang2024glamle, rastelli2024computationally, rastelli2023continuous}.
We refer to \textcite{LPNM_2023} for a recent overview of the existing literature on LPMs.

One key feature of this modeling approach is that it embeds the network into a latent space endowed with a basic geometric structure, and then uses the distances between nodes to determine the connections.
The geometry of the latent space can naturally support homophily, clustering, and other motifs that are often observed in social networks \parencite{hoff2002latent}.
In addition, this latent space can be used to provide clear graphical visualizations of the network data, or to derive model-based statistics that summarize the network's complex topology.
However, the geometrical properties of the latent space can also create limitations in capturing several common features of the observed networks.
In fact, actors can often exhibit different connectivity patterns based on who they interact with, or, more generally, based on the social context of an interaction.
This leads to a characterization of the edges that can go far beyond the encoding of all the node's information into a single latent position.

An example for this arises in networks with directed edges.
Since the LPM is constructed using the pairwise distances between nodes, it generates networks that are inherently symmetric.
This means that, in the case of directed networks, the direction of the edge is not relevant for determining the associated statistical distribution.
However, this property is rather inconsistent with the topologies of observed networks.
The patterns of how nodes send edges can be remarkably different from how they receive edges.
For example, in an email network, an account sending out newsletters may send out many emails while receiving few or none.
In the presence of a core-periphery structure, it can be difficult to find a configuration of latent positions that is consistent with the connectivity profiles that the nodes exhibit.\\

In this work, we focus on a new formulation of the LPM that aims at identifying and disentangling the multiple social profiles that a node may express, using latent positions.
Instead of relying on a single latent position to fully describe the individual features of a node, we rather associate a distribution of latent positions to each node.
This distribution is used to widen the range of social profiles that the node can exhibit, while still maintaining a parsimonious and interpretable framework.
Thanks to this new model structure, we are able to gain more flexibility in how the directed edges are characterized, allowing for asymmetric network structures and disassortative patterns.

Our model extension is connected to some of the literature on LPMs in the context of directed edges, which includes \textcite{sewell2015latent,sewell2016latent}. Similarly to these two works, we focus on non-negatively weighted networks, as these can exhibit asymmetric edges to a much greater extent, compared to binary networks. However, one difference with respect to these works is that we handle the asymmetry of the relations using latent positions, rather than the intercept term.
A similar statement can be made for the work of \textcite{krivitsky2009representing}, where instead the asymmetry is captured using sender and receiver random effects terms on the nodes.\\

Another aspect that we consider in this research is the clustering of the nodes.
The formulation that we adopt follows closely the pioneering work on the Latent Position Cluster Model (LPCM) of \textcite{handcock2007model}, where a finite Gaussian mixture model prior is used on the latent positions to promote the emergence of community structures.
One important research question that arises in this clustering context concerns the choice of the optimal number of groups.
For this task, we develop a novel model-based criterion, which we name the Partially Integrated Complete Likelihood (PICL).
This criterion is inspired by the literature on penalized likelihood methods, which include the Akaike Information Criterion (AIC) \parencite{akaike1974new}, the Bayesian Information Criterion (BIC) of \textcite{schwarz1978estimating}, the Integrated Complete Likelihood (ICL) of \textcite{biernacki2002assessing} and the Exact ICL $(\text{ICL}_{ex})$ of \textcite{biernacki2010exact,come2015model}.
Similarly to the $\text{ICL}_{ex}$, we marginalize the complete likelihood with respect to the model parameters using the conjugacy of the prior distributions.
The idea of integrating out the model parameters is widely utilized in the literature, both in a posterior sampling and optimization contexts.
Some relevant works include \textcite{wyse2012block,come2015model,ryan2017bayesian,legramanti2022extended,lu2024zero}.
However, in our case, unlike the exact ICL framework, not all of the model parameters can be integrated out, leaving us with a ``partially collapsed'' model-choice criterion, which we use to select the number of groups. \\

As regards parameter inference, the most common approach for parameter inference in the LPM is based on Markov chain Monte Carlo (MCMC) sampling from the posterior distribution of the model.
In this paper, we derive a variational Bayes approach \parencite{latouche2012variational,blei2017variational}, previously validated in the context of LPMs by various works including \textcite{salter2013variational, gollini2016joint, rastelli2018sparse}.
Differently from the more common MCMC methods, the variational Bayes framework aims at approximating the posterior distribution using an alternative ``variational'' distribution that satisfies some convenient independence properties.
It is worthwhile to note that, in a typical variational Bayes approach, various numerical approximations are usually required to write down the variational posterior analytically. These additional approximations can include, for example, a Jensen's inequality step as used in \textcite{jordan1999introduction,salter2013variational,gollini2016joint}.
In our framework, thanks to the model assumptions, we are able to bypass this issue and avoid any further numerical approximations, other than the one implied by the variational Bayes method itself.
Our implementation of our variational Bayes algorithm for the MLPCM is publicly available from our GitHub repository\footnote{\url{https://github.com/Chaoyi-Lu/Mixed-Latent-Position-Cluster-Model}.}.\\

This paper is organized as follows.
In Section~\ref{Model}, the mixed latent position cluster model is defined and the corresponding simulation algorithm is introduced.
The detailed variational Bayes procedures for the inference of our model is provided in Section~\ref{VB} and Section~\ref{sec:max}, with an outline of all the update steps that we define and their corresponding equations.
We describe the model selection criterion in Section~\ref{Algorithm_and_PICL}, along with an initialization method for the inferential procedure.
Section~\ref{SS} contains two simulation studies that illustrate the performance of our proposed variational Bayes algorithm, for data generated with our own model and under model misspecification.
As an alternative and competing methodology, we use the framework of \textcite{lu2025zero}, and provide a brief sensitivity analysis to verify how results may change under different initial settings and priors.
Finally, we utilize the methodology on a network of international arms transfers in Section~\ref{RDA} and draw some conclusions in Section~\ref{Conclusion}.


\section{Model}
\label{Model}
In this work, we focus on directed networks with non-negative discrete weighted edges.
We denote with $\bm{Y}:= \{y_{ij} \in \mathbbm{N}:i,j=1,\dots,N;i\neq j\}$ an observed interaction matrix, where each $y_{ij}$ is a non-negative integer.
Without loss of generality, we can assume that this value represents the intensity of the interaction from $i$ to $j$, where an intensity equal to zero is equivalent to the absence of an edge or interaction.
Self-edges are not allowed, thus $y_{ii} = 0$, $\forall i=1,\dots,N$.
Since the network is directed, $y_{ij}$ corresponds to the interaction from node $i$ to node $j$, whereas $y_{ji}$ represents the interaction from node $j$ to node $i$ and it may take a different value from $y_{ij}$.
Some common datasets that can be represented within this framework include networks counting the number of phone calls interactions, the number of emails sent, or packages deliveries.

In real directed weighted networks, we usually observe significant discrepancy between $y_{ij}$ and $y_{ji}$, for example, one of them can be very small or even a zero while the other exhibits a large value.
This can be common in networks that display disassortative mixing or core-periphery structures, as it would be typical in the presence of star patterns, i.e. core nodes that connect to many isolated nodes.
In these scenarios, a formulation following the principles of a distance-based LPM would struggle to fit the data well.
In fact, the geometric construction of the LPM favors assortative mixing structures that arise due to homophily \parencite{hoff2002latent} or clustering \parencite{handcock2007model}.
More importantly, since the canonical LPM is based on the distance between nodes (which is a symmetric operator), it does not permit a separate characterization of $y_{ij}$ and $y_{ji}$.
To deal with this issue, we introduce a novel extension of the LPCM to non-negative discrete weighted directed networks as follows:
\begin{equation}
\label{MixedLPCM1}
\begin{split}
&y_{ij}\sim\text{Pois}(\lambda_{ij}),\\
&\log(\lambda_{ij})=\beta-||\bm{u}_i-\bm{v}_{j\leftarrow i}||^2,\\
&\bm{u}_i|z_i=k\sim\text{MVN}_{d}(\bm{\mu}_{k},1/\tau_k\mathbbm{I}_d),\\
&\bm{v}_{j\leftarrow i}\sim\text{MVN}_{d}(\bm{u}_j,1/\gamma_j\mathbbm{I}_d), \\
&z_i\sim\text{categorical}(\pi_1,\dots,\pi_K).
\end{split}
\end{equation}
for $i,j = 1,\dots,N;\;i\neq j$.
We now clarify the notation starting from the bottom line of Eq.~\eqref{MixedLPCM1} upwards.
Each node is characterized by a clustering variable $z_i$, arising from the mixing proportions $\pi_1,\dots,\pi_K$.
This clustering variable determines the allocations for a Gaussian finite mixture prior on the latent positions $\bm{u}_i \in \mathbb{R}^d$, with $\bm{\mu}_{k} \in \mathbb{R}^d$ and $\tau_k > 0$ being the mean and precision associated to component $k$, respectively.
We refer to $\bm{u}_i$ as the ``covert'' position of $i$: this is a node feature that determines the behavior of $i$ since it represents its true, hidden, social characteristics.
This covert position directly drives how $i$ sends edges, and this individual information is not known by the other nodes.
Based on the covert positions $\bm{u}_1,\dots,\bm{u}_N$ we create a number of ``overt'' positions $\bm{v}_{j\leftarrow i} \in \mathbb{R}^d$, for all $i$s and $j$s.
These are points that are scattered nearby the covert positions, and they represent how nodes are perceived by other nodes.
For example, $\bm{v}_{j\leftarrow i}$ indicates how $i$ perceives $j$, which may be different from $\bm{v}_{j\leftarrow k}$, representing how $k$ perceives the same node $j$.
However, both positions $\bm{v}_{j\leftarrow i}$ and $\bm{v}_{j\leftarrow k}$ are expected to be located nearby $\bm{u}_j$, meaning that the true, hidden social characteristics of node $j$ will generally be similar to how $j$ is actually perceived.
The value $\gamma_j > 0$ is the precision parameter that regulates the dispersion of the overt positions around the corresponding covert position.
This parameter can be interpreted as a measure of how authentic the behavior of a node is, as it describes the range of ways in which a node can be perceived.
A node with a high $\gamma$ will generally be perceived in very similar ways by all other nodes, and its patterns of sending and receiving edges will generally coincide.
In this case, the behavior of the node could be considered very authentic and its edges will tend to be symmetric.
As a special case, if $\gamma \rightarrow \infty$ for every node, then we recover a basic LPCM.
By contrast, a node with a small $\gamma$ will exhibit high social flexibility, in that it may be perceived in very different ways by other nodes, thus permitting connections that would otherwise be very unlikely to appear.\\

We then assume that an edge weight is determined using an intercept $\beta \in \mathbb{R}$, and the squared Euclidean distance between the covert position of the sender and the overt position of the receiver.
In other words, the intensity of the interaction is determined by $i$'s true, hidden, social traits, and how $i$ perceives the social traits of $j$.
On the contrary, the reverse interaction $y_{ji}$ is determined by the covert latent position $\bm{u}_j$ of individual $j$ and the overt latent position $\bm{v}_{i\leftarrow j}$ of individual $i$ as it is perceived by individual $j$.
This permits the characterization of two different distributions for $y_{ij}$ and $y_{ji}$, respectively, while maintaining a marginal dependence between the two variables as implied by the relation between the covert and overt positions of each node.\\

The latent space is $d$-dimensional, for $d \in \{1,2,\dots\}$.
While our notation remains general and considers latent spaces of an arbitrary number of dimensions, in this work we only focus on the case $d=2$.
We note that inference on the number of latent dimension is a critical and difficult research question that has been studied by the recent literature.
Since this research aspect is not a main focus for our work, we choose $d=2$ for simplicity and to obtain more straightforward visualizations.
But we emphasize that our approach can be easily framed into more than two dimensions, and future work may address the problem of selecting the ideal number of dimensions for for this model, given a dataset.\\

We denote the collection of all covert latent positions with a $d \times N$ matrix $\bm{U}:=(\bm{u}_1,\dots,\bm{u}_N)$, where each $\bm{u}_i$ is a $d \times 1$ vector for $i=1,2,\dots,N$.
Similarly, the collection of all overt latent positions is denoted as $\bm{V}:=\{\bm{v}_{j\leftarrow i}:i,j=1,2,\dots,N;i\neq j\}$, which is interpreted as a $N \times N$ matrix with each $ij$th entry being the $d \times 1$ vector $\bm{v}_{j\leftarrow i}$, that is, $\bm{V}$ is a $d \times N \times N$ array.
The $N \times 1$ vector $\bm{\gamma}:=(\gamma_1,\dots,\gamma_N)^T$ is used to denote the collection of all the precision of overt latent positions.
Following \textcite{gollini2016joint}, we use the squared distance in place of the simple Euclidean distance to encourage the sensitivity of the data distribution to the change of distance in the latent space, bringing clearer data visualization.
Furthermore, as we show in Section \ref{VB}, this leads to useful simplifications in the development of our inferential algorithm, in that no further numerical approximations are required in the variational Bayes optimization.\\

We term the new model arising from Eq.~\eqref{MixedLPCM1} as the Mixed Latent Position Cluster Model (MLPCM).
The name of this model hints at its connection with another renowned networks approach: the mixed-membership stochastic block model (MMSBM) proposed by \textcite{airoldi2008mixed}.
As in the MMSBM, each node is characterized by a set of individual latent characteristics.
Using a data-augmentation perspective, these latent characteristics are augmented using an edge-specific latent variable.
In the context of the MMSBM, this variable is the cluster membership that is used to generate a particular edge.
Since for each edge of an MMSBM we draw a new cluster membership, we obtain a ``mixed membership'' framework.
In the context of our MLPCM model, this augmented variable is the overt latent position $\bm{v}_{\cdot}$.
In analogy, by drawing a new overt position for each edge of a MLPCM, we obtain a ``mixed latent position'' structure, where multiple latent positions can simultaneously characterize the profile of a node.\\

The MLPCM can also be connected to the random effects network models of \parencite{hoff2005bilinear,krivitsky2009representing}.
That is, the model in Eq.~\eqref{MixedLPCM1} is equivalent to:
\begin{equation}
\label{MixedLPCM_RandomEffects}
\begin{split}
 y_{ij}\sim\text{Pois}(\lambda_{ij})&,\;\;\log(\lambda_{ij})=\beta-||\bm{u}_i-(\bm{u}_j+\bm{\epsilon}_{ij})||^2,\\
\bm{\epsilon}_{ij}&\sim\text{MVN}_{d}(\bm{0},1/\gamma_j\mathbbm{I}_d),
\end{split}
\end{equation}
where $\bm{\epsilon}_{ij}$ can be seen as an edge-specific random effect which possibly embeds sender or receiver effects.
However, a critical difference between the MLPCM and the random effects models is the characterization of the overt position as an alteration of the covert one.
This leads to a different interpretation of the latent positions, and of their role in defining the social profile of a node.

\begin{figure}[htbp!]
\centering
\includegraphics[scale=0.525]{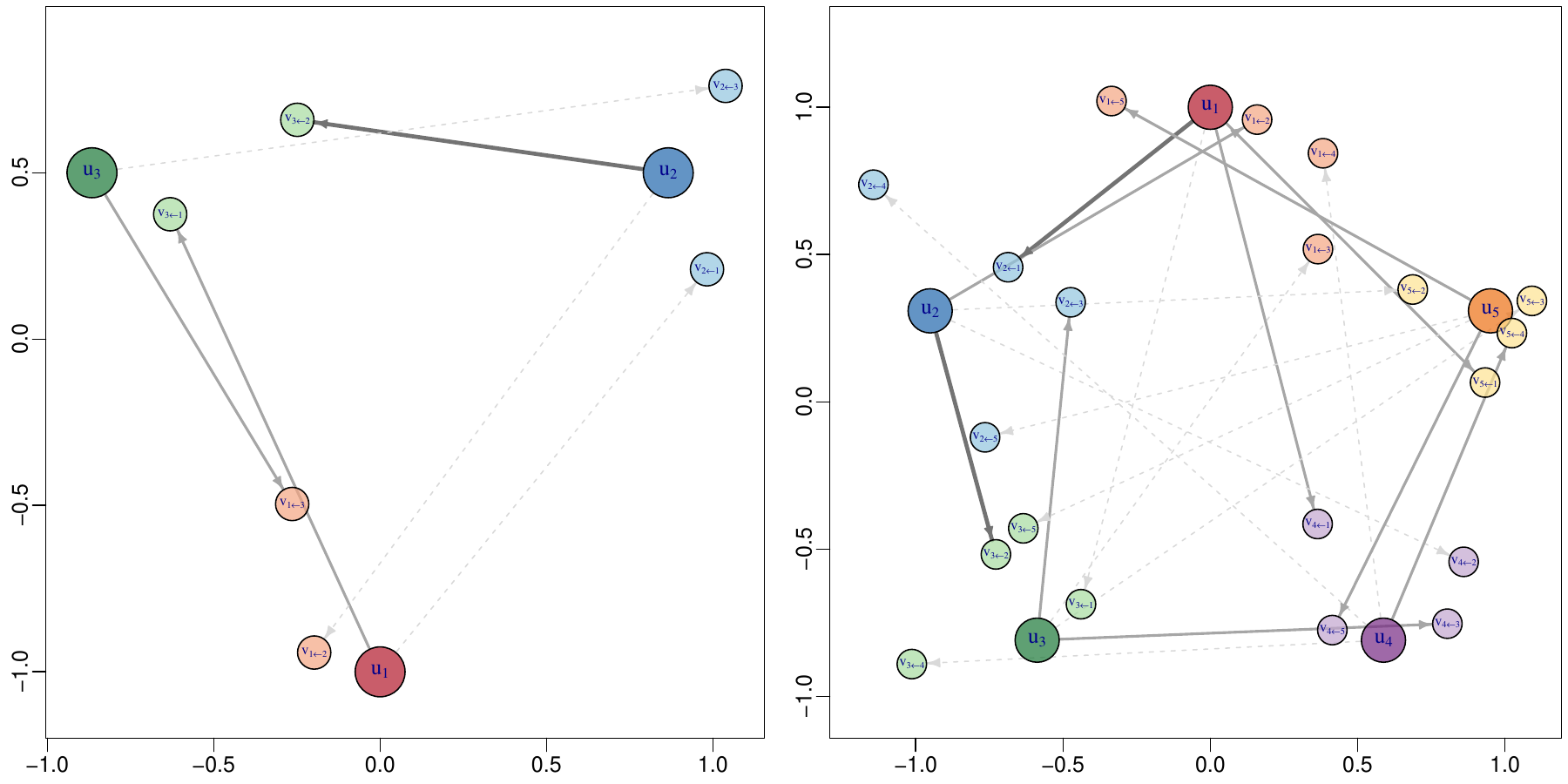}
\caption{Two motivating network examples generated from the MLPCM. The network on the left has $3$ actors, whereas the one on the right has $5$. Each dark-color node represents the covert latent position of each individual, while the corresponding light-color nodes are the overt latent positions. The edges in both networks have non-negative discrete weights ranging from 0 to 2, indicated by the edge widths. The dashed lines between nodes indicate a zero edges, whereas the solid lines correspond to non-zero edges.}
\label{MixedLPCM_2Examples}
\end{figure}

Generating artificial data for the MLPCM using Eq.~\eqref{MixedLPCM1} is straightforward, as we outline in Algorithm~\ref{MixedLPCMSimulation}.
\begin{algorithm}[htbp!]
\caption{Simulation from the MLPCM.}
\label{MixedLPCMSimulation}
\begin{algorithmic} 
\State \textbf{Input}: $\beta, \bm{\mu}, \bm{\tau}, \bm{\gamma}$ and either $\bm{z}$ or $\bm{\Pi}$.
\If{$\bm{z}$ is not provided}
\For {$i=1,\dots,N$} 
\State \textcolor{red}{1.} $z_i \sim \text{cat}(\bm{\Pi})$.
\EndFor
\EndIf

\For {$j=1,\dots,N$} 
\State \textcolor{red}{2.} $\bm{u}_j \sim \text{MVN}_d(\bm{\mu}_{z_j},1/\tau_{z_j}\mathbbm{I}_d)$.
\For {$i=1,\dots,N;\; i\neq j$} 
\State \textcolor{red}{3.} $\bm{v}_{j\leftarrow i} \sim \text{MVN}_d(\bm{u}_j,1/\gamma_j\mathbbm{I}_d)$.
\EndFor
\EndFor

\For {$i,j=1,\dots,N;\; i\neq j$} 
\State \textcolor{red}{4.} $y_{ij} \sim \text{Pois}\left[\exp (\beta-||\bm{u}_i-\bm{v}_{j\leftarrow i}||^2)\right]$.
\EndFor

\State \textbf{Output}: $\bm{Y}, \bm{U}, \bm{V}, \bm{z}$.
\end{algorithmic}
\end{algorithm}
To complete our model overview, we illustrate two simulated examples in Figure~\ref{MixedLPCM_2Examples}, exemplifying how the interactions are determined by the distance between different pairs of covert and overt latent positions.
Different colors (red, blue, green, purple and orange) correspond to different individuals existing in the network, with darker nodes representing the covert latent positions, and lighter colors indicating the overt latent positions.
For example, the dark-red node represents the covert latent position $\bm{u}_1$ of the individual $1$, while the light-red nodes surrounding around the $\bm{u}_1$ are the overt latent positions $\{\bm{v}_{1\leftarrow 2},\bm{v}_{1\leftarrow 3},\dots\}$ that arise for the dyads going from $1$ to the other nodes.\\

We further extend our model using a hierarchical structure, with the following prior distributions:
\begin{equation}
\label{Priors}
\begin{split}
\beta &\sim \text{Normal}(\eta,\rho^2),\\
\bm{\mu}_k &\sim \text{MVN}_d(\bm{0},\omega^2\mathbbm{I}_d),\;\;\text{for}\;k=1,\dots,K,\\
\tau_k&\sim\text{Gamma}(\xi,\psi),\;\;\text{for}\;k=1,\dots,K,\\
\gamma_i&\sim\text{Gamma}(a,b),\;\;\text{for}\;i=1,\dots,N,\\
\bm{\Pi}&\sim\text{Dirichlet}(\delta_1,\dots,\delta_K).
\end{split}
\end{equation}
As a consequence, the posterior distribution of the MLPCM reads as:
\begin{equation*}
\label{MLPCM_Posterior}
\resizebox{1.0\hsize}{!}{$
\begin{split}
& \pi(\bm{V},\bm{U},\bm{z}, \beta, \bm{\gamma},\bm{\mu}, \bm{\tau}, \bm{\Pi}|\bm{Y}) 
\propto p(\bm{Y}|\bm{V},\bm{U}, \beta)p(\bm{V}|\bm{U}, \bm{\gamma})p(\bm{U}|\bm{\mu}, \bm{\tau}, \bm{z})p(\bm{z}|\bm{\Pi})\times\pi(\beta)\pi(\bm{\gamma})\pi(\bm{\mu})\pi(\bm{\tau})\pi(\bm{\Pi})\\
&= \prod^N_{\substack{i,j=1, \\ i\neq j}}f_{\text{Pois}}\left(y_{ij}|\exp\left(\beta-||\bm{u}_i-\bm{v}_{j\leftarrow i}||^2\right)\right)
\prod^N_{\substack{i,j=1, \\ i\neq j}} f_{\text{MVN}_d}(\bm{v}_{j\leftarrow i}|\bm{u}_j,1/\gamma_j\mathbbm{I}_d)
\prod^N_{i=1} f_{\text{MVN}_d}(\bm{u}_i|\bm{\mu}_{z_i},1/\tau_{z_i}\mathbbm{I}_d)\prod^N_{i =1} f_{\text{cat}}(z_i|\bm{\Pi})\\
&\hspace{1em}\times f_{\text{Normal}}(\beta|\eta,\rho^2) \left[\prod^N_{i=1} f_{\text{Ga}}(\gamma_i|a,b) \prod^K_{k=1} f_{\text{MVN}_d}(\bm{\mu}_k|\bm{0},\omega^2\mathbbm{I}_d) \prod^K_{k=1} f_{\text{Ga}}(\tau_k|\xi,\psi)\right] f_{\text{Dirichlet}}(\bm{\Pi}|\bm{\delta}).
\end{split}$}
\end{equation*}
where $\bm{\delta}:=(\delta_1,\dots,\delta_K)$ are non-negative Dirichlet weights, $a,b,\rho,\xi,\psi>0$ and $\eta\in\mathbb{R}$.

As default hyperparameters for the MLPCM, we set $\eta=\rho^2=\omega^2=\xi=\psi=b=\delta_1=\dots=\delta_K=1$, and $a=10$ in all applications.
The motivation for setting $a$ as a higher value is that we want to favor a small dispersion of the overt positions around their covert positions.
This is meaningful in a context of model parsimony, since we are trying to penalize a potentially overfitted model where the overt positions are not constrained to be close to their covert position, and so they may be identified just based on the value of the edge they correspond to.
This can be seen as a case of overfitting, since the number of overt positions grows with the number of edges, and these parameters do not have any meaningful constraint on them.
By contrast, a $Ga(10,1)$ prior on the $\boldsymbol{\gamma}$ ensures that the overt positions are pushed closer to their corresponding covert position.
From an applied point of view, this assumption describes a situation where, as a baseline, nodes tend to behave in the same way as they are perceived.
Different choices of hyperparameters are perfectly possible, especially in situations where additional information is available: we explore this more in detail in Section~\ref{SS2}.


\section{Variational approximation}
\label{VB}
We develop a variational Bayesian inference procedure to approximate the posterior distribution of the MLPCM.
We denote the collection of model parameters with $\bm{\Theta}:=\{\bm{U},\bm{V},\bm{z}, \beta, \bm{\mu}, \bm{\tau}, \bm{\gamma}, \bm{\Pi}\}$.
The model evidence is written as:
\begin{equation}
\label{Evidence}
p(\bm{Y})=p(\bm{Y})\frac{\pi(\bm{Y},\bm{\Theta})}{\pi(\bm{Y},\bm{\Theta})}=\frac{\pi(\bm{Y},\bm{\Theta})}{\pi(\bm{\Theta}|\bm{Y})}=\frac{\pi(\bm{Y},\bm{\Theta})}{q(\bm{\Theta})}
\frac{q(\bm{\Theta})}{\pi(\bm{\Theta}|\bm{Y})},
\end{equation}
where $q(\bm{\Theta})$ is a so-called variational distribution that is meant to approximate the posterior distribution $\pi(\bm{\Theta}|\bm{Y})$.
A common assumption is that $q(\cdot)$ should belong to a family of distributions that are in simpler forms than the posterior distribution $\pi(\bm{\Theta}|\bm{Y})$.
By taking the expectation with respect to $q(\bm{\Theta})$ on the log transformation of Eq.~\eqref{Evidence}, we obtain:
\begin{equation*}
\label{ElogEvidence}
\log p(\bm{Y})=\mathbbm{E}_q[\log \pi(\bm{Y},\bm{\Theta})]-\mathbbm{E}_q[\log q(\bm{\Theta})]+
\mathbbm{E}_q[\log q(\bm{\Theta})]-\mathbbm{E}_q[\log \pi(\bm{\Theta}|\bm{Y})],
\end{equation*}
which corresponds to the sum of the so-called Evidence Lower Bound (ELBO) and the Kullback-Leibler (KL) divergence between $q(\bm{\Theta})$ and $\pi(\bm{\Theta}|\bm{Y})$.
More explicitly: $\text{ELBO}=\mathbbm{E}_q[\log \pi(\bm{Y},\bm{\Theta})]-\mathbbm{E}_q[\log q(\bm{\Theta})]$, and $\text{KL}[q(\bm{\Theta})||\pi(\bm{\Theta}|\bm{Y})]=\mathbbm{E}_q[\log q(\bm{\Theta})]-\mathbbm{E}_q[\log \pi(\bm{\Theta}|\bm{Y})]$.
It is clear that the log-evidence $\log p(\bm{Y})$ is fixed when the observed data $\bm{Y}$ is provided.
Thus minimizing the $\text{KL}[q(\bm{\Theta})||p(\bm{\Theta}|\bm{Y})]$ is equivalent to maximizing the ELBO above.


\subsection{Derivation of the ELBO}

We consider a set of typical choices of variational distributions for $q(\bm{\Theta})$ that satisfy a mean-field assumption, as in the equation below:
\begin{equation}
\label{Variational_distributions}
\begin{split}
q_{\bm{U}}(\bm{u}_i|\bm{\tilde{u}}_i,\tilde{\sigma}^2_i) &\sim \text{MVN}_d(\bm{\tilde{u}}_i,\tilde{\sigma}^2_i\mathbbm{I}_d)\;\;\text{for}\;i=1,\dots,N,\\
q_{\bm{V}}(\bm{v}_{j\leftarrow i}|\bm{\tilde{v}}_{j\leftarrow i},\tilde{\varphi}^2_{ij}) &\sim \text{MVN}_d(\bm{\tilde{v}}_{j\leftarrow i},\tilde{\varphi}^2_{ij}\mathbbm{I}_d)\;\;\text{for}\;i,j=1,\dots,N;\;i\neq j,\\
q_{\bm{z}}(z_i|\tilde{\pi}_{i1},\dots,\tilde{\pi}_{iK}) &\sim \text{categorical}(\tilde{\pi}_{i1},\dots,\tilde{\pi}_{iK})\;\;\text{for}\;i=1,\dots,N,\\
q_{\beta}(\beta|\tilde{\eta},\tilde{\rho}^2) &\sim \text{Normal}(\tilde{\eta},\tilde{\rho}^2),\\
q_{\bm{\mu}}(\bm{\mu}_k|\bm{\tilde{\mu}}_k,\tilde{\omega}^2_k) &\sim \text{MVN}_d(\bm{\tilde{\mu}}_k,\tilde{\omega}^2_k\mathbbm{I}_d)\;\;\text{for}\;k=1,\dots,K,\\
q_{\bm{\tau}}(\tau_k|\tilde{\xi}_k,\tilde{\psi}_k)&\sim\text{Gamma}(\tilde{\xi}_k,\tilde{\psi}_k)\;\;\text{for}\;k=1,\dots,K,\\
q_{\bm{\gamma}}(\gamma_j|\tilde{a}_j,\tilde{b}_j)&\sim\text{Gamma}(\tilde{a}_j,\tilde{b}_j)\;\;\text{for}\;j=1,\dots,N,\\
q_{\bm{\Pi}}(\pi_1,\dots,\pi_K|\tilde{\delta}_1,\dots,\tilde{\delta}_K)&\sim\text{Dirichlet}(\tilde{\delta}_1,\dots,\tilde{\delta}_K).
\end{split}
\end{equation}
We use $\bm{\tilde{\Theta}}:=\{\bm{\tilde{U}},\bm{\tilde{\sigma}^2},\bm{\tilde{V}},\bm{\tilde{\varphi}^2},\bm{\tilde{\Pi}},\tilde{\eta},\tilde{\rho}^2,\bm{\tilde{\mu}},\bm{\tilde{\omega}^2},\bm{\tilde{\xi}},\bm{\tilde{\psi}},\bm{\tilde{a}},\bm{\tilde{b}},\bm{\tilde{\delta}}\}$ to denote the collection of all variational parameters, where $\bm{\tilde{U}}:=(\bm{\tilde{u}}_1,\dots,\bm{\tilde{u}}_N)$ is a $d \times N$ matrix, while $\bm{\tilde{\sigma}^2}:=(\tilde{\sigma}^2_1,\dots,\tilde{\sigma}^2_N)^T$, $\bm{\tilde{a}}:=(\tilde{a}_1,\dots,\tilde{a}_N)^T$ and $\bm{\tilde{b}}:=(\tilde{b}_1,\dots,\tilde{b}_N)^T$ are $N\times 1$ vectors.
The object $\bm{\tilde{V}}:=\{\bm{\tilde{v}}_{j\leftarrow i}:i,j=1,\dots,N;i\neq j\}$ consists of a $d \times N \times N$ array, and $\bm{\tilde{\varphi}^2}:=\{\tilde{\varphi}^2_{ij}:i,j=1,\dots,N;i\neq j\}$ is a $N \times N$ matrix with diagonal elements being undefined.
Similarly, $\bm{\tilde{\Pi}}_i:=(\tilde{\pi}_{i1},\dots,\tilde{\pi}_{iK})^T$ is a $K \times 1$ vector constituting the $K \times N$ matrix: $\bm{\tilde{\Pi}}:=(\bm{\tilde{\Pi}}_1,\dots,\bm{\tilde{\Pi}}_N)$.
The collection of variational parameters $\bm{\tilde{\mu}}:=(\bm{\tilde{\mu}}_1,\dots,\bm{\tilde{\mu}}_K)$ is a $d \times K$ matrix, while
$\bm{\tilde{\omega}^2}:=(\tilde{\omega}^2_1,\dots,\tilde{\omega}^2_K)^T$, 
$\bm{\tilde{\xi}}:=(\tilde{\xi}_1,\dots,\tilde{\xi}_K)^T$,
$\bm{\tilde{\psi}}:=(\tilde{\psi}_1,\dots,\tilde{\psi}_K)^T$ and $\bm{\tilde{\delta}}:=(\tilde{\delta}_1,\dots,\tilde{\delta}_K)^T$ are $K\times 1$ vectors.\\

Let $\mathcal{F}(\bm{\tilde{\Theta}})$ denote the ELBO, then $\mathcal{F}(\bm{\tilde{\Theta}})$ can be decomposed into a number of terms that can be derived separately:
\begin{equation*}
\begin{split}
&\text{ELBO}=\mathcal{F}(\bm{\tilde{\Theta}})=\mathbbm{E}_q[\text{log}\;\pi(\bm{Y},\bm{\Theta})]-\mathbbm{E}_q[\text{log}\;q(\bm{\Theta})]\\
&= \mathbbm{E}_q[\text{log}\;p(\bm{Y}|\bm{U},\bm{V}, \beta)]+
\mathbbm{E}_q[\text{log}\;p(\bm{U}|\bm{\mu}, \bm{\tau}, \bm{z})]+
\mathbbm{E}_q[\text{log}\;p(\bm{V}|\bm{U}, \bm{\gamma})]+
\mathbbm{E}_q[\text{log}\;p(\bm{z}|\bm{\Pi})]\\
&+\mathbbm{E}_q[\text{log}\;\pi(\beta)]+
\mathbbm{E}_q[\text{log}\;\pi(\bm{\mu})]+
\mathbbm{E}_q[\text{log}\;\pi(\bm{\tau})]+
\mathbbm{E}_q[\text{log}\;\pi(\bm{\gamma})]+
\mathbbm{E}_q[\text{log}\;\pi(\bm{\Pi})]\\
&-\mathbbm{E}_q[\text{log}\;q_{\bm{U}}(\bm{U}|\bm{\tilde{U}},\bm{\tilde{\sigma}^2})]-
\mathbbm{E}_q[\text{log}\;q_{\bm{V}}(\bm{V}|\bm{\tilde{V}},\bm{\tilde{\varphi}^2})]-
\mathbbm{E}_q[\text{log}\;q_{\bm{z}}(\bm{z}|\bm{\tilde{\Pi}})]-
\mathbbm{E}_q[\text{log}\;q_{\beta}(\beta|\tilde{\eta},\tilde{\rho}^2)]\\
&-\mathbbm{E}_q[\text{log}\;q_{\bm{\mu}}(\bm{\mu}|\bm{\tilde{\mu}},\bm{\tilde{\omega}^2})]-
\mathbbm{E}_q[\text{log}\;q_{\bm{\tau}}(\bm{\tau}|\bm{\tilde{\xi}},\bm{\tilde{\psi}})]-
\mathbbm{E}_q[\text{log}\;q_{\bm{\gamma}}(\bm{\gamma}|\bm{\tilde{a}},\bm{\tilde{b}})]-
\mathbbm{E}_q[\text{log}\;q_{\bm{\Pi}}(\bm{\Pi}|\bm{\tilde{\delta}})].
\end{split}
\end{equation*}

In many frameworks, the ELBO cannot be obtained analytically, and often a Jensen's inequality approximation is leveraged to calculate this quantity. Some related works that use this approximation include \textcite{salter2013variational,gollini2016joint,jordan1999introduction}.
By contrast, one advantage of the MLPCM modeling framework is that the ELBO can be calculated exactly using the propositions below.
\begin{proposition}
Under the variational framework of Eq.~\eqref{Variational_distributions}, the first term $\mathbbm{E}_q[\text{\normalfont{log}}\;p(\bm{Y}|\bm{U},\bm{V}, \beta)]$ of the ELBO for the MLPCM is derived without any approximation as:
\begin{equation*}
\resizebox{1\hsize}{!}{$
\begin{split}
&\mathbbm{E}_q[\log p(\bm{Y}|\bm{U},\bm{V}, \beta)]\\
&=\sum^N_{\substack{i,j=1, \\ i\neq j}}\left\{ y_{ij}\left[\tilde{\eta}-||\bm{\tilde{u}}_i-\bm{\tilde{v}}_{j\leftarrow i}||^2-d\left(\tilde{\sigma}^2_i+\tilde{\varphi}^2_{ij}\right)\right]-\frac{\exp \left(\tilde{\eta}+\frac{\tilde{\rho}^2}{2}-\frac{||\bm{\tilde{u}}_i-\bm{\tilde{v}}_{j\leftarrow i}||^2}{1+2\tilde{\sigma}^2_i+2\tilde{\varphi}^2_{ij}}\right)}{(1+2\tilde{\sigma}^2_i+2\tilde{\varphi}^2_{ij})^{d/2}} - \log(y_{ij}!) \right\}.
\end{split}$}
\end{equation*}
\end{proposition}

\begin{proposition}
Under the variational framework of Eq.~\eqref{Variational_distributions}, the ELBO for the MLPCM can be written as:
\begin{equation}
\label{ELBO_MLPCM}
\begin{split}
&\mathcal{F}(\bm{\tilde{\Theta}})=
\sum^N_{\substack{i,j=1, \\ i\neq j}}\left\{ y_{ij}\left[\tilde{\eta}-||\bm{\tilde{u}}_i-\bm{\tilde{v}}_{j\leftarrow i}||^2-d\left(\tilde{\sigma}^2_i+\tilde{\varphi}^2_{ij}\right)\right]-\frac{\exp \left(\tilde{\eta}+\frac{\tilde{\rho}^2}{2}-\frac{||\bm{\tilde{u}}_i-\bm{\tilde{v}}_{j\leftarrow i}||^2}{1+2\tilde{\sigma}^2_i+2\tilde{\varphi}^2_{ij}}\right)}{(1+2\tilde{\sigma}^2_i+2\tilde{\varphi}^2_{ij})^{d/2}} \right\}\\
&+\sum^N_{\substack{i,j=1, \\ i\neq j}}\left\{\frac{d}{2}\left[\bm{\Psi}(\tilde{a}_j)-\log(\tilde{b}_j)\right]-\frac{1}{2}\frac{\tilde{a}_j}{\tilde{b}_j}[||\bm{\tilde{v}}_{j\leftarrow i}-\bm{\tilde{u}}_j||^2+d(\tilde{\varphi}^2_{ij}+\tilde{\sigma}^2_j)]+\frac{d}{2}\log (\tilde{\varphi}^2_{ij})\right\}\\
&+\sum^N_{i=1}\sum^K_{k=1}\tilde{\pi}_{ik}
\left\{\frac{d}{2}\left[\bm{\Psi}(\tilde{\xi}_k)-\log (\tilde{\psi}_k)\right]-\frac{1}{2}\frac{\tilde{\xi}_k}{\tilde{\psi}_k}\left[||\bm{\tilde{u}}_i-\bm{\tilde{\mu}}_k||^2+d(\tilde{\sigma}^2_i+\tilde{\omega}^2_k)\right]\right\}\\
&+\sum^N_{i=1}\sum^K_{k=1}\tilde{\pi}_{ik}\left[\bm{\Psi}\left(\tilde{\delta}_k\right)-\bm{\Psi}\left(\sum^K_{g=1}\tilde{\delta}_g\right)-\log(\tilde{\pi}_{ik})\right]\\
&+\sum^K_{k=1}\left\{(\delta_k-\tilde{\delta}_k)\left[\bm{\Psi}\left(\tilde{\delta}_{k}\right)-\bm{\Psi}\left(\sum^K_{g=1}\tilde{\delta}_{g}\right)\right]+\log\Gamma(\tilde{\delta}_k)
-\frac{1}{2\omega^2}\left(\bm{\tilde{\mu}}_k^T\bm{\tilde{\mu}}_k+d\tilde{\omega}^2_k\right)+\frac{d}{2}\log(\tilde{\omega}^2_k)\right\}\\
&+\sum^K_{k=1}\left\{(\xi-\tilde{\xi}_k)\bm{\Psi}\left(\tilde{\xi}_k\right)-\xi\log\tilde{\psi}_k-\psi\frac{\tilde{\xi}_k}{\tilde{\psi}_k}+\log\Gamma(\tilde{\xi}_k)+\tilde{\xi}_k\right\}\\
&+\sum^N_{i=1}\left\{(a-\tilde{a}_i)\bm{\Psi}\left(\tilde{a}_i\right)-a\log\tilde{b}_i-b\frac{\tilde{a}_i}{\tilde{b}_i}+\log\Gamma(\tilde{a}_i)+\tilde{a}_i+\frac{d}{2}\log (\tilde{\sigma}^2_i)\right\}\\
&-\frac{1}{2\rho^2}\left[(\tilde{\eta}-\eta)^2+\tilde{\rho}^2\right]+\frac{1}{2}\log(\tilde{\rho}^2)-\log \Gamma\left(\sum^K_{k=1}\tilde{\delta}_k\right)+\normalfont{\texttt{const}}.
\end{split}
\end{equation}
\end{proposition}
The detailed proofs of the calculations for each term of the ELBO are provided in Appendix~\ref{MLPCM_ELBO_Derivation_Appendix}.


\section{Maximization of the ELBO}
\label{sec:max}

In order to maximize the ELBO, we propose an algorithmic framework that combines closed form updates, standard gradient ascent and natural gradient ascent updates.
For some variational parameters, closed form updates are available.
For all the other parameters, we use standard gradient ascent or the Natural Gradient Ascent \parencite{amari1998natural,martens2020new}.


\subsection{Closed form updates}

In the following proposition, we provide the closed form update steps of the variational parameters: $\{\bm{\tilde{\Pi}}, \bm{\tilde{\mu}},\bm{\tilde{\omega}^2},\bm{\tilde{\xi}},\bm{\tilde{\psi}},\bm{\tilde{a}},\bm{\tilde{b}}\}$, which correspond to the variational distributions: $\{q_{\bm{z}}(z_i|\bm{\tilde{\Pi}}_i)\},\{q_{\bm{\mu}}(\bm{\mu}_k|\bm{\tilde{\mu}}_k,\tilde{\omega}^2_k)\},\{q_{\bm{\tau}}(\tau_k|\tilde{\xi}_k,\tilde{\psi}_k)\},\{q_{\bm{\gamma}}(\gamma_j|\tilde{a}_j,\tilde{b}_j)\}$ illustrated in Eq.\eqref{Variational_distributions}, respectively.
The detailed proofs are provided in Appendix~\ref{Proof_Proposition_Pi_MuOmega2_XiPsi_ab}.

\begin{proposition}
\label{Proposition_Pi_MuOmega2_XiPsi_ab}
For any values of $\bm{\tilde{\Theta}}$,
\begin{enumerate}
\item the closed form update of $\tilde{\pi}_{jk}$, for $j=1,2,\dots,N$ and for $k=1,2,\dots,K$, is written as: $\tilde{\pi}_{jk}^\star=\frac{\exp(A_{jk})}{\sum^K_{g=1}\exp(A_{jg})}$, where
\begin{equation}
\label{Pi_Analytical_Sol}
\resizebox{1\hsize}{!}{$
\begin{split}
&A_{jk}=\frac{d}{2}\left[\bm{\Psi}\left(\tilde{\xi}_k\right)-\log \left(\tilde{\psi}_k\right)\right]-\frac{1}{2}\frac{\tilde{\xi}_k}{\tilde{\psi}_k}\left[||\bm{\tilde{u}}_j-\bm{\tilde{\mu}}_k||^2+d(\tilde{\sigma}^{2}_j+\tilde{\omega}^{2}_k)\right]+\bm{\Psi}\left(\tilde{\delta}_k\right)-\bm{\Psi}\left(\sum^K_{g=1}\tilde{\delta}_g\right).
\end{split}$}
\end{equation}

\item the closed form updates of $\bm{\tilde{\mu}}_k$ and $\tilde{\omega}^{2}_k$, for $k=1,2,\dots,K$, are, respectively, written as:
\begin{equation}
\label{MuOmega2_Analytical_Sol}
\begin{split}
&\bm{\tilde{\mu}}_k^\star=
\frac{\omega^2\tilde{\xi}_k\left(\sum^N_{i=1}\tilde{\pi}_{ik}\bm{\tilde{u}}_i\right)}{\omega^2\tilde{\xi}_k\left(\sum^N_{i=1}\tilde{\pi}_{ik}\right)+\tilde{\psi}_k},\qquad\tilde{\omega}^{2^\star}_k=
\frac{\omega^2\tilde{\psi}_k}{\tilde{\psi}_k+\omega^2\tilde{\xi}_k\sum^N_{i=1}\tilde{\pi}_{ik}}.
\end{split}
\end{equation}

\item the closed form updates of $\tilde{\xi}_k$ and $\tilde{\psi}_k$, for $k=1,2,\dots,K$, are written as:
\begin{equation}
\label{XiPsi_Analytical_Sol}
\begin{split}
&\tilde{\xi}_k^\star=\xi+\frac{d}{2}\sum^N_{i=1}\tilde{\pi}_{ik},\qquad
\tilde{\psi}_k^\star=\psi+\frac{1}{2}\sum^N_{i=1}\tilde{\pi}_{ik}\left[||\bm{\tilde{u}}_i-\bm{\tilde{\mu}}_k||^2+d(\tilde{\sigma}^{2}_i+\tilde{\omega}^{2}_k)\right].
\end{split}
\end{equation}

\item the closed form updates of $\tilde{a}_j$ and $\tilde{b}_j$, for $j=1,2,\dots,N$, are written as:
\begin{equation}
\label{ab_Analytical_Sol}
\begin{split}
&\tilde{a}_j^\star=a+\frac{d}{2}(N-1),\qquad
\tilde{b}_j^\star=b+\frac{1}{2}\sum^N_{\substack{i=1, \\ i\neq j}}\left[||\bm{\tilde{v}}_{j\leftarrow i}-\bm{\tilde{u}}_j||^2+d(\tilde{\varphi}^{2}_{ij}+\tilde{\sigma}^{2}_j)\right].
\end{split}
\end{equation}
\end{enumerate}
\end{proposition}


\subsection{Natural gradient ascent and standard gradient ascent updates}

The standard gradient ascent is implemented as $\bm{\tilde{\theta}}^\star=\bm{\tilde{\theta}}^\dagger+\epsilon\grad\mathcal{F}(\bm{\tilde{\theta}}^\dagger)$. 
Here, $\bm{\tilde{\theta}}^\dagger$ denotes the current state of the variational parameters, whereas $\bm{\tilde{\theta}}^\star$ denotes their new updated state corresponding to a better value of the ELBO.
The natural gradient ascent further incorporates a Fisher information matrix $\mathcal{I}(\bm{\tilde{\theta}})$ which corrects the gradient by accounting for the underlying geometry of the parameter space, leading to more efficient optimizations.
Provided that the variational parameters $\bm{\tilde{\theta}}$ are raised from the corresponding variational distribution $q(\bm{\theta}|\bm{\tilde{\theta}})$ following Eqs.~\eqref{Variational_distributions}, the natural gradient ascent is implemented as $\bm{\tilde{\theta}}^\star=\bm{\tilde{\theta}}^\dagger+\epsilon\mathcal{I}^{-1}(\bm{\tilde{\theta}}^\dagger)\grad\mathcal{F}(\bm{\tilde{\theta}}^\dagger)$, where 
$$\mathcal{I}(\bm{\tilde{\theta}})=-\mathbbm{E}_{q(\bm{\theta}|\bm{\tilde{\theta}})}\left[\grad^2_{\bm{\tilde{\theta}}}\log q(\bm{\theta}|\bm{\tilde{\theta}})\right]
\equiv \mathbbm{E}_{q(\bm{\theta}|\bm{\tilde{\theta}})}\left\{\left[\grad_{\bm{\tilde{\theta}}}\log q(\bm{\theta}|\bm{\tilde{\theta}})\right]\left[\grad_{\bm{\tilde{\theta}}}\log q(\bm{\theta}|\bm{\tilde{\theta}})\right]^T\right\}.$$
In our case, the natural gradient steps are implemented following the propositions below to optimize the variational parameters $\{\bm{\tilde{U}},\bm{\tilde{\sigma}^2},\bm{\tilde{V}},\bm{\tilde{\varphi}^2},\tilde{\eta},\tilde{\rho}^2\}$, which correspond to the variational distributions $\{q_{\bm{U}}(\bm{u}_i|\bm{\tilde{u}}_i,\tilde{\sigma}^2_i)\},\{q_{\bm{V}}(\bm{v}_{j\leftarrow i}|\bm{\tilde{v}}_{j\leftarrow i},\tilde{\varphi}^2_{ij})\}$ and $q_{\beta}(\beta|\tilde{\eta},\tilde{\rho}^2)$, respectively.

\begin{proposition}
\label{Proposition_USigma2_VVarphi2_EtaRho2}
For any values of $\bm{\tilde{\Theta}}$ and for a small enough $\epsilon>0$,
\begin{enumerate}
\item the natural gradient ascent updates of $(\bm{\tilde{u}}_i,\tilde{\sigma}^2_i)$, for $i=1,2,\dots,N$, do not decrease the ELBO for the MLPCM, and are defined by:
\begin{equation}
\label{USigma2_NGA}
\begin{cases} 
\bm{\tilde{u}}_i^\star=\bm{\tilde{u}}_i^\dagger+\epsilon\tilde{\sigma}^{2^\dagger}_i\frac{\partial\mathcal{F}}{\partial\bm{\tilde{u}}_i}\left(\bm{\tilde{u}}_i^\dagger,\tilde{\sigma}^{2^\dagger}_i,\dots\right);\\  
\tilde{\sigma}^{2^\star}_i=\tilde{\sigma}^{2^\dagger}_i\exp[\epsilon\frac{2}{d}\tilde{\sigma}^{2^\dagger}_i\frac{\partial\mathcal{F}}{\partial\tilde{\sigma}^2_i}\left(\bm{\tilde{u}}_i^\dagger,\tilde{\sigma}^{2^\dagger}_i,\dots\right)],
\end{cases}
\end{equation}
where $(\cdot)^\star$ and $(\cdot)^\dagger$ denotes the new and current states of the variational parameters, respectively, and,
\begin{equation}
\label{USigma2_partial}
\resizebox{1\hsize}{!}{$
\begin{split}
&\frac{\partial\mathcal{F}}{\partial\bm{\tilde{u}}_i}
=\sum^N_{\substack{j=1, \\ j\neq i}}\left\{-2\left[y_{ij}-\frac{\exp \left(\tilde{\eta}+\frac{\tilde{\rho}^2}{2}-\frac{||\bm{\tilde{u}}_i-\bm{\tilde{v}}_{j\leftarrow i}||^2}{1+2\tilde{\sigma}^2_i+2\tilde{\varphi}^2_{ij}}\right)}{(1+2\tilde{\sigma}^2_i+2\tilde{\varphi}^2_{ij})^{d/2+1}}\right](\bm{\tilde{u}}_i-\bm{\tilde{v}}_{j\leftarrow i})+\frac{\tilde{a}_i}{\tilde{b}_i}(\bm{\tilde{v}}_{i\leftarrow j}-\bm{\tilde{u}}_i)\right\}-\sum^K_{k=1}\left[\tilde{\pi}_{ik}\frac{\tilde{\xi}_k}{\tilde{\psi}_k}(\bm{\tilde{u}}_i-\bm{\tilde{\mu}}_k)\right],\\
&\frac{\partial\mathcal{F}}{\partial\tilde{\sigma}^2_i}
=\frac{d}{2}\cdot\left[-2\left(\sum^N_{\substack{j=1, \\ j\neq i}}y_{ij}\right)-(N-1)\frac{\tilde{a}_i}{\tilde{b}_i}-\left(\sum^K_{k=1}\tilde{\pi}_{ik}\frac{\tilde{\xi}_k}{\tilde{\psi}_k}\right)+\frac{1}{\tilde{\sigma}^2_i}\right]-\sum^N_{\substack{j=1, \\ j\neq i}}
\frac{\exp \left(\tilde{\eta}+\frac{\tilde{\rho}^2}{2}-\frac{||\bm{\tilde{u}}_i-\bm{\tilde{v}}_{j\leftarrow i}||^2}{1+2\tilde{\sigma}^2_i+2\tilde{\varphi}^2_{ij}}\right)}{(1+2\tilde{\sigma}^2_i+2\tilde{\varphi}^2_{ij})^{d/2+1}}\left[2\frac{||\bm{\tilde{u}}_i-\bm{\tilde{v}}_{j\leftarrow i}||^2}{1+2\tilde{\sigma}^2_i+2\tilde{\varphi}^2_{ij}}-d\right].
\end{split}$}
\end{equation}

\item the natural gradient ascent updates of $(\bm{\tilde{v}}_{j\leftarrow i},\tilde{\varphi}^{2}_{ij})$, for $i,j=1,2,\dots,N$ and $i\neq j$, do not decrease the ELBO for the MLPCM, and are defined by:
\begin{equation}
\label{VVarphi2_NGA}
\begin{cases} 
\bm{\tilde{v}}_{j\leftarrow i}^\star=\bm{\tilde{v}}_{j\leftarrow i}^\dagger+\epsilon\tilde{\varphi}^{2^\dagger}_{ij}\frac{\partial\mathcal{F}}{\partial\bm{\tilde{v}}_{j\leftarrow i}}\left(\bm{\tilde{v}}_{j\leftarrow i}^\dagger,\tilde{\varphi}^{2^\dagger}_{ij},\dots\right);\\  
\tilde{\varphi}^{2^\star}_{ij}=\tilde{\varphi}^{2^\dagger}_{ij}\exp[\epsilon\frac{2}{d}\tilde{\varphi}^{2^\dagger}_{ij}\frac{\partial\mathcal{F}}{\partial\tilde{\varphi}^2_{ij}}\left(\bm{\tilde{v}}_{j\leftarrow i}^\dagger,\tilde{\varphi}^{2^\dagger}_{ij},\dots\right)],
\end{cases}
\end{equation}
where,
\begin{equation*}
\begin{split}
&\frac{\partial\mathcal{F}}{\partial\bm{\tilde{v}}_{j\leftarrow i}}
=\left\{2\left[y_{ij}-\frac{\exp \left(\tilde{\eta}+\frac{\tilde{\rho}^2}{2}-\frac{||\bm{\tilde{u}}_i-\bm{\tilde{v}}_{j\leftarrow i}||^2}{1+2\tilde{\sigma}^2_i+2\tilde{\varphi}^2_{ij}}\right)}{(1+2\tilde{\sigma}^2_i+2\tilde{\varphi}^2_{ij})^{d/2+1}}\right](\bm{\tilde{u}}_i-\bm{\tilde{v}}_{j\leftarrow i})-\frac{\tilde{a}_j}{\tilde{b}_j}(\bm{\tilde{v}}_{j\leftarrow i}-\bm{\tilde{u}}_j)\right\},\\
&\frac{\partial\mathcal{F}}{\partial\tilde{\varphi}^2_{ij}}=
\frac{d}{2}\left(-2y_{ij}-\frac{\tilde{a}_j}{\tilde{b}_j}+\frac{1}{\tilde{\varphi}^2_{ij}}\right)
-\frac{
\exp \left(\tilde{\eta}+\frac{\tilde{\rho}^2}{2}-\frac{||\bm{\tilde{u}}_i-\bm{\tilde{v}}_{j\leftarrow i}||^2}{1+2\tilde{\sigma}^2_i+2\tilde{\varphi}^2_{ij}}\right)\left[2\frac{||\bm{\tilde{u}}_i-\bm{\tilde{v}}_{j\leftarrow i}||^2}{1+2\tilde{\sigma}^2_i+2\tilde{\varphi}^2_{ij}}-d\right]
}{(1+2\tilde{\sigma}^2_i+2\tilde{\varphi}^2_{ij})^{d/2+1}}.
\end{split}
\end{equation*}

\item the natural gradient ascent updates of $(\tilde{\eta},\tilde{\rho}^{2})$ do not decrease the ELBO for the MLPCM, and are defined by:
\begin{equation}
\label{EtaRho2_NGA}
\begin{split}
\begin{cases} 
\tilde{\eta}^\star=\tilde{\eta}^\dagger+\epsilon\tilde{\rho}^{2^\dagger}\frac{\partial\mathcal{F}}{\partial\tilde{\eta}}\left(\tilde{\eta}^\dagger,\tilde{\rho}^{2^\dagger},\dots\right);\\  
\tilde{\rho}^{2^\star}=\tilde{\rho}^{2^\dagger}\exp[\epsilon2\tilde{\rho}^{2^\dagger}\frac{\partial\mathcal{F}}{\partial\tilde{\rho}^2}\left(\tilde{\eta}^\dagger,\tilde{\rho}^{2^\dagger},\dots\right)],
\end{cases}
\end{split}
\end{equation}
where,
\begin{equation*}
\begin{split}
&\frac{\partial\mathcal{F}}{\partial\tilde{\eta}}
=\left(\sum^N_{\substack{i,j=1, \\ i\neq j}}y_{ij}\right)-\frac{\tilde{\eta}-\eta}{\rho^2}-
\sum^N_{\substack{i,j=1, \\ i\neq j}}\frac{\exp \left(\tilde{\eta}+\frac{\tilde{\rho}^2}{2}-\frac{||\bm{\tilde{u}}_i-\bm{\tilde{v}}_{j\leftarrow i}||^2}{1+2\tilde{\sigma}^2_i+2\tilde{\varphi}^2_{ij}}\right)}{(1+2\tilde{\sigma}^2_i+2\tilde{\varphi}^2_{ij})^{d/2}},\\
&\frac{\partial\mathcal{F}}{\partial\tilde{\rho}^2}
=-\frac{1}{2\rho^2}+\frac{1}{2\tilde{\rho}^2}-
\frac{1}{2}\sum^N_{\substack{i,j=1, \\ i\neq j}}\frac{\exp \left(\tilde{\eta}+\frac{\tilde{\rho}^2}{2}-\frac{||\bm{\tilde{u}}_i-\bm{\tilde{v}}_{j\leftarrow i}||^2}{1+2\tilde{\sigma}^2_i+2\tilde{\varphi}^2_{ij}}\right)}{(1+2\tilde{\sigma}^2_i+2\tilde{\varphi}^2_{ij})^{d/2}}.
\end{split}
\end{equation*}

\end{enumerate}
\end{proposition}

The proofs of the above propositions are detailed in Appendix~\ref{Proof_Proposition_USigma2_VVarphi2_EtaRho2}.
Note that the natural gradient ascent requires the evaluation of the inverse Fisher information.
This can be easily derived if the Fisher information matrix is diagonal, which is not always the case.
Thus, the calculation of the inverse matrix poses a computational problem as it can be very time consuming.
For those updates where the Fisher information matrix is not diagonal, we resort to a standard gradient ascent for the optimization.
This is for example the case for the variational parameter $\bm{\tilde{\delta}}$, associated to the variational distribution $q_{\bm{\Pi}}(\bm{\Pi}|\bm{\tilde{\delta}})$, which we specifically address in the proposition here below (the proofs are provided in Appendix~\ref{Proof_Proposition_delta}).

\begin{proposition}
\label{Proposition_delta}
For any values of $\bm{\tilde{\Theta}}$ and for a small enough $\epsilon>0$, the standard gradient ascent updates of $(\tilde{\delta}_1,\tilde{\delta}_2,\dots,\tilde{\delta}_K)$ do not decrease the ELBO for the MLPCM, and are defined by:
\begin{equation}
\label{delta_SGA}
\begin{split}
&\tilde{\delta}_k^\star=\tilde{\delta}_k^\dagger\exp[\epsilon\tilde{\delta}_k^\dagger\frac{\partial\mathcal{F}}{\partial\tilde{\delta}_k}\left(\tilde{\delta}_1^\dagger,\tilde{\delta}_2^\dagger,\dots,\tilde{\delta}_K^\dagger,\dots\right)],\;\text{for}\;k=1,2,\dots,K,
\end{split}
\end{equation}
where $(\cdot)^\star$ and $(\cdot)^\dagger$ denotes the new and current states of the variational parameters, respectively, and,
\begin{equation*}
\begin{split}
&\frac{\partial\mathcal{F}}{\partial\tilde{\delta}_k}
=\bm{\Psi}'\left(\tilde{\delta}_k\right)\cdot\left(\delta_k-\tilde{\delta}_k+\sum^N_{i=1}\tilde{\pi}_{ik}\right)-
\bm{\Psi}'\left(\sum^K_{g=1}\tilde{\delta}_g\right)\cdot
\left[\sum^K_{g=1}\left(\delta_g-\tilde{\delta}_g+\sum^N_{i=1}\tilde{\pi}_{ig}\right)\right].
\end{split}
\end{equation*}
\end{proposition}


\section{Optimization algorithm and model selection}
\label{Algorithm_and_PICL}

\subsection{Optimization}

The variational Bayes optimization algorithm that results from the variational updates is outlined in Algorithm~\ref{VB_algorithm_MLPCM}.
\begin{algorithm}[htpb!]
\caption{The variational Bayes inference algorithm for the MLPCM.}
\label{VB_algorithm_MLPCM}
\begin{algorithmic} 
\State \textbf{Input}: $\bm{Y},d,K,a,b,\eta,\rho^2,\omega^2,\xi,\psi,\bm{\delta}$,
the tolerance level \texttt{tol}, 
the gradient ascent step sizes $\{\epsilon_{\bm{u}_i}\}^N_{i=1},\{\epsilon_{\bm{v}_{j\leftarrow i}}\}^N_{i,j=1,i\neq j},
\epsilon_{\beta},\epsilon_{\bm{\Pi}}$,
and the initial state: $$\bm{\tilde{\Theta}}^{(0)}=\{\bm{\tilde{U}}^{(0)},\bm{\tilde{\sigma}^{2^{\textnormal{(0)}}}},\bm{\tilde{V}}^{(0)},\bm{\tilde{\varphi}^{2^{\textnormal{(0)}}}},\bm{\tilde{\Pi}}^{(0)},\tilde{\eta}^{(0)},\tilde{\rho}^{2^{\textnormal{(0)}}},\bm{\tilde{\mu}}^{(0)},\bm{\tilde{\omega}^{2^{\textnormal{(0)}}}},\bm{\tilde{\xi}}^{(0)},\bm{\tilde{\psi}}^{(0)},\bm{\tilde{a}}^{(0)},\bm{\tilde{b}}^{(0)},\bm{\tilde{\delta}}^{(0)}\}.$$
Set \texttt{stop}=\texttt{FALSE} and $t=0$. 
Calculate the ELBO at the initial state: $\mathcal{F}(\bm{\tilde{\Theta}}^{(0)})$. 

\While {\texttt{stop} $\neq$ \texttt{TRUE}}

\For {$i = 1,\dots,N$}
\State \textcolor{red}{1.} Update $(\bm{\tilde{u}}_i^{(t+1)},\tilde{\sigma}_i^{2^{(t+1)}})$ following Eq.~\eqref{USigma2_NGA} with $\epsilon=\epsilon_{\bm{u}_i}$.
\If {$\mathcal{F}(\bm{\tilde{u}}_i^{(t+1)},\tilde{\sigma}^{2^{(t+1)}}_i,\dots)<\mathcal{F}(\bm{\tilde{u}}_i^{(t)},\tilde{\sigma}^{2^{(t)}}_i,\dots)$}
$\epsilon_{\bm{u}_i}=\epsilon_{\bm{u}_i}/2$; rerun Step~\textcolor{red}{1}.
\EndIf
\EndFor

\For {$i,j = 1,\dots,N$ and $i\neq j$}
\State \textcolor{red}{2.} Update $(\bm{\tilde{v}}_{j\leftarrow i}^{(t+1)},\tilde{\varphi}_{ij}^{2^{(t+1)}})$ following Eq.~\eqref{VVarphi2_NGA} with $\epsilon=\epsilon_{\bm{v}_{j\leftarrow i}}$.
\If {$\mathcal{F}(\bm{\tilde{v}}_{j\leftarrow i}^{(t+1)},\tilde{\varphi}_{ij}^{2^{(t+1)}},\dots)<\mathcal{F}(\bm{\tilde{v}}_{j\leftarrow i}^{(t)},\tilde{\varphi}_{ij}^{2^{(t)}},\dots)$}
$\epsilon_{\bm{v}_{j\leftarrow i}}=\epsilon_{\bm{v}_{j\leftarrow i}}/2$; rerun Step~\textcolor{red}{2}.
\EndIf
\EndFor

\For {$j = 1,\dots,N$}
\For {$k = 1,\dots,K$}
\State \textcolor{red}{3.} Update $\tilde{\pi}_{jk}^{(t+1)}$ following Eq.~\eqref{Pi_Analytical_Sol}.
\EndFor
\State \textcolor{red}{4.} Update $(\tilde{a}_j^{(t+1)},\tilde{b}_j^{(t+1)})$ following Eq.~\eqref{ab_Analytical_Sol}.
\EndFor

\State \textcolor{red}{5.} Update $(\tilde{\eta}^{(t+1)},\tilde{\rho}^{2^{(t+1)}})$ following Eq.~\eqref{EtaRho2_NGA} with $\epsilon=\epsilon_{\beta}$.
\If {$\mathcal{F}(\tilde{\eta}^{(t+1)},\tilde{\rho}^{2^{(t+1)}},\dots)<\mathcal{F}(\tilde{\eta}^{(t)},\tilde{\rho}^{2^{(t+1)}},\dots)$}
$\epsilon_{\beta}=\epsilon_{\beta}/2$, Step~\textcolor{red}{5}.
\EndIf

\For {$k = 1,\dots,K$}
\State \textcolor{red}{6.} Update $\bm{\tilde{\mu}}_k^{(t+1)}$ and $\tilde{\omega}_k^{2^{(t+1)}}$ following Eq.~\eqref{MuOmega2_Analytical_Sol}.
\State \textcolor{red}{7.} Update $(\tilde{\xi}_k^{(t+1)},\tilde{\psi}_k^{(t+1)})$ following Eq.~\eqref{XiPsi_Analytical_Sol}.
\State \textcolor{red}{8.} Update $\tilde{\delta}_k^{(t+1)}$ following Eq.~\eqref{delta_SGA} with $\epsilon=\epsilon_{\bm{\Pi}}$.
\EndFor
\If {$\mathcal{F}(\bm{\tilde{\delta}}^{(t+1)},\dots)<\mathcal{F}(\bm{\tilde{\delta}}^{(t)},\dots)$}
$\epsilon_{\bm{\Pi}}=\epsilon_{\bm{\Pi}}/2$, rerun Step~\textcolor{red}{8} for $k = 1,\dots,K$.
\EndIf

\If {$\mathcal{F}(\bm{\tilde{\Theta}}^{(t+1)})-\mathcal{F}(\bm{\tilde{\Theta}}^{(t)})<\texttt{tol}$}
\texttt{stop}=\texttt{TRUE}
\EndIf

\hspace{-0.8em} Set $t=t+1$.
\EndWhile

\State \textbf{Output}:
$\{\bm{\tilde{\Theta}}^{(s)},\mathcal{F}(\bm{\tilde{\Theta}}^{(s)}): s = 1,2,\dots,t\}$.
\end{algorithmic}
\end{algorithm}
The learning rates for each of the variational parameters are initially set an arbitrary large value.
Subsequently, the value of a learning rate gets halved whenever a natural gradient update leads to a decrease in the ELBO.
In this way, the procedure in Algorithm~\ref{VB_algorithm_MLPCM} ensures the non-decreasing of the ELBO, and a maximum ELBO value can be obtained if the tolerance level \texttt{tol} is set to be small enough.
However, such a maximization is not guaranteed to lead to a global maximum of the ELBO, so multiple implementations with random initializations of the parameters can be beneficial in order to validate the output.\\

\subsection{Model choice}

Algorithm~\ref{VB_algorithm_MLPCM} runs on a predetermined number of clusters, $K$, as input.
In order to infer the best value for $K$, we run the algorithm with different candidate values of $K$, and then compare the solutions to select the best one.
As a criterion to choose the ideal number of groups, we develop a novel Partially Integrated Complete Likelihood (PICL) score for the model selection.
Such a criterion combines the idea of the Integrated Complete Likelihood (ICL) as well as that of the Exact ICL ($\text{ICL}_{ex}$) of \textcite{biernacki2010exact,come2015model}.
The PICL is suitable when the conjugacy of the likelihoods and priors can lead to a partially ``collapsed'' posterior, where some of the model parameters have been integrated out.

To be more specific, we start by considering the integrated log-likelihood of $\bm{Y},\bm{U}, \bm{V}, \bm{z}|K$ for the MLPCM, which reads as follows:
\begin{equation}
\label{MLPCM_Integrated_Log_likelihood}
\begin{split}
&\log p(\bm{Y},\bm{U}, \bm{V}, \bm{z}|K)=
\log[p(\bm{Y}|\bm{U}, \bm{V}, \bm{z},K)p(\bm{U}, \bm{V}, \bm{z}|K)]\\
&=\log[p(\bm{Y}|\bm{U}, \bm{V})p(\bm{V}|\bm{U}, \bm{z},K)p(\bm{U}, \bm{z}|K)]=
\log[p(\bm{Y}|\bm{U}, \bm{V})p(\bm{V}|\bm{U})p(\bm{U}|\bm{z})p(\bm{z}|K)]\\
&=\log\int_{\beta}p(\bm{Y}|\bm{U}, \bm{V},\beta)\pi(\beta)\text{d}\beta +
\log \int_{\bm{\gamma}}p(\bm{V}|\bm{U},\bm{\gamma})\pi(\bm{\gamma})\text{d}\bm{\gamma}\\
&\hspace{1em} + 
\log\int_{\bm{\tau}}\int_{\bm{\mu}}p(\bm{U}|\bm{z},\bm{\mu},\bm{\tau})\pi(\bm{\mu})\pi(\bm{\tau})\text{d}\bm{\mu}\text{d}\bm{\tau}+ 
\log\int_{\bm{\Pi}}p(\bm{z}|\bm{\Pi})\pi(\bm{\Pi}|K)\text{d}\bm{\Pi}.
\end{split}
\end{equation}
From Eq.~\ref{MLPCM_Integrated_Log_likelihood}, we note that: (i) the parameter $\beta$ in the first term cannot be easily integrated out, and thus we apply a Bayesian Information Criterion (BIC) approximation \parencite{schwarz1978estimating} to evaluate this term;
(ii) the model parameters $\bm{\gamma}$ and $\bm{\Pi}$ in the second and fourth terms can be easily integrated out thanks to conjugate priors;
(iii) $\bm{\mu}$ or $\bm{\tau}$ may be integrated out, so we propose to marginalize with respect to the parameter $\bm{\mu}$, which has higher dimension than $\bm{\tau}$, and then we can apply the BIC approximation on the remaining partially integrated log-likelihood term.

After the marginalizations of some of the parameters, we obtain a partially integrated ICL which we can use for model choice. The equation for this criterion is shown in the proposition below.

\begin{proposition}
\label{Proposition_PICL}
Given point estimates of the model parameters $\bm{\hat{U}}, \bm{\hat{V}}, \bm{\hat{z}}$ for the MLPCM, the PICL criteria of Eq.~\eqref{MLPCM_Integrated_Log_likelihood} is equal to:
\begin{equation*}
\begin{split}
&PICL=\normalfont{\texttt{const}}-\left(\sum^N_{\substack{i,j=1, \\ i\neq j}}y_{ij}\right)\log\left[\sum^N_{\substack{i,j=1, \\ i\neq j}}\exp(-||\bm{\hat{u}}_i-\bm{\hat{v}}_{j\leftarrow i}||^2)\right]
-\sum^N_{\substack{i,j=1, \\ i\neq j}}\left(y_{ij}||\bm{\hat{u}}_i-\bm{\hat{v}}_{j\leftarrow i}||^2\right)\\
&-\left[a+\frac{d}{2}(N-1)\right]\sum^N_{j=1}\log(b+\frac{1}{2}\sum^N_{\substack{i=1, \\ i\neq j}}||\bm{\hat{v}}_{j\leftarrow i}-\bm{\hat{u}}_j||^2) -\frac{1}{2}K\log(N)\\
&+\max_{\bm{\tau}} \sum^K_{g=1} \left[
  \frac{d}{2}\hat{n}_g\log(\tau_g)-\frac{d}{2}\log\left(\tau_g\hat{n}_g\omega^2+1\right)+
  \frac{1}{2} \frac{\tau_g^2\omega^2}{\tau_g\hat{n}_g\omega^2+1}\norm{\sum_{\hat{z}_i=g}\bm{\hat{u}}_i}^2
  - \frac{\tau_g}{2} \left(\sum_{\hat{z}_i=g}||\bm{\hat{u}}_i||^2\right)
\right]\\
&+\left[\sum^K_{k=1}\log\Gamma(\hat{n}_k+\delta)\right]-\log\Gamma\left(N+K\delta\right)+\log\Gamma\left(K\delta\right)-K\log\Gamma(\delta),
\end{split}
\end{equation*}
where,
\begin{equation*}
\begin{split}
&\normalfont{\texttt{const}}=\left(\sum^N_{\substack{i,j=1, \\ i\neq j}}y_{ij}\right) 
\left[\log(\sum^N_{\substack{i,j=1, \\ i\neq j}}y_{ij})-1\right]
-\left[\sum^N_{\substack{i,j=1, \\ i\neq j}}\log(y_{ij}!)\right] - \frac{1}{2}\log[N(N-1)]\\
&\hspace{1em}+\sum^N_{j=1}\left[-\frac{d}{2}(N-1)\log(2\pi)+a\log(b)-\log\Gamma(a)+\log\Gamma\left(a+\frac{d}{2}(N-1)\right)\right] - \frac{d}{2}N\log(2\pi),
\end{split}
\end{equation*}
and the $\max_{\bm{\tau}}(\cdot)$ term is obtained by the standard gradient ascent for $\{\tau_g:g=1,2,\dots,K\}$:
\begin{equation*}
\begin{split}
&\tau_g^\star=\tau_g^\dagger\exp[\epsilon\tau_g^\dagger\frac{\partial h}{\partial \tau_g}\left(\tau_g^\dagger\right)].
\end{split}
\end{equation*}
Here, let $\hat{n}_g:=\sum_{i=1}^N\mathbbm{1}(\hat{z}_i=g)$, then the partial derivative is written as
\begin{equation*}
\begin{split}
&\frac{\partial h(\bm{\tau})}{\partial \tau_g}=\frac{d}{2}\hat{n}_g\left(\frac{1}{\tau_g}-\frac{\omega^2}{\tau_g\hat{n}_g\omega^2+1}\right)
+\frac{\tau_g\omega^2\left(\tau_g\hat{n}_g\omega^2+2\right)}{2\left(\tau_g\hat{n}_g\omega^2+1\right)^2}\norm{\sum_{\hat{z}_i=g}\bm{\hat{u}}_i}^2
-\frac{1}{2} \left(\sum_{\hat{z}_i=g}||\bm{\hat{u}}_i||^2\right).
\end{split}
\end{equation*}
\end{proposition}

Following from the PICL definition, and thanks to its close connection to other widely used penalized likelihood model-choice criteria, we select the optimal value $K$ as the one that maximizes the PICL.

\subsection{Initialization}\label{sec:initialization}
Variational Bayes methods can suffer from convergence issues, since the optimization procedure may just lead to an approximate and local posterior mode.
Re-running the algorithm multiple times using different random starting points may help with convergence issues, however this comes at a higher computational cost, since the algorithm has to be run many times. This can become impractical in the case of a complex model with a high dimensional parameter space, because randomization will often propose poor initial states, making the inference algorithm more likely to converge to some local posterior modes.
Thus, a good initialization method is essential for efficient explorations of the solution space and to obtain good inference results.
In this paper, we suggest the following default initialization approach which yields good performance in both simulated and real data applications.

The variational parameter $\bm{\tilde{U}}$ is initialized by applying the classical metric Multidimensional Scaling (MDS) of \textcite{gower1966some} on the geodesic matrix of $\bm{Y}$, whereas $\{\bm{\tilde{v}}_{j\leftarrow i}:i=1,2,\dots,N;i\neq j\}$ are initialized to be in common with the corresponding $\bm{\tilde{u}}_j$ for $j=1,2,\dots,N$.
By applying the k-means algorithm on the initial $\bm{\tilde{U}}$ with 1000 random starts, an initial approximate grouping is obtained based on minimized sum of squares from individuals to the corresponding group centres, and the variational parameter $\bm{\tilde{\mu}}$ is thus initialized as the corresponding group centers based on $\bm{\tilde{U}}$.
The initial values of the variational parameters $\{\tilde{\eta},\tilde{\rho}^{2},\bm{\tilde{\omega}^{2}},\bm{\tilde{\xi}},\bm{\tilde{\psi}},\bm{\tilde{a}},\bm{\tilde{b}},\bm{\tilde{\delta}}\}$ are set to align with the corresponding prior settings $\{\eta,\rho^{2},\bm{\omega^{2}},\bm{\xi},\bm{\psi},\bm{a},\bm{b},\bm{\delta}\}$, while the elements of initial $\{\bm{\tilde{\sigma}^{2}},\bm{\tilde{\varphi}^{2}}\}$ are set to a default value of $1$.
The variational parameter $\bm{\tilde{\Pi}}_i=\{\tilde{\pi}_{i1},\tilde{\pi}_{i2},\dots,\tilde{\pi}_{i\bar{K}}\}$ is initialized for $i=1,2,\dots,N$ by applying Step $3$ of Algorithm~\ref{VB_algorithm_MLPCM} conditional on all other initial parameters.


\section{Simulation studies}
\label{SS}

We propose two simulation studies to explore the performance of the newly proposed MLPCM as well as the variational Bayes inference method.
In this section, along with our MLPCM, we make use of another type of LPM: the PoisLPCM of \textcite{lu2025zero}.
This will serve as a comparison tool but also as an alternative model to generate data from.
In order to make a fairer comparison between the estimation algorithms for MLPCM and PoisLPCM, we derived a variational Bayes for the PoisLPCM to replace the Markov chain Monte Carlo methodology of \textcite{lu2025zero}.
This alternative estimation framework for the PoisLPCM is outlined in Appendix~\ref{Poisson_LPCM_Section}.

The first simulation study is designed to have two different scenarios: the synthetic network in scenario $1$ is randomly generated from a MLPCM with multiple groups which share the same value of precision, $\gamma_j$, for each $j=1,2,\dots,N$.
By contrast, the network in scenario $2$ is randomly generated from a PoisLPCM which has model settings that are similar to those used in scenario $1$ but without relying on the mixed latent positions framework.
The second simulation study instead focuses on a more challenging situation: we randomly generate an artificial network from a MLPCM where some groups show different degrees of overlap, making it more difficult to distinguish them.
In addition, the dispersion parameters $\boldsymbol{\gamma}$ are set to be different for each of the clusters.

We leverage our newly proposed PICL criteria provided in Propositions~\ref{Proposition_PICL} and \ref{Proposition_PICL_PoisLPCM} 
to choose the best $K$ compared to the true number of clusters $K^*$ used for generating those artificial networks.
Throughout this paper, the tolerance level \texttt{tol} is set to be 0.01 for all the variational Bayes inference by default.


\subsection{Simulation study 1}
\label{SS1}

In all simulation studies we focus on networks of $N=100$ nodes, and we assume that these nodes are evenly clustered into $K^*=4$ groups in block-order from node $1$ to node $100$.
So, the true clustering $\bm{z}^*$ leads to group sizes: $n_1=n_2=n_3=n_4=25$.
The synthetic network in scenario 1 is randomly simulated from a MLPCM with settings: $d=2, \beta^*=1, \bm{\tau}^*=(8,6,4,2)^T$, and with $\gamma_j^*=10$ for each $j=1,2,\dots,N$. The group centers are proposed to be:
\begin{equation*}
\boldsymbol{\mu}^* = \left[
\begin{pmatrix}
1.25 \\1.25
\end{pmatrix},
\begin{pmatrix}
1.25 \\-1.25
\end{pmatrix},
\begin{pmatrix}
-1.25 \\-1.25
\end{pmatrix},
\begin{pmatrix}
-1.25 \\1.25
\end{pmatrix}\right].
\end{equation*}
From here onwards, we use $(\cdot)^*$ to denote the reference values of the model parameters or variables used for generating the networks to be compared to the point estimates we obtain from the variational Bayes algorithm output.

\begin{figure}[htbp!]
\centering
\includegraphics[scale=0.525]{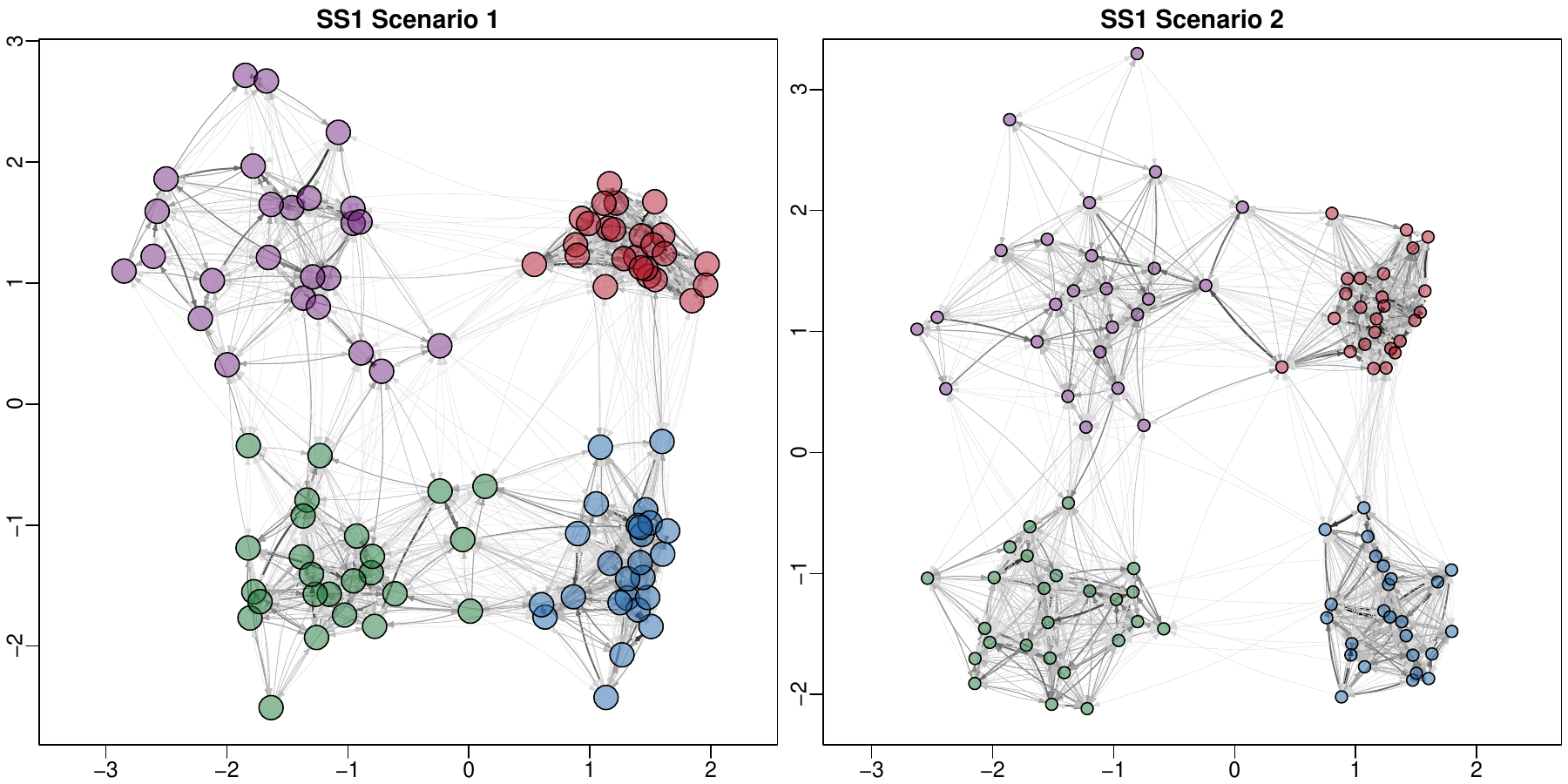}
\caption{Simulation study 1 synthetic networks. Node positions correspond to the latent positions $\bm{U}^*$ used for simulating the networks. Left panel: scenario 1 simulated through a MLPCM, where node sizes are proportional to $\{1/\gamma_j^*\}$. Right panel: scenario 2 simulated through a PoisLPCM, where node sizes are constant. Edge widths and colors are proportional to edge weights.}
\label{SS1LatentPositionsRef}
\end{figure}

\begin{figure}[htbp!]
\centering
\includegraphics[scale=0.525]{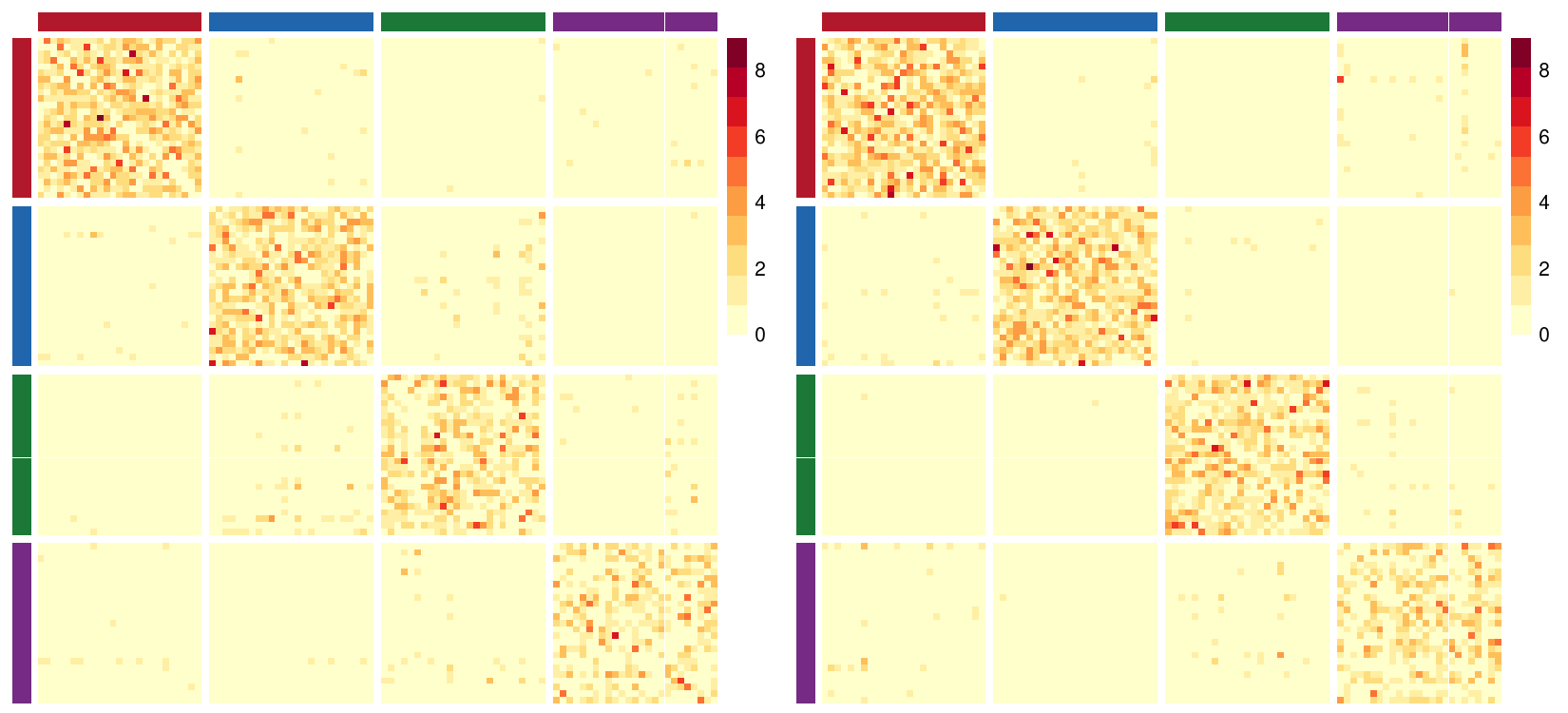}
\caption{Simulation study 1 adjacency matrices. Darker squares correspond to higher edge weights. The gaps, which are used for separating blocks, and the sidebar colors illustrate the true clustering $\bm{z}^*$. Left panel: scenario 1 simulated through a MLPCM. Right panel: scenario 2 simulated through a PoisLPCM.}
\label{SS1_adj_heatmaps}
\end{figure}

Similarly, for scenario $2$ we create a PoisLPCM network with the same settings as above, excluding the mixed latent positions' precision $\bm{\gamma}^*$ since this model does not include this feature.
The latent space representations and the adjacency matrices of the scenarios $1$ and $2$ are illustrated in Figures~\ref{SS1LatentPositionsRef} and \ref{SS1_adj_heatmaps}, respectively.
Networks simulated from a MLPCM tend to be generally sparser than those simulated from a PoisLPCM model with similar settings.
This is a consequence of the fact that the overt latent positions are created as perturbed covert positions, and so they add an additional layer of dispersion which translates into larger distances between nodes.
Although not immediately obvious, this is actually reflected in Figures~\ref{SS1LatentPositionsRef} and \ref{SS1_adj_heatmaps}, where the PoisLPCM network has a mean weight of $0.3734$ compared to a mean weight of $0.3403$ for the MLPCM.

\begin{figure}[htbp!]
\centering
\includegraphics[scale=0.625]{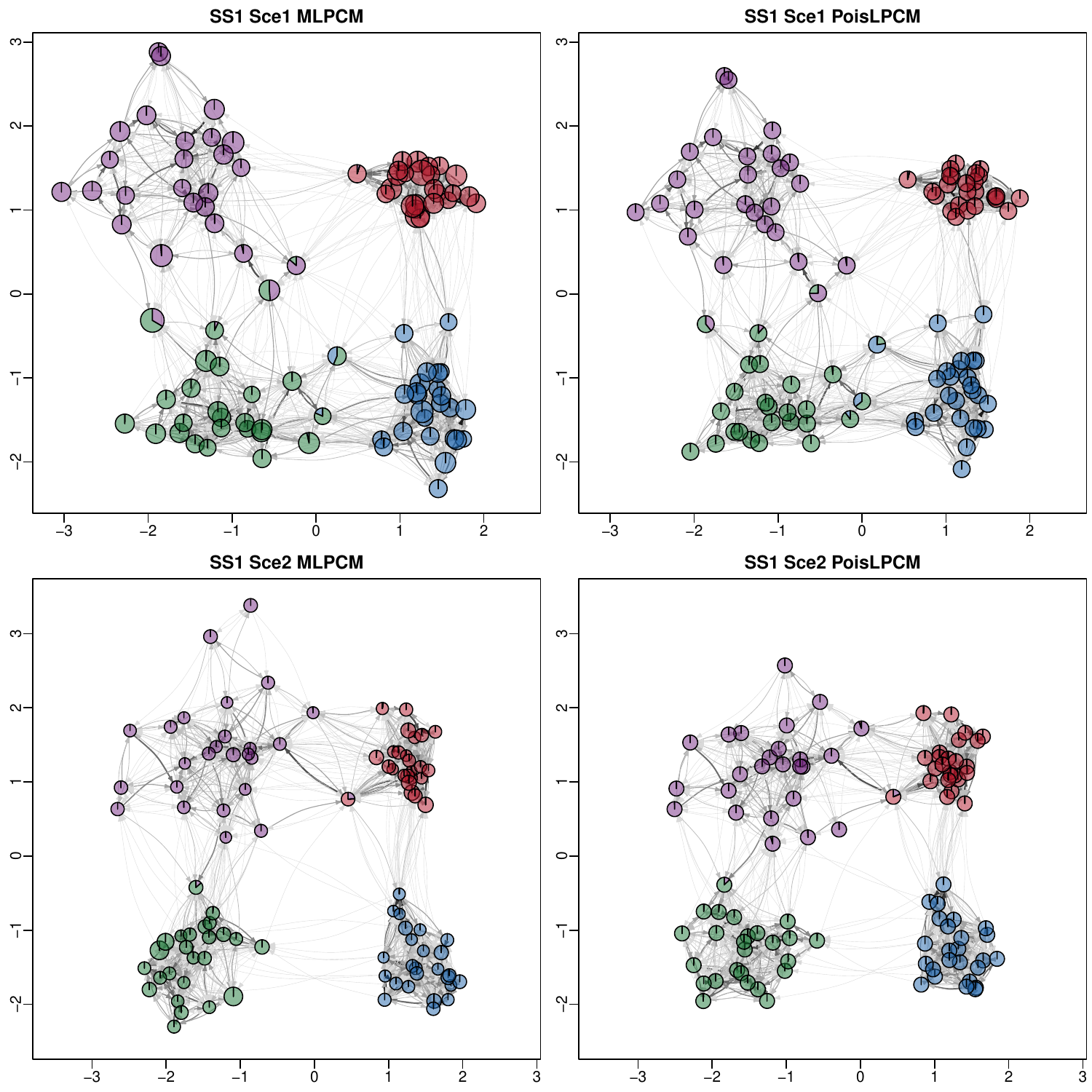}
\caption{Simulation study 1. Inferred latent spaces provided by the MLPCM (left) and the PoisLPCM (right) for scenario $1$ (top) and $2$ (bottom). Nodes are positioned according to $\bm{\hat{U}}$, and each node pie chart illustrates the corresponding inferred variational probability of group assignments. For the MLPCM plots, the size of nodes is proportional to the inferred $\{1/\hat{\gamma}_j=\tilde{b}_j/\tilde{a}_j\}$ that approximates the dispersion of the overt positions. Node sizes for PoisLPCM are proportional to a constant. Edge widths and colors are proportional to edge weights.
}
\label{SS1_InferredLatentRepresentations}
\end{figure}

As candidate values for the number of clusters, we consider $K=2,3,4,5,6$ for both the variational Bayes algorithms, in both scenarios.
We use the approximate posterior means of the models' parameters as point estimates.
The PICL values shown in Table~\ref{SS1PICLtable} are evaluated using the point estimates $\bm{\hat{U}}$ and $\bm{\hat{V}}$ for each fixed $K$ and for each model in scenarios 1 and 2.
\setlength{\tabcolsep}{12pt}
\begin{table}[htbp!]
\renewcommand{\arraystretch}{0.9}
\centering
\caption{\footnotesize{Simulation study 1. The PICL criteria values for number of groups ranging from $K=2$ to $K=6$. The bold value in each row indicates the highest PICL value.}}
\begin{adjustbox}{width=0.75\textwidth,center=\textwidth}
\begin{tabular}[c]{l|c|c|c|c|c}
\midrule
  & K=2 & K=3 & K=4 & K=5 & K=6     \\
\midrule
Scenario 1 MLPCM & 13,457 & 14,705  & \textbf{14,724} & 14,722  & 14,720\\
Scenario 1 PoisLPCM & -4197.2 & -4161.3  & \textbf{-4136.8} & -4141.1  & -4146.5\\
\midrule
Scenario 2 MLPCM & 14,750 & 14,782  & \textbf{14,810} & 14,806  & 14,802\\
Scenario 2 PoisLPCM & -4452.6 & -4167.5 & \textbf{-4134.8} & -4140.3  & -4226.2\\
\midrule
\end{tabular}
\end{adjustbox}
\label{SS1PICLtable}
\end{table}
The estimate of $\bm{\hat{z}}$ corresponds to the maximum a posteriori cluster memberships.
Table~\ref{SS1PICLtable} shows that the PICL criteria successfully recovers the correct number of clusters $K^*=4$ for both the MLPCM and the PoisLPCM in both scenarios.
Thus, from here onwards we only focus on the outputs from the runs of the algorithms where $K=4$, and we show the model fitting results in Table~\ref{SS1table} and in Figure~\ref{SS1_InferredLatentRepresentations}.
\setlength{\tabcolsep}{12pt}
\begin{table}[htbp!]
\renewcommand{\arraystretch}{0.9}
\centering
\caption{\footnotesize{Simulation study 1. Performance of MLPCM and PoisLPCM in scenarios 1 and 2.
The summary statistics are
(\lowerromannumeral{1}) $(\tilde{\eta}, \tilde{\rho}^2)$: the inferred variational parameters from $q_{\beta}$;
(\lowerromannumeral{2}) $\mathbbm{E}(\{|{\hat{d}_{ij}}^2-{d^{*2}_{ij}}|\})${\scriptsize [\text{sd}]}: the mean absolute error for squared pairwise distances with the corresponding standard deviation;
(\lowerromannumeral{3}) $\mathbbm{E}(\{\tilde{\sigma}^2_i\})${\scriptsize [\text{sd}]}: the mean and sd of the variational variance associated to the covert positions;
(\lowerromannumeral{4}) $\mathbbm{E}(\{\tilde{\varphi}^2_{ij}\})${\scriptsize [\text{sd}]}: the mean and sd of the variational variance associated to the overt positions;
(\lowerromannumeral{5}) $\mathbbm{E}(\{\tilde{b}_j/\tilde{a}_j\})${\scriptsize [\text{sd}]}: the mean and sd of the approximations $\{\tilde{b}_j/\tilde{a}_j\}$ for $\{1/\gamma_j\}$;
(\lowerromannumeral{6}) $\text{VI}(\hat{\boldsymbol{z}},\boldsymbol{z}^*)$: the VI distance between the true and estimated clustering.
}}
\begin{adjustbox}{width=1.00\textwidth,center=\textwidth}
\begin{tabular}[c]{l|c|c|c|c|c|c}
\midrule

 & $(\tilde{\eta}, \tilde{\rho}^2)$
 & $\mathbbm{E}(\{|{\hat{d}_{ij}}^2-{d^{*2}_{ij}}|\})${\scriptsize [\text{sd}]}
 & $\mathbbm{E}(\{\tilde{\sigma}^2_i\})${\scriptsize [\text{sd}]}
 & $\mathbbm{E}(\{\tilde{\varphi}_{ij}\})${\scriptsize [\text{sd}]}
 & $\mathbbm{E}(\{\tilde{b}_j/\tilde{a}_j\})${\scriptsize [\text{sd}]}
 & $\text{VI}(\hat{\boldsymbol{z}},\boldsymbol{z}^*)$
 \\
\midrule
Sce 1 MLPCM
& (0.9655,\;0.0003)
& {1.5760{\scriptsize[1.6027]}}
& {0.0010{\scriptsize[0.0001]}}
& {0.1011{\scriptsize[0.0132]}}
& {0.1041{\scriptsize[0.0094]}}
& 0.1217
\\
Sce 1 MLPCM  \texttt{tol}=0.001
& (0.9681,\;0.0003)
& {1.5128{\scriptsize[1.5261]}}
& {0.0010{\scriptsize[0.0001]}}
& {0.1006{\scriptsize[0.0132]}}
& {0.1036{\scriptsize[0.0094]}}
& 0.1217
\\
Sce 1 PoisLPCM
& (0.8211,\;0.0005)
& {1.8047{\scriptsize[1.9058]}}
& {0.0143{\scriptsize[0.0048]}}
& --
& --
& 0.1217
\\
\midrule
Sce 2 MLPCM
& (1.0505,\;0.0003)
& {1.5332{\scriptsize[1.7009]}}
& {0.0010{\scriptsize[0.0001]}}
& {0.0996{\scriptsize[0.0137]}}
& {0.1025{\scriptsize[0.0104]}}
& 0.0000
\\
Sce 2 MLPCM a=20
& (0.9800,\;0.0003)
& {0.9939{\scriptsize[1.1074]}}
& {0.0005{\scriptsize[0.0001]}}
& {0.0507{\scriptsize[0.0033]}}
& {0.0515{\scriptsize[0.0017]}}
& 0.0000
\\
Sce 2 PoisLPCM
& (0.9212,\;0.0003)
& {1.0420{\scriptsize[2.1206]}}
& {0.0126{\scriptsize[0.0031]}}
& --
& --
& 0.0000
\\
\midrule
\end{tabular}
\end{adjustbox}
\label{SS1table}
\end{table}
In order to compare the estimated latent spaces that are obtained with the two different models, we run a Procrustes transformation \parencite{hoff2002latent} to match the two solutions with the ground truth.
In both scenarios, the MLPCM and PoisLPCM provide comparable performance for the clustering and inferred latent positions $\bm{\hat{U}}$, which are also in good agreement with the reference ones used for generating the networks.

Table~\ref{SS1table} shows various summaries for both scenarios and models, including estimates for the intercept parameter, errors in estimating the latent pairwise distances, and the Variation of Information (VI) advocated by \textcite{wade2018bayesian} as a distance between the true and estimated clustering partitions.
Overall, the results indicate relatively good model fits overall, but the MLPCM tends to perform better both when the data is generated according to the LPCM (scenario $1$) and when the data is generated according to the PoisLPCM (scenario $2$).
In particular, the table shows that the MLPCM provides a significantly better estimate for $\hat{\beta}$, which we can determine by comparing its variational parameter $\hat{\eta}$ to the ground truth of $1$.
As a sensitivity test, we also fitted our MLPCM reducing the convergence tolerance from $\texttt{tol}=0.01$ to $\texttt{tol}=0.001$.
This shows that the estimate of $\hat{\beta}$ could be further improved but the computing time on a laptop equipped with a Intel Core i7 1.80GHz CPU would increase from $15$ to $25$ minutes, approximately.
We performed a second sensitivity test on the parameter $a$, by testing out the value $a=20$ in the MLPCM, for scenario $2$.
This allows for more dispersion of the overt positions around the covert ones.
As regards the results for this case in Table~\ref{SS1table}, the model fit notably improves, and, in particular, we obtain smaller variational variance parameters $\bm{\tilde{\sigma}^2}$ and $\bm{\tilde{\varphi}}$.
This indicates that the posterior uncertainty around the latent positions is much reduced, suggesting a better fit.
The results from this table show that, while the PoisLPCM cannot fully characterize the features of a generated MLPCM, the MLPCM method is flexible enough to characterize all the features of a generated PoisLPCM.
Finally, we note that the MLPCM and the PoisLPCM provide similar inference performance of other variational parameters including $\bm{\tilde{\delta}}, \bm{\tilde{\omega}^2}, \bm{\tilde{\xi}}, \bm{\tilde{\psi}}$, of which the point estimates generally agree well with the corresponding reference values in both scenarios.

We close this first simulation study by performing a robustness experiment by running the MLPCM algorithm for $10$ different times for the case $K=4$.
The $10$ runs all use the default settings of our framework, but with small perturbations applied to each of the parameters.
The outputs are shown in Table~\ref{SS1tableRandomInitial}.
\setlength{\tabcolsep}{12pt}
\begin{table}[htbp!]
\renewcommand{\arraystretch}{0.9}
\centering
\caption{\footnotesize{Simulation study 1. Scenario 1 performance of the MLPCM across $10$ independent runs with random initializations around the default settings.
The maximum, median, minimum and the standard deviation of the corresponding summary statistics over multiple implementations' outputs are provided.
}}
\begin{adjustbox}{width=1.00\textwidth,center=\textwidth}
\begin{tabular}[c]{l|c|c|c|c|c|c}
\midrule
 & $(\tilde{\eta}, \tilde{\rho}^2)$
 & $\mathbbm{E}(\{|{\hat{d}_{ij}}^2-{d^{*2}_{ij}}|\})${\scriptsize [\text{sd}]}
 & $\mathbbm{E}(\{\tilde{\sigma}^2_i\})${\scriptsize [\text{sd}]}
 & $\mathbbm{E}(\{\tilde{\varphi}_{ij}\})${\scriptsize [\text{sd}]}
 & $\mathbbm{E}(\{\tilde{b}_j/\tilde{a}_j\})${\scriptsize [\text{sd}]}
 & $\text{VI}(\hat{\boldsymbol{z}},\boldsymbol{z}^*)$
 \\
\midrule
Maximum
& (1.0800,\;0.0011)
& {2.8866{\scriptsize[2.9869]}}
& {0.0012{\scriptsize[0.0005]}}
& {0.1277{\scriptsize[0.0709]}}
& {0.1347{\scriptsize[0.0753]}}
& 0.2429
\\
Median
& (0.8782,\;0.0003)
& {1.5985{\scriptsize[1.6549]}}
& {0.0010{\scriptsize[0.0001]}}
& {0.1012{\scriptsize[0.0145]}}
& {0.1042{\scriptsize[0.0113]}}
& 0.1217
\\
Minimum
& (0.8209,\;0.0003)
& {1.5191{\scriptsize[1.5531]}}
& {0.0010{\scriptsize[0.0001]}}
& {0.0996{\scriptsize[0.0126]}}
& {0.1024{\scriptsize[0.0086]}}
& 0.1217
\\
\midrule
Standard deviation
& (0.0916,\;0.0003)
& {0.4587{\scriptsize[0.4734]}}
& {0.0001{\scriptsize[0.0001]}}
& {0.0088{\scriptsize[0.0189]}}
& {0.0102{\scriptsize[0.0215]}}
& 0.0404
\\
\midrule
\end{tabular}
\end{adjustbox}
\label{SS1tableRandomInitial}
\end{table}
These results show that the results are fairly robust in that at convergence the error scores are comparable across runs.


\subsection{Simulation study 2}
\label{SS2}

\begin{figure}[htbp!]
\centering
\includegraphics[scale=0.525]{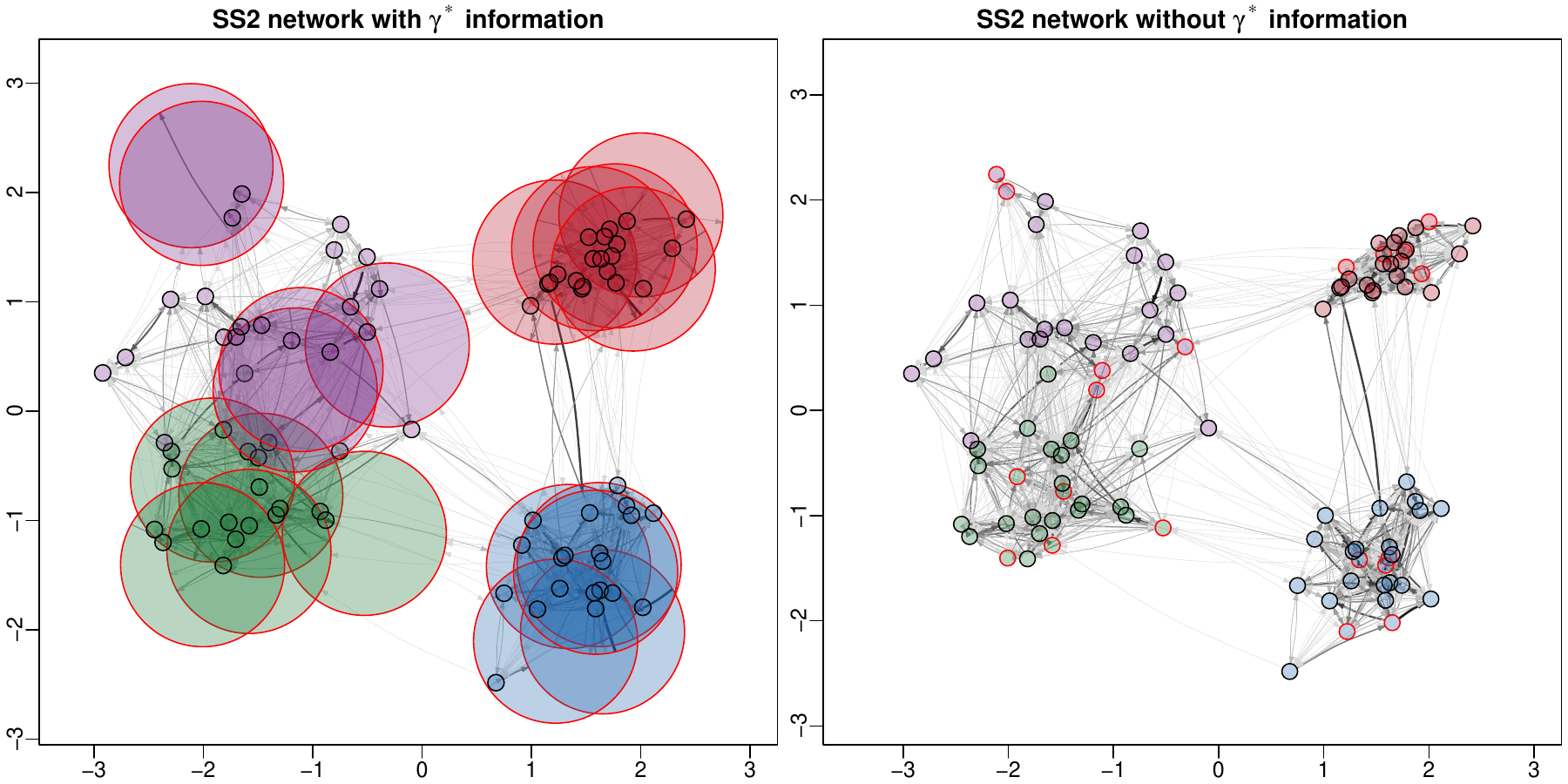}
\caption{Simulation study 2. Node positions correspond to the covert latent positions $\bm{U}^*$ used for simulating the network. Edge widths and colors are proportional to edge weights. Nodes with red contour have larger dispersion of the overt positions. Left panel: node sizes are proportional to $\{1/\gamma_j^*\}$. Right panel: node sizes are constant to make plot more clear.}
\label{SS2LatentPositionsRef}
\end{figure}

In this simulation study, we consider a more challenging situation, where the variances of mixed latent positions are no longer equal across groups, and two of the four groups are positioned significantly closer to each other.
The synthetic network in this second simulation study is randomly generated from a MLPCM with most of the model settings being similar to those in simulation study $1$.
The key differences are: (i) five nodes in each cluster have significantly more dispersed overt positions, that is, we set $\gamma_j^*=1$ for these special nodes as opposed to $\gamma_j^*=10$ for the rest of the nodes; (ii) the cluster means are set as:
\begin{equation*}
\boldsymbol{\mu}^* = \left[
\begin{pmatrix}
1.5 \\1.5
\end{pmatrix},
\begin{pmatrix}
1.5 \\-1.5
\end{pmatrix},
\begin{pmatrix}
-1.5 \\-1.0
\end{pmatrix},
\begin{pmatrix}
-1.5 \\1.0
\end{pmatrix}\right],
\end{equation*}
with increasing levels of dispersion of the covert positions around their cluster centers as the group label goes from $1$ to $4$.
The randomly simulated network is shown in Figure~\ref{SS2LatentPositionsRef}, using colors and nodes sizes in different ways to represent dispersion at the cluster level and at the node level.

\begin{figure}[htbp!]
\centering
\includegraphics[scale=0.625]{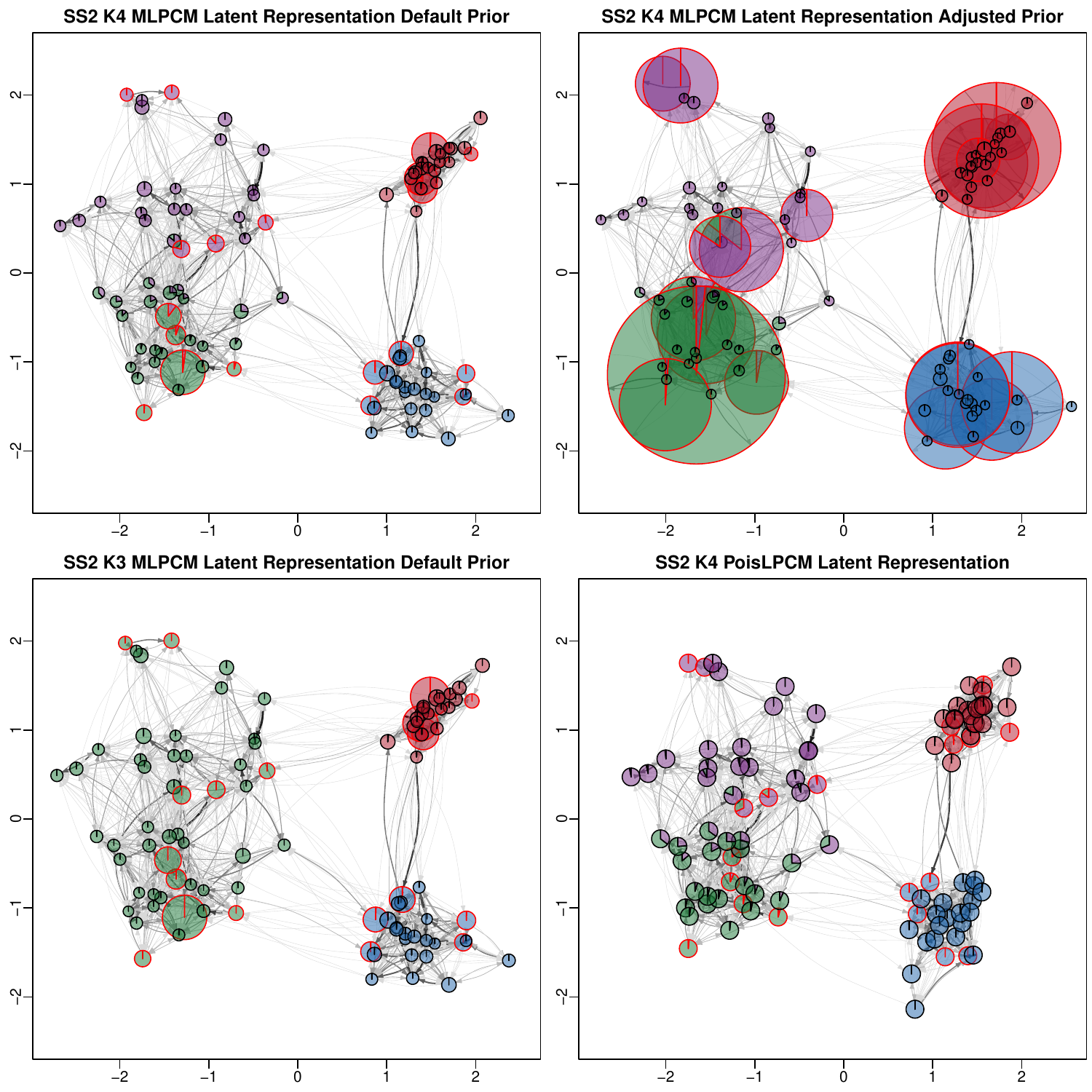}
\caption{Simulation study 2. Inferred latent space representations for various models and settings. Nodes' sizes are proportional to the dispersion of their overt positions. Node pie charts illustrate the inferred variational probability of group assignments. Nodes' contour colors indicate which nodes have a larger dispersion of overt positions, according to ground truth. Edge widths and colors are proportional to edge weights. Top-left: MLPCM for $K=4$ and default priors. Top-right: MLPCM for $K=4$ and informative priors. Bottom-left: MLPCM for $K=3$ and default priors. Bottom-right: PoisLPCM for $K=4$.}
\label{SS2_InferredLatentRepresentations}
\end{figure}

The same implementation settings as the default ones used in Section~\ref{SS1} are applied, with candidate values of $K$ being $\{2,3,4,5,6\}$ for both the MLPCM and the PoisLPCM.
The model-choice results are provided in Table~\ref{SS2PICLtable}.
\setlength{\tabcolsep}{12pt}
\begin{table}[htbp!]
\renewcommand{\arraystretch}{0.9}
\centering
\caption{\footnotesize{Simulation study 2. The PICL criteria values for number of groups ranging from $K=2$ to $K=6$. The bold value in each row indicates the highest PICL value.}}
\begin{adjustbox}{width=0.75\textwidth,center=\textwidth}
\begin{tabular}[c]{l|c|c|c|c|c}
\midrule
  & K=2 & K=3 & K=4 & K=5 & K=6     \\
\midrule
MLPCM & 13,709 & \textbf{13,766}  & \textbf{13,765} & 13,735  & 13,738\\
PoisLPCM & -4575.0 & \textbf{-4518.5}  & \textbf{-4518.4} & -4524.0  & -4529.7\\
\midrule
\end{tabular}
\end{adjustbox}
\label{SS2PICLtable}
\end{table}
The PICL favors $K=3$ for both models, however the PICL differences with $K=4$ are minimal.
Figure~\ref{SS2_InferredLatentRepresentations}) shows a graphical representation of the results, from which we can see that the two almost-overlapping groups are merged in the inferred latent spaces.
The left plots of this figure show the MLPCM results for $K=3$ and $K=4$.
The two solutions are pretty much identical if not for the two overlapping clusters being joined together.
The figure highlights with a different contour color those special nodes that were chosen to have more contrast between overt and covert positions.
If we sort the estimated values $\hat{\bm{\gamma}}$, we find that among the top $20$ nodes by $\hat{\bm{\gamma}}$, $17$ are in the set of special nodes with higher overt dispersion.
The model is able to identify these well, however the approximation of the dispersion around the covert positions is underestimated.
This is perhaps connected to some known limitations of variational Bayes in approximating posterior variances \parencite{blei2017variational}.
As regards the PoisLPCM results, we note that the latent positions of the special nodes are visibly different in the PoisLPCM estimates, compared to the MLPCM.
This indicates that the complexity of the dataset biased the latent space estimates for the PoisLPCM.

By means of example, we perform one additional run of the MLPCM in a more informative setting, where we provide the model with some additional information using the ground truth data.
In particular, we tune the $\text{Ga}(a,b)$ prior to match the information provided by $\bm{\gamma}^*$.
While this is unrealistic due to this information not generally being available, we use this setup as a demonstration of how a different priors could impact the results.
Figure~\ref{SS2_InferredLatentRepresentations} illustrates that, under this more informative prior, we tend to get the same estimated latent space, but the dispersion of the overt positions becomes more accurate.
In this case, the model is capable of clearly identifying the nodes with different sender and receiver profiles, and is more accurate in quantifying the asymmetric traits of their profiles.
We can conclude that the LPCM can effectively recover the latent space with a high degree of accuracy, and it can generally identify nodes with an asymmetric social profile.
However, the accurate estimation of overt positions associated to a node can highly depend on its priors, so it is critical that a practitioner considers various prior specifications and their effect on the results.

For completeness, we show in Table~\ref{SS2table} a number of different error measures for the different models and priors.
\setlength{\tabcolsep}{12pt}
\begin{table}[htbp!]
\renewcommand{\arraystretch}{0.9}
\centering
\caption{\footnotesize{Simulation study 2. Performance of various algorithms and settings. The summary statistics are the same as those used in Table~\ref{SS1table}
}}
\begin{adjustbox}{width=1.00\textwidth,center=\textwidth}
\begin{tabular}[c]{l|c|c|c|c|c|c}
\midrule

 & $(\tilde{\eta}, \tilde{\rho}^2)$ 
 & $\mathbbm{E}(\{|{\hat{d}_{ij}}^2-{d^{*2}_{ij}}|\})${\scriptsize [\text{sd}]}     
 & $\mathbbm{E}(\{\tilde{\sigma}^2_i\})${\scriptsize [\text{sd}]}   
 & $\mathbbm{E}(\{\tilde{\varphi}_{ij}\})${\scriptsize [\text{sd}]}   
 & $\mathbbm{E}(\{\tilde{b}_j/\tilde{a}_j\})${\scriptsize [\text{sd}]}  
 & $\text{VI}(\hat{\boldsymbol{z}},\boldsymbol{z}^*)$ 
 \\
\midrule
MLPCM K=4 default prior
& (0.7613,\;0.0003)  
& {2.9426{\scriptsize[3.5967]}}
& {0.0011{\scriptsize[0.0004]}} 
& {0.1161{\scriptsize[0.0509]}} 
& {0.1190{\scriptsize[0.0479]}} 
& 0.2423 
\\
MLPCM K=4 adjusted prior
& (0.8123,\;0.0003)  
& {2.6054{\scriptsize[3.1895]}}
& {0.0021{\scriptsize[0.0022]}} 
& {0.2509{\scriptsize[0.3449]}} 
& {0.2630{\scriptsize[0.3496]}} 
& 0.2423 
\\
MLPCM K=3 default prior 
& (0.7613,\;0.0003)  
& {2.9393{\scriptsize[3.5898]}}
& {0.0011{\scriptsize[0.0004]}}
& {0.1158{\scriptsize[0.0499]}} 
& {0.1186{\scriptsize[0.0468]}}
& 0.5000
\\
PoisLPCM K=4 default prior
& (0.5965,\;0.0012)  
& {3.5545{\scriptsize[4.0102]}}
& {0.0148{\scriptsize[0.0052]}} 
& -- 
& --
& 0.2423 
\\
\midrule
\end{tabular}
\end{adjustbox}
\label{SS2table}
\end{table}
We highlight that, compared to simulation study $1$, the performance of the PoisLPCM is significantly worse than that of the MLPCM.
This is expected since it is much more challenging for the PoisLPCM to fit the network when interactions have very asymmetric patterns.
The poor performance is reflected by worse values on all statistics shown in Table~\ref{SS2}


\section{Real data application}
\label{RDA}

As a real data application, we focus on the International Arms Transfers Database, which is publicly available from the Stockholm International Peace Research Institute (SIPRI)\footnote{\url{https://www.sipri.org/databases/armstransfers}}.
The database includes all transfers of major conventional arms between countries, regions, subregions or non-state actors from 1950 to the most recent full calendar year.
We focus on a subset of this dataset, corresponding to the ego-centric network of the United States for 2024.
Without loss of generality, we define that a directed weighted edge exists from country $i$ to country $j$ if country $i$ ordered arms transfers from country $j$, and the interaction weight corresponds to the volume of such arms transfers in 2024 that is recorded in millions of SIPRI Trend-Indicator Values (TIVs).
In other words, the edge direction follows the direction of the order, thus it is opposite to how the arms are moving.
The SIPRI TIV is a novel common unit developed by SIPRI aiming to regularize the data of deliveries of different weapons and to identify general trends of these arms transfers\footnote{More details at \url{https://www.sipri.org/databases/armstransfers/sources-and-methods}}.

\begin{figure}[htbp!]
\centering
\includegraphics[scale=0.775]{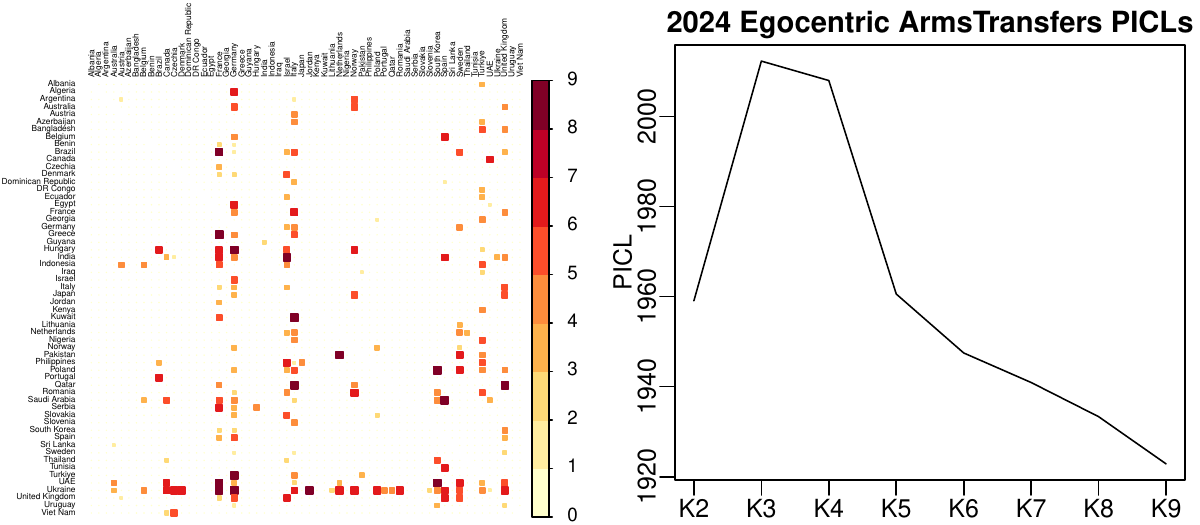}
\caption{2024 Egocentric ArmsTransfers real data application. Left: interaction matrix $\bm{Y}$ of the observed egocentric network with 58 countries. Darker entry color corresponds to higher volume of the transfers. This matrix shows the value of the order made, meaning that the arms are moving from column-node to row-node. Right: PICL values obtained with the MLPCM for $K=2,3,\dots, 9$.}
\label{RDA_adj_PICL}
\end{figure}

We begin by listing all the countries in 2024 based on the most recent SIPRI Military Expenditure Database\footnote{\url{https://www.sipri.org/databases/milex}}.
There are 163 countries which have records of military expenditures from 2001 to 2023, and, among these 163 countries, there are 110 countries which have records of arms transfers in 2024.
To obtain the egocentric network for the US, we extract the interaction data in which the US are involved.
After this step, we remove the US and all the countries that only have arms transfers with the US.
The remaining network consists of all the interactions that the neighbors of the US have between themselves, excluding the US and any isolated node.
This effectively corresponds to the network as it is seen from the point of view of the US.
The network is composed of 58 nodes and the corresponding interaction matrix is illustrated as the left plot of Figure~\ref{RDA_adj_PICL}.
Note that the original recorded values for TIVs are either non-negative integers or \texttt{NULL}s, where a \texttt{NULL} corresponds to the case of no arms transfer observed.
This network is a very sparse one with 94.22\% entries in the adjacency matrix being \texttt{NULL}s, while the remaining 191 interactions have discrete weights ranging from 0 to 716.
By treating \texttt{NULL}s as true zeros (to be distinguished from the recorded zero millions of TIVs), the resulting distribution of the network interaction weights is extremely right-skewed.
Thus, in practice, we process the original data by applying a transformation $f(x)=\lfloor\log_2(x+1)+0.5\rfloor$ on all the recorded values that are not \texttt{NULL}s, where the formula $g(s)=\lfloor s+0.5\rfloor$ denotes the general rounding approach which rounds the value $s$ to the nearest integer.
This transformation helps in reducing the extreme discrepancies between different interaction weights and at the same time it preserves the general patterns of the data.
For those \texttt{NULL}s which correspond to zero arms transfers, we assign the true zeros $\log_2(1)=0$ in the transformed dataset.
The resulting processed data thus contains only non-negative integers as illustrated in the left plot of Figure~\ref{RDA_adj_PICL}.

\begin{figure}[htbp!]
\centering
\includegraphics[scale=0.45]{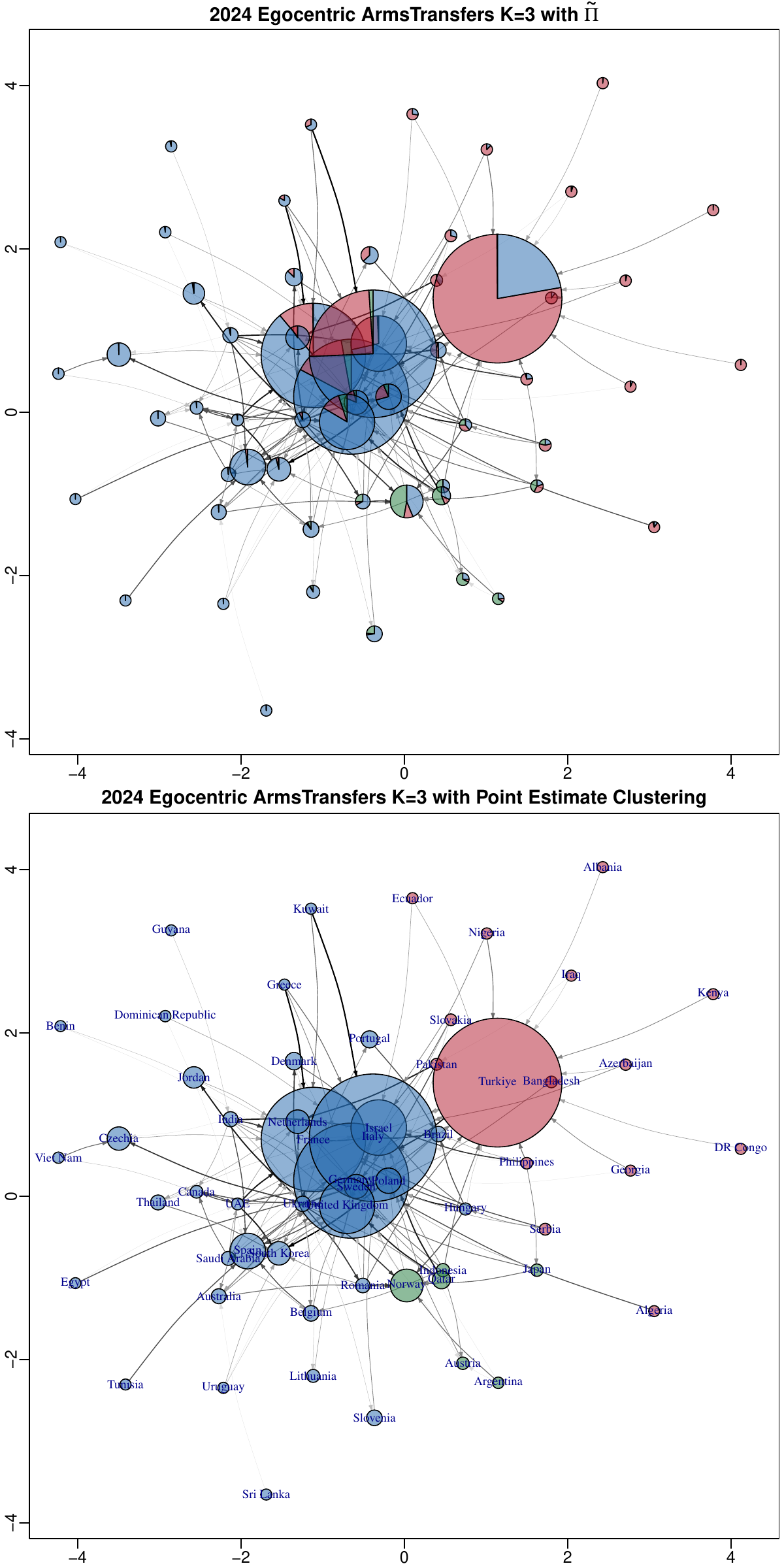}
\caption{2024 Egocentric ArmsTransfers real data application: inferred latent space for $K=3$. For both plots, node positions correspond to $\bm{\hat{U}}$ and node sizes are proportional to the inferred $\{1/\hat{\gamma}_j=\tilde{b}_j/\tilde{a}_j\}$, while edge widths and colors are proportional to edge weights.
Top: each node pie chart illustrates the corresponding inferred variational probability of group assignment. Bottom: the node colors correspond to the maximimum a posteriori clustering; each node is labeled by the corresponding country name.}
\label{RDA_K3LatentRepresentations}
\end{figure}

In order to fit the MLPCM to this real egocentric arms transfers network, we apply the proposed variational Bayes Algorithm~\ref{VB_algorithm_MLPCM}.
After testing various hyperparameters configurations, we adopt the default settings that were proposed in Section~\ref{sec:initialization}, with the exception of prior parameters $\{\bm{\omega^{2}},\bm{\xi},\bm{\delta}\}$ which are set to be $3$ instead of $1$.
This further encourages the within-group cohesion and the separation of different groups in the latent space.
We consider $K=2,3,\dots,9$ as candidate values for the number of groups, and we obtain the PICL results shown in the right panel of Figure~\ref{RDA_adj_PICL}.
The PICL criterion chooses $K=3$ as the best model, and the corresponding inferred latent space visualization of this arms transfers network is provided in Figure~\ref{RDA_K3LatentRepresentations}.
The results support quite strongly the presence of three different groups in this network, and the bottom plot in Figure~\ref{RDA_K3LatentRepresentations} provides a point estimate of the inferred clustering.
The blue group has several significantly centralized countries including France, Germany, Israel, Italy, Sweden, Poland, UK, Ukraine, and the Netherlands.
Some of these nodes have a remarkable overt dispersion, for example France, Germany, Italy, Israel and the UK.
This high dispersion means that they are perceived in very diverse ways, and so they can send arms to different areas of the latent space.
In other words, these countries have more reach, exhibit a diverse connectivity profile, and show more asymmetric connectivity patterns.
Ukraine also has a fairly central position in the blue cluster, but a small overt dispersion. This is expected since the country primarily imported arms, which is is not reflected in the dispersion of the latent positions.

The most influential country over the whole network is Turkey which appears to take a central position in the red group.
Due to its high dispersion, Turkey sends arms to almost every red group member.
However, based on the top plot of Figure~\ref{RDA_K3LatentRepresentations}, there is still a non-negligible probability that Turkey could be clustered together with those central countries from the blue group.
The green group is small and exhibits fewer connections: all of green countries have significant chances to be clustered into either the blue group or the red group.
A significant center of this green group is Norway which has a relatively higher dispersion and higher tendency to receive edges.

In Appendix~\ref{RDA_K5LatentRepresentations_Appendix}, we also provide the inferred latent space visualization for the $K=5$ model, as this shows some interesting patterns.
The general outlook of the network is similar to that of $K=3$ shown in Figure~\ref{RDA_K3LatentRepresentations}, but the blue group is further split into three subgroups with the central nodes forming a cluster of hubs.
The resulting network exhibits a typical core-periphery structure where the most active actors are all clustered together.


\section{Conclusions}
\label{Conclusion}

In this paper, we developed a new mixed latent position cluster model that is well-suited for directed social networks with weighted edges.
The model can identify nodes that exhibit a multitude of different connectivity patterns, and it can better address the asymmetry that is often observed in directed networks.
Within a clustering framework, we have demonstrated that the variational Bayes procedure can lead to accurate estimates of the nodes' partition and of the latent space.
Our model extends the mixed-membership concept, which is widely used in the context of stochastic blockmodels, to latent position network modeling.
This mimics a random effects framework, while providing more clear graphical representations and interpretations.
Through simulated and real data, we showed a variety of results obtained with our methodology, assessing its accuracy in recovering good solutions, and its robustness to varying initial settings and priors.

In terms of further work, this contribution opens up various possible future directions.
As regards the modeling, one may consider adding layers to the model so that both covert and overt positions are characterized by parametric distributions.
This can make the model more flexible since nodes could exhibit a wider range of social connectivity patterns.
For inference, the traditional approach for LPMs generally relies on Markov chain Monte Carlo sampling from the posterior.
This could be a perfectly viable alternative to our variational Bayes approach, where one could get a more detailed characterization of the posterior distribution of the model, perhaps at the cost of a slower algorithm.
In terms of computing, an interesting direction could be to consider methods that choose the number of clusters automatically, in the style of \textcite{nobile2007bayesian,miller2018mixture,legramanti2022extended}.
This could provide a substantial speedup as the variational Bayes would not need to be run for every candidate value of number of groups.
Finally, natural extensions of this model could be considered for temporal or multiview network data \textcite{sarkar2005dynamic,sewell2016latent,kim2018review} whereby covert and overt positions could change across time or network layers.


\section*{Funding Statement}
Insight Research Ireland Centre for Data Analytics is supported by Science Foundation Ireland under Grant Number 12/RC/2289$\_$P2.

\section*{Competing Interests}
The authors declare no competing interests.

\section*{Code and Data}
The MLPCM software is publicly available from our GitHub repository \url{https://github.com/Chaoyi-Lu/Mixed-Latent-Position-Cluster-Model}. The data for application on the arms transfer network is publicly available from \url{https://www.sipri.org/databases/armstransfers}.

\printbibliography


\newpage

\appendix

\small

\section{Appendix}


\subsection{Derivation of the ELBO for the MLPCM}
\label{MLPCM_ELBO_Derivation_Appendix}

Recall here that the ELBO, $\mathcal{F}(\bm{\tilde{\Theta}})$, under the MLPCM can be decomposed into a number of terms that can be derived separately:
\begin{equation*}
\begin{split}
&\mathcal{F}(\bm{\tilde{\Theta}})= \mathbbm{E}_q[\text{log}\;p(\bm{Y}|\bm{U},\bm{V}, \beta)]+
\mathbbm{E}_q[\text{log}\;p(\bm{U}|\bm{\mu}, \bm{\tau}, \bm{z})]+
\mathbbm{E}_q[\text{log}\;p(\bm{V}|\bm{U}, \bm{\gamma})]+
\mathbbm{E}_q[\text{log}\;p(\bm{z}|\bm{\Pi})]\\
&+\mathbbm{E}_q[\text{log}\;\pi(\beta)]+
\mathbbm{E}_q[\text{log}\;\pi(\bm{\mu})]+
\mathbbm{E}_q[\text{log}\;\pi(\bm{\tau})]+
\mathbbm{E}_q[\text{log}\;\pi(\bm{\gamma})]+
\mathbbm{E}_q[\text{log}\;\pi(\bm{\Pi})]\\
&-\mathbbm{E}_q[\text{log}\;q_{\bm{U}}(\bm{U}|\bm{\tilde{U}},\bm{\tilde{\sigma}^2})]-
\mathbbm{E}_q[\text{log}\;q_{\bm{V}}(\bm{V}|\bm{\tilde{V}},\bm{\tilde{\varphi}^2})]-
\mathbbm{E}_q[\text{log}\;q_{\bm{z}}(\bm{z}|\bm{\tilde{\Pi}})]-
\mathbbm{E}_q[\text{log}\;q_{\beta}(\beta|\tilde{\eta},\tilde{\rho}^2)]\\
&-\mathbbm{E}_q[\text{log}\;q_{\bm{\mu}}(\bm{\mu}|\bm{\tilde{\mu}},\bm{\tilde{\omega}^2})]-
\mathbbm{E}_q[\text{log}\;q_{\bm{\tau}}(\bm{\tau}|\bm{\tilde{\xi}},\bm{\tilde{\psi}})]-
\mathbbm{E}_q[\text{log}\;q_{\bm{\gamma}}(\bm{\gamma}|\bm{\tilde{a}},\bm{\tilde{b}})]-
\mathbbm{E}_q[\text{log}\;q_{\bm{\Pi}}(\bm{\Pi}|\bm{\tilde{\delta}})].
\end{split}
\end{equation*}
In the following sections, we derive the above ELBO term by term.


\subsubsection[$Y,U,V$]{Evaluate $\mathbbm{E}_q[\text{\normalfont{log}}\;p(\bm{Y}|\bm{U},\bm{V}, \beta)]$}
\label{ELBO_pY_Section_Appendix}

\begin{equation*}
\begin{split}
&\mathbbm{E}_q[\log p(\bm{Y}|\bm{U},\bm{V}, \beta)]=
\mathbbm{E}_q\left[\log \prod^N_{\substack{i,j=1, \\ i\neq j}}f_{\text{Pois}}\left(y_{ij}\middle|\exp (\beta-||\bm{u}_i-\bm{v}_{j\leftarrow i}||^2)\right)\right]\\
&=\mathbbm{E}_q\left[\log\prod^N_{\substack{i,j=1, \\ i\neq j}}\frac{\left[\exp(\beta-||\bm{u}_i-\bm{v}_{j\leftarrow i}||^2)\right]^{y_{ij}}\exp[-\exp (\beta-||\bm{u}_i-\bm{v}_{j\leftarrow i}||^2)]}{y_{ij}!}\right]\\
&=\mathbbm{E}_q\left[\sum^N_{\substack{i,j=1, \\ i\neq j}}\left[y_{ij}\left(\beta-||\bm{u}_i-\bm{v}_{j\leftarrow i}||^2\right)-\exp (\beta-||\bm{u}_i-\bm{v}_{j\leftarrow i}||^2)-\log(y_{ij}!)\right]\right]\\
&=\sum^N_{\substack{i,j=1, \\ i\neq j}}\left\{y_{ij}\left[\mathbbm{E}_q(\beta)-\mathbbm{E}_q\left(||\bm{u}_i-\bm{v}_{j\leftarrow i}||^2\right)\right]-\mathbbm{E}_q\left[\exp (\beta-||\bm{u}_i-\bm{v}_{j\leftarrow i}||^2)\right]\right\}-
\sum^N_{\substack{i,j=1, \\ i\neq j}}\log(y_{ij}!)\\
&=\sum^N_{\substack{i,j=1, \\ i\neq j}}\left\{ y_{ij}\left[\tilde{\eta}-||\bm{\tilde{u}}_i-\bm{\tilde{v}}_{j\leftarrow i}||^2-d\left(\tilde{\sigma}^2_i+\tilde{\varphi}^2_{ij}\right)\right]-\frac{\exp \left(\tilde{\eta}+\frac{\tilde{\rho}^2}{2}-\frac{||\bm{\tilde{u}}_i-\bm{\tilde{v}}_{j\leftarrow i}||^2}{1+2\tilde{\sigma}^2_i+2\tilde{\varphi}^2_{ij}}\right)}{(1+2\tilde{\sigma}^2_i+2\tilde{\varphi}^2_{ij})^{d/2}} \right\}+\texttt{const}.
\end{split}
\end{equation*}\;\hfill$\square$\\

Here, \texttt{const} denotes the constant term(s).
The exact analytical solutions of the three expectations above can all be derived.
The 1st expectation, $\mathbbm{E}_q(\beta)$, can be easily obtained, that is, $\tilde{\eta}$.
The detailed proof for the $\mathbbm{E}_q\left(||\bm{u}_i-\bm{v}_{j\leftarrow i}||^2\right)$ and the $\mathbbm{E}_q\left[\exp (\beta-||\bm{u}_i-\bm{v}_{j\leftarrow i}||^2)\right]$ are provided next:
\begin{itemize}
\item Here we first prove for the general form of the $\mathbbm{E}_q\left(||\bm{u}_i-\bm{v}_{j\leftarrow i}||^2\right)$.

\textbf{* Find} $\mathbbm{E}(||\bm{x}||^2)$, \textbf{where} $\bm{x}:=(x_1,\dots,x_d)^T\sim\text{MVN}_d(\bm{m},\bm{\Sigma})$:
Then there exists a matrix $\bm{A}\in \mathbbm{R}^{d\times d'}$ such that $\bm{x}=\bm{A}\bm{w}+\bm{m}$ where $\bm{w}:=(w_1,\dots,w_{d'})^T\sim\text{MVN}_{d'}(\bm{0}_{d'},\mathbbm{I}_{d'})$ is the standard MVN distribution with $d'$ dimension that is not necessarily equal to $d$.
Here the matrix $\bm{A}$ is actually the squared root matrix of $\bm{\Sigma}$, that is, $\bm{\Sigma}=\bm{A}\bm{A}^T$.

Then we have
\begin{equation*}
\begin{split}
&\mathbbm{E}(||\bm{x}||^2)=\mathbbm{E}(||\bm{A}\bm{w}+\bm{m}||^2)=\mathbbm{E}[(\bm{A}\bm{w}+\bm{m})^T(\bm{A}\bm{w}+\bm{m})]\\
&=\mathbbm{E}(\bm{w}^T\bm{A}^T\bm{A}\bm{w}+\bm{w}^T\bm{A}^T\bm{m}+\bm{m}^T\bm{A}\bm{w}+\bm{m}^T\bm{m})=
\mathbbm{E}(\bm{w}^T\bm{A}^T\bm{A}\bm{w})+\bm{m}^T\bm{m}.
\end{split}
\end{equation*}
Here, since $\mathbbm{E}(\bm{w}^T)=\bm{0}_{d'}^T$, we have $\mathbbm{E}(\bm{w}^T\bm{A}^T\bm{m})=\mathbbm{E}(\bm{w}^T)\bm{A}^T\bm{m}=0$. Similarly for the $\mathbbm{E}(\bm{m}^T\bm{A}\bm{w})$.
Then we decompose the rest expectation term as:
$$\mathbbm{E}(\bm{w}^T\bm{A}^T\bm{A}\bm{w})=\mathbbm{E}\left[\sum^{d'}_{i,j=1}(\bm{A}^T\bm{A})_{ij}w_iw_j\right]=\sum^{d'}_{i,j=1}(\bm{A}^T\bm{A})_{ij}\mathbbm{E}(w_iw_j),$$
where $\mathbbm{E}(w_iw_j)=0$ if $i\neq j$, whereas $\mathbbm{E}(w^2_i)=\text{Var}(w_i)+\mathbbm{E}(w_i)^2=1$. Thus we have:
$$\mathbbm{E}(||\bm{x}||^2)=\mathbbm{E}(\bm{w}^T\bm{A}^T\bm{A}\bm{w})+\bm{m}^T\bm{m}=\bm{m}^T\bm{m}+\sum^{d'}_{i=1}(\bm{A}^T\bm{A})_{ii}=||\bm{m}||^2+\text{tr}(\bm{\Sigma}),$$
provided the fact that $\sum^{d'}_{i=1}(\bm{A}^T\bm{A})_{ii}=\text{tr}(\bm{A}^T\bm{A})=\text{tr}(\bm{A}\bm{A}^T)=\text{tr}(\bm{\Sigma})$.\;\hfill$\square$\\

Thus it remains to show the distribution of $(\bm{u}_i-\bm{v}_{j\leftarrow i})$ provided that $q_{\bm{U}}(\bm{u}_i|\bm{\tilde{u}}_i,\tilde{\sigma}^2_i) \sim \text{MVN}_d(\bm{\tilde{u}}_i,\tilde{\sigma}^2_i\mathbbm{I}_d)$ and $q_{\bm{V}}(\bm{v}_{j\leftarrow i}|\bm{\tilde{v}}_{j\leftarrow i},\tilde{\varphi}^2_{ij}) \sim \text{MVN}_d(\bm{\tilde{v}}_{j\leftarrow i},\tilde{\varphi}^2_{ij}\mathbbm{I}_d)$.
We prove this by the characteristic functions of multivariate normal distributions.

\textbf{* Find the distribution of} $\bm{(u+v)}$: Here we provide the proof of a general case, where we assume that $\bm{u}\sim\text{MVN}_d(\bm{\mu}_{\bm{u}},\bm{\Sigma}_{\bm{u}})$ and $\bm{v}\sim\text{MVN}_d(\bm{\mu}_{\bm{v}},\bm{\Sigma}_{\bm{v}})$, then
\begin{equation*}
\begin{split}
& \bm{\phi}_{\bm{u}+\bm{v}}(\bm{t}) = 
\bm{\phi}_{\bm{u}}(\bm{t})\bm{\phi}_{\bm{v}}(\bm{t})=
\mathbbm{E}[\exp (i\bm{t}^T\bm{u})]\mathbbm{E}[\exp (i\bm{t}^T\bm{v})]\\
&=\exp (i\bm{t}^T\bm{\mu}_{\bm{u}}-\bm{t}^T\bm{\Sigma}_{\bm{u}}\bm{t})\exp (i\bm{t}^T\bm{\mu}_{\bm{v}}-\bm{t}^T\bm{\Sigma}_{\bm{v}}\bm{t})\\
&=\exp (i\bm{t}^T(\bm{\mu}_{\bm{u}}+\bm{\mu}_{\bm{v}})-\bm{t}^T(\bm{\Sigma}_{\bm{u}}+\bm{\Sigma}_{\bm{v}})\bm{t}),
\end{split}
\end{equation*}
which is the characteristic function of $\text{MVN}_d(\bm{\mu}_{\bm{u}}+\bm{\mu}_{\bm{v}},\bm{\Sigma}_{\bm{u}}+\bm{\Sigma}_{\bm{v}})$.
So $\bm{u}+\bm{v} \sim \text{MVN}_d(\bm{\mu}_{\bm{u}}+\bm{\mu}_{\bm{v}},\bm{\Sigma}_{\bm{u}}+\bm{\Sigma}_{\bm{v}})$.
Considering the fact that $-\bm{v}\sim\text{MVN}_d(-\bm{\mu}_{\bm{v}},\bm{\Sigma}_{\bm{v}})$, we have $\bm{u}-\bm{v} \sim \text{MVN}_d(\bm{\mu}_{\bm{u}}-\bm{\mu}_{\bm{v}},\bm{\Sigma}_{\bm{u}}+\bm{\Sigma}_{\bm{v}})$.\;\hfill$\square$\\

Combining the above two proofs, we can obtain that $\bm{u}_i-\bm{v}_{j\leftarrow i}\sim\text{MVN}_d(\bm{\tilde{u}}_i-\bm{\tilde{v}}_{j\leftarrow i},\tilde{\sigma}^2_i\mathbbm{I}_d+\tilde{\varphi}^2_{ij}\mathbbm{I}_d)$, 
and thus $\mathbbm{E}_q\left(||\bm{u}_i-\bm{v}_{j\leftarrow i}||^2\right)=||\bm{\tilde{u}}_i-\bm{\tilde{v}}_{j\leftarrow i}||^2+\text{tr}[(\tilde{\sigma}^2_i+\tilde{\varphi}^2_{ij})\mathbbm{I}_d]=||\bm{\tilde{u}}_i-\bm{\tilde{v}}_{j\leftarrow i}||^2+d(\tilde{\sigma}^2_i+\tilde{\varphi}^2_{ij})$.\;\hfill$\square$\\

\item Note that $\mathbbm{E}_q\left[\exp (\beta-||\bm{u}_i-\bm{v}_{j\leftarrow i}||^2)\right]=\mathbbm{E}_{q(\beta)}[\exp (\beta)]\mathbbm{E}_{q(\bm{U}),q(\bm{V})}[\exp (-||\bm{u}_i-\bm{v}_{j\leftarrow i}||^2)]$. 
So we first prove the 1st expectation here.

\textbf{* Find} $\mathbbm{E}_q[\exp (\beta)]$, \textbf{where} $q_{\beta}(\beta|\tilde{\eta},\tilde{\rho}^2) \sim \text{Normal}(\tilde{\eta},\tilde{\rho}^2)$: 
\begin{equation*}
\resizebox{1\hsize}{!}{$
\begin{split}
& \mathbbm{E}_q[\exp (\beta)]=
\int^\infty_{-\infty}\exp (\beta)\frac{\exp \left[-\frac{(\beta-\tilde{\eta})^2}{2\tilde{\rho}^2}\right]}{\sqrt{2\pi\tilde{\rho}^2}}\text{d}\beta=
\int^\infty_{-\infty}\frac{\exp \left[-\frac{\beta^2}{2\tilde{\rho}^2}+\frac{\tilde{\eta}\beta}{\tilde{\rho}^2}-\frac{\tilde{\eta}^2}{2\tilde{\rho}^2}+\beta\right]}{\sqrt{2\pi\tilde{\rho}^2}}\text{d}\beta\\
&=\int^\infty_{-\infty}\frac{\exp \left[-\frac{\beta^2}{2\tilde{\rho}^2}+\frac{(\tilde{\eta}+\tilde{\rho}^2)\beta}{\tilde{\rho}^2}-\frac{\tilde{\eta}^2}{2\tilde{\rho}^2}\right]}{\sqrt{2\pi\tilde{\rho}^2}}\text{d}\beta=
\int^\infty_{-\infty}\frac{\exp \left[-\frac{[\beta-(\tilde{\eta}+\tilde{\rho}^2)]^2}{2\tilde{\rho}^2}\right]}{\sqrt{2\pi\tilde{\rho}^2}}\text{d}\beta\cdot\exp \left[\frac{(\tilde{\eta}+\tilde{\rho}^2)^2}{2\tilde{\rho}^2}-\frac{\tilde{\eta}^2}{2\tilde{\rho}^2}\right] \\
&= \exp \left[\frac{\tilde{\eta}^2+2\tilde{\eta}\tilde{\rho}^2+\tilde{\rho}^4}{2\tilde{\rho}^2}-\frac{\tilde{\eta}^2}{2\tilde{\rho}^2}\right]=
\exp \left(\tilde{\eta}+\frac{\tilde{\rho}^2}{2}\right).
\end{split}$}
\end{equation*}\;\hfill$\square$\\

We next move to prove the general form of the 2nd expectation above.

\textbf{* Find} $\mathbbm{E}[\exp (-||\bm{x}||^2)]$, \textbf{where} $\bm{x}:=(x_1,\dots,x_d)^T\sim\text{MVN}_d(\bm{m},\bm{\Sigma})$:
\begin{equation*}
\resizebox{1\hsize}{!}{$
\begin{split}
& \mathbbm{E}[\exp (-||\bm{x}||^2)]=
\mathbbm{E}[\exp (-\bm{x}^T\bm{x})]=
\int_{\mathbbm{R}^2}\exp (-\bm{x}^T\bm{x})\frac{\exp \left[-\frac{1}{2}(\bm{x}-\bm{m})^T\bm{\Sigma}^{-1}(\bm{x}-\bm{m})\right]}{\sqrt{(2\pi)^d|\bm{\Sigma}|}}\text{d}\bm{x}\\
&=\int_{\mathbbm{R}^2}\frac{\exp \left[-\frac{1}{2}(\bm{x}^T\bm{\Sigma}^{-1}\bm{x}-\bm{x}^T\bm{\Sigma}^{-1}\bm{m}-\bm{m}^T\bm{\Sigma}^{-1}\bm{x}+\bm{m}^T\bm{\Sigma}^{-1}\bm{m}+2\bm{x}^T\bm{x})\right]}{\sqrt{(2\pi)^d|\bm{\Sigma}|}}\text{d}\bm{x}\\
&=\int_{\mathbbm{R}^2}\frac{\exp \left[-\frac{1}{2}\left(\bm{x}^T(\bm{\Sigma}^{-1}+2\mathbbm{I}_d)\bm{x}-2\bm{x}^T\bm{\Sigma}^{-1}\bm{m}\right)-\frac{1}{2}\bm{m}^T\bm{\Sigma}^{-1}\bm{m}\right]}{\sqrt{(2\pi)^d|\bm{\Sigma}|}}\text{d}\bm{x}\\
&=\int_{\mathbbm{R}^2}\frac{\exp \left[-\frac{1}{2}\left(\bm{x}^T\bm{\Sigma}^{-1}(\mathbbm{I}_d+2\bm{\Sigma})\bm{x}-2\bm{x}^T\bm{\Sigma}^{-1}(\mathbbm{I}_d+2\bm{\Sigma})(\mathbbm{I}_d+2\bm{\Sigma})^{-1}\bm{m}\right)-\frac{1}{2}\bm{m}^T\bm{\Sigma}^{-1}\bm{m}\right]}{\sqrt{(2\pi)^d|\bm{\Sigma}|}}\text{d}\bm{x}\\
&=\int_{\mathbbm{R}^2}\frac{\exp \left[-\frac{1}{2}(\bm{x}-(\mathbbm{I}_d+2\bm{\Sigma})^{-1}\bm{m})^T\bm{\Sigma}^{-1}(\mathbbm{I}_d+2\bm{\Sigma})(\bm{x}-(\mathbbm{I}_d+2\bm{\Sigma})^{-1}\bm{m})
\right]}{\sqrt{(2\pi)^d|(\bm{\Sigma}^{-1}(\mathbbm{I}_d+2\bm{\Sigma}))^{-1}|}}\text{d}\bm{x}\\
&\hspace{0.5em}\times\frac{\sqrt{|(\bm{\Sigma}^{-1}(\mathbbm{I}_d+2\bm{\Sigma}))^{-1}|}}{\sqrt{|\bm{\Sigma}|}} \exp \left[\frac{1}{2}((\mathbbm{I}_d+2\bm{\Sigma})^{-1}\bm{m})^T\bm{\Sigma}^{-1}(\mathbbm{I}_d+2\bm{\Sigma})((\mathbbm{I}_d+2\bm{\Sigma})^{-1}\bm{m})-
\frac{1}{2}\bm{m}^T\bm{\Sigma}^{-1}\bm{m}\right]\\
&=\frac{\sqrt{|(\mathbbm{I}_d+2\bm{\Sigma})^{-1}\bm{\Sigma}|}}{\sqrt{|\bm{\Sigma}|}}\exp \left[\frac{1}{2}((\mathbbm{I}_d+2\bm{\Sigma})^{-1}\bm{m})^T\bm{\Sigma}^{-1}\bm{m}-
\frac{1}{2}\bm{m}^T\bm{\Sigma}^{-1}\bm{m}\right]\\
&=\frac{\sqrt{|(\mathbbm{I}_d+2\bm{\Sigma})|^{-1}}\sqrt{|\bm{\Sigma}|}}{\sqrt{|\bm{\Sigma}|}}\exp \left[\frac{1}{2}\bm{m}^T(\mathbbm{I}_d+2\bm{\Sigma})^{-T}\bm{\Sigma}^{-1}\bm{m}-
\frac{1}{2}\bm{m}^T\bm{\Sigma}^{-1}\bm{m}\right]\\
&=\frac{\exp \left[-\frac{1}{2}\left[\bm{m}^T(\mathbbm{I}_d-(\mathbbm{I}_d+2\bm{\Sigma})^{-T})\bm{\Sigma}^{-1}\bm{m}\right]\right]}{\sqrt{|\mathbbm{I}_d+2\bm{\Sigma}|}}.
\end{split}$}
\end{equation*}\;\hfill$\square$\\

In the case that $\bm{\Sigma}=c\mathbbm{I}_d$ where $c$ is some constant, the above formula can be reduced to:
\begin{equation*}
\begin{split}
&=\frac{\exp \left[-\frac{1}{2}\left[\bm{m}^T(\mathbbm{I}_d-(\mathbbm{I}_d+2c\mathbbm{I}_d)^{-T})(c\mathbbm{I}_d)^{-1}\bm{m}\right]\right]}{\sqrt{|\mathbbm{I}_d+2c\mathbbm{I}_d|}}=
\frac{\exp \left[-\frac{1}{2}\left[\bm{m}^T\left(\mathbbm{I}_d-\frac{1}{1+2c}\mathbbm{I}_d\right)\frac{1}{c}\mathbbm{I}_d\bm{m}\right]\right]}{\sqrt{|(1+2c)\mathbbm{I}_d|}}\\
&=\frac{\exp \left[-\frac{1}{2}\frac{2c}{1+2c}\frac{1}{c}\bm{m}^T\bm{m}\right]}{\sqrt{|(1+2c)\mathbbm{I}_d|}}=
\frac{\exp \left[-(1+2c)^{-1}||\bm{m}||^2\right]}{(1+2c)^{\frac{d}{2}}}.
\end{split}
\end{equation*}
Note that the above result is equivalent to the form:
\begin{equation*}
\resizebox{1\hsize}{!}{$
\begin{split}
&=\frac{\exp \left[-(1+2c)^{-1}\bm{m}^T\bm{m}\right]}{\sqrt{|(1+2c)\mathbbm{I}_d|}}=
\frac{\exp \left[-\bm{m}^T(\mathbbm{I}_d+2c\mathbbm{I}_d)^{-1}\bm{m}\right]}{\sqrt{|\mathbbm{I}_d+2c\mathbbm{I}_d|}}=
\frac{\exp \left[-\bm{m}^T(\mathbbm{I}_d+2\bm{\Sigma})^{-1}\bm{m}\right]}{\sqrt{|\mathbbm{I}_d+2\bm{\Sigma}|}}.
\end{split}$}
\end{equation*}
To show the $\mathbbm{E}_{q(\bm{U}),q(\bm{V})}[\exp (-||\bm{u}_i-\bm{v}_{j\leftarrow i}||^2)]$, recall here that $\bm{u}_i-\bm{v}_{j\leftarrow i}\sim\text{MVN}_d(\bm{\tilde{u}}_i-\bm{\tilde{v}}_{j\leftarrow i},(\tilde{\sigma}^2_i+\tilde{\varphi}^2_{ij})\mathbbm{I}_d)$.
Incorporating such information into the formula gives:
\begin{equation*}
\begin{split}
& \mathbbm{E}[\exp (-||\bm{u}_i-\bm{v}_{j\leftarrow i}||^2)]=
\frac{\exp \left[-(1+2(\tilde{\sigma}^2_i+\tilde{\varphi}^2_{ij}))^{-1}||\bm{\tilde{u}}_i-\bm{\tilde{v}}_{j\leftarrow i}||^2\right]}{(1+2(\tilde{\sigma}^2_i+\tilde{\varphi}^2_{ij}))^{\frac{d}{2}}}.
\end{split}
\end{equation*}\;\hfill$\square$\\

\end{itemize}


\subsubsection[$U,\mu,\tau,z$]{Evaluate $\mathbbm{E}_q[\text{\normalfont{log}}\;p(\bm{U}|\bm{\mu}, \bm{\tau}, \bm{z})]$}
\label{ELBO_pU_Section_Appendix}

\begin{equation*}
\resizebox{1\hsize}{!}{$
\begin{split}
&\mathbbm{E}_q[\log p(\bm{U}|\bm{\mu}, \bm{\tau}, \bm{z})]=
\mathbbm{E}_q\left[\log \prod^N_{i=1}f_{\text{MVN}_d}(\bm{u}_i|\bm{\mu}_{z_i},1/\tau_{z_i}\mathbbm{I}_d)\right]\\
&=\mathbbm{E}_q\left[\log \prod^N_{i=1}\frac{|\tau_{z_i}\mathbbm{I}_d|^{\frac{1}{2}}}{(2\pi)^{\frac{d}{2}}}\exp \left[-\frac{1}{2}(\bm{u}_i-\bm{\mu}_{z_i})^T\tau_{z_i}\mathbbm{I}_d(\bm{u}_i-\bm{\mu}_{z_i})\right]\right]\\
&=\mathbbm{E}_q\left\{\sum^N_{i=1}\left[-\frac{d}{2}\log (2\pi)+\frac{d}{2}\log (\tau_{z_i})-\frac{1}{2}(\bm{u}_i-\bm{\mu}_{z_i})^T\tau_{z_i}\mathbbm{I}_d(\bm{u}_i-\bm{\mu}_{z_i})\right]\right\}\\
&=-\left[\sum^N_{i=1}\frac{d}{2}\log (2\pi)\right]+\mathbbm{E}_q\left\{\sum^N_{i=1}\left[\frac{d}{2}\log (\tau_{z_i})-\frac{1}{2}\tau_{z_i}||\bm{u}_i-\bm{\mu}_{z_i}||^2\right]\right\}\\
&\equiv -\left[\sum^N_{i=1}\frac{d}{2}\log (2\pi)\right]+\mathbbm{E}_q\left\{\sum^N_{i=1}\sum^K_{k=1}z_{ik}\left[\frac{d}{2}\log (\tau_k)-\frac{1}{2}\tau_k||\bm{u}_i-\bm{\mu}_k||^2\right]\right\}\\
&=-\left[\sum^N_{i=1}\frac{d}{2}\log (2\pi)\right]+
\sum^N_{i=1}\sum^K_{k=1}\mathbbm{E}_q(z_{ik})\left[\frac{d}{2}\mathbbm{E}_q[\log (\tau_k)]-\frac{1}{2}\mathbbm{E}_q(\tau_k)\mathbbm{E}_q\left(||\bm{u}_i-\bm{\mu}_k||^2\right)\right]\\
&=-\left[\sum^N_{i=1}\frac{d}{2}\log (2\pi)\right]+
\sum^N_{i=1}\sum^K_{k=1}\tilde{\pi}_{ik}
\left\{\frac{d}{2}\left[\bm{\Psi}(\tilde{\xi}_k)-\log (\tilde{\psi}_k)\right]-\frac{1}{2}\frac{\tilde{\xi}_k}{\tilde{\psi}_k}\left[||\bm{\tilde{u}}_i-\bm{\tilde{\mu}}_k||^2+d(\tilde{\sigma}^2_i+\tilde{\omega}^2_k)\right]\right\}\\
&=\texttt{const}+\sum^N_{i=1}\sum^K_{k=1}\tilde{\pi}_{ik}
\left\{\frac{d}{2}\left[\bm{\Psi}(\tilde{\xi}_k)-\log (\tilde{\psi}_k)\right]-\frac{1}{2}\frac{\tilde{\xi}_k}{\tilde{\psi}_k}\left[||\bm{\tilde{u}}_i-\bm{\tilde{\mu}}_k||^2+d(\tilde{\sigma}^2_i+\tilde{\omega}^2_k)\right]\right\},
\end{split}$}
\end{equation*}
where $\bm{\Psi}(\theta):=\frac{\partial}{\partial \theta}\log \Gamma(\theta)$ for some parameter $\theta$.\;\hfill$\square$\\

Here we use a trick to replace all the $z_i$ by the alternative equivalent notation $\bm{z}_i=(z_{i1},\dots,z_{iK})^T$ in line 5 above, so that the expectations in the formula can be easily separated.
Since we have that $q_{\bm{z}}(z_i|\tilde{\pi}_{i1},\dots,\tilde{\pi}_{iK}) \sim \text{categorical}(\tilde{\pi}_{i1},\dots,\tilde{\pi}_{iK})$, this implies that $q_{\bm{Z}}(\bm{z}_i|\tilde{\pi}_{i1},\dots,\tilde{\pi}_{iK}) \sim \text{multinomial}(1;\tilde{\pi}_{i1},\dots,\tilde{\pi}_{iK})$, and thus $\mathbbm{E}_q(z_{ik})=\tilde{\pi}_{ik}$ in the last line above.
Provided that $q_{\bm{\mu}}(\bm{\mu}_k|\bm{\tilde{\mu}}_k,\tilde{\omega}^2_k) \sim \text{MVN}_d(\bm{\tilde{\mu}}_k,\tilde{\omega}^2_k\mathbbm{I}_d)$ and combining with the proofs in Appendix~\ref{ELBO_pY_Section_Appendix}, we have that $(\bm{u}_i-\bm{\mu}_k)\sim\text{MVN}_d(\bm{\tilde{u}}_i-\bm{\tilde{\mu}}_k,\tilde{\sigma}^2_i\mathbbm{I}_d+\tilde{\omega}^2_k\mathbbm{I}_d)$, and thus $\mathbbm{E}_q\left(||\bm{u}_i-\bm{\mu}_k||^2\right)=||\bm{\tilde{u}}_i-\bm{\tilde{\mu}}_k||^2+d(\tilde{\sigma}^2_i+\tilde{\omega}^2_k)$.
Since $q_{\bm{\tau}}(\tau_k|\tilde{\xi}_k,\tilde{\psi}_k)\sim\text{Gamma}(\tilde{\xi}_k,\tilde{\psi}_k)$, we have $\mathbbm{E}_q(\tau_k)=\tilde{\xi}_k/\tilde{\psi}_k$.
So it remains to show the $\mathbbm{E}_q[\log (\tau_k)]$ as illustrated below.
Here we prove for the general form:

\textbf{* Find} $\mathbbm{E}[\log (x)]$, \textbf{where} $x \sim \text{Gamma}(a,b)$:
\begin{equation*}
\begin{split}
& \mathbbm{E}[\log (x)]=\int_0^{\infty}\log (x)\frac{b^a}{\Gamma(a)}x^{a-1}\exp(-bx)\text{d}x,\\
& \text{let}\;y=bx\;\text{implying that}\;\frac{\partial y}{\partial x}=b\text{, and then,}\\
&=\int_0^{\infty}\log \left(\frac{y}{b}\right)\frac{b^a}{\Gamma(a)}\left(\frac{y}{b}\right)^{a-1}\exp(-y)\frac{\text{d}y}{b}
=\int_0^{\infty}[\log (y)-\log (b)]\frac{1}{\Gamma(a)}y^{a-1}\exp(-y)\text{d}y\\
&=\int_0^{\infty}\frac{1}{\Gamma(a)}\log (y)y^{a-1}\exp(-y)\text{d}y-\int_0^{\infty}\frac{1}{\Gamma(a)}\log (b)y^{a-1}\exp(-y)\text{d}y\\
&=\frac{1}{\Gamma(a)}\int_0^{\infty}\frac{\partial}{\partial a}[y^{a-1}\exp(-y)]\text{d}y-\frac{\log (b)}{\Gamma(a)}\int_0^{\infty}y^{a-1}\exp(-y)\text{d}y\\
&=\frac{1}{\Gamma(a)}\frac{\partial}{\partial a}\int_0^{\infty}y^{a-1}\exp(-y)\text{d}y-\frac{\log (b)}{\Gamma(a)}\Gamma(a)
=\frac{\frac{\partial}{\partial a}\Gamma(a)}{\Gamma(a)}-\log (b)\\
&=\left[\frac{\partial}{\partial a}\log\Gamma(a)\right]-\log (b)=\bm{\Psi}(a)-\log(b).
\end{split}
\end{equation*}
Thus $\mathbbm{E}_q[\log (\tau_k)]=\bm{\Psi}(\tilde{\xi}_k)-\log(\tilde{\psi}_k)$.\;\hfill$\square$\\


\subsubsection[$V,U,\gamma$]{Evaluate $\mathbbm{E}_q[\text{\normalfont{log}}\;p(\bm{V}|\bm{U}, \bm{\gamma})]$}

\begin{equation*}
\begin{split}
&\mathbbm{E}_q[\log p(\bm{V}|\bm{U}, \bm{\gamma})]=
\mathbbm{E}_q\left[\log \prod^N_{\substack{i,j=1, \\ i\neq j}}f_{\text{MVN}_d}(\bm{v}_{j\leftarrow i}|\bm{u}_j,1/\gamma_j\mathbbm{I}_d)\right]\\
&=\mathbbm{E}_q\left[\log \prod^N_{\substack{i,j=1, \\ i\neq j}}\frac{|\gamma_j\mathbbm{I}_d|^{\frac{1}{2}}}{(2\pi)^{\frac{d}{2}}}\exp \left[-\frac{1}{2}(\bm{v}_{j\leftarrow i}-\bm{u}_j)^T\gamma_j\mathbbm{I}_d(\bm{v}_{j\leftarrow i}-\bm{u}_j)\right]\right]\\
&=\mathbbm{E}_q\left\{\sum^N_{\substack{i,j=1, \\ i\neq j}}\left[-\frac{d}{2}\log (2\pi)+\frac{d}{2}\log (\gamma_j)-\frac{1}{2}(\bm{v}_{j\leftarrow i}-\bm{u}_j)^T\gamma_j\mathbbm{I}_d(\bm{v}_{j\leftarrow i}-\bm{u}_j)\right]\right\}\\
&=-\left[\sum^N_{\substack{i,j=1, \\ i\neq j}}\frac{d}{2}\log (2\pi)\right]+\mathbbm{E}_q\left\{\sum^N_{i=1}\left[\frac{d}{2}\log (\gamma_j)-\frac{1}{2}\gamma_j||\bm{v}_{j\leftarrow i}-\bm{u}_j||^2\right]\right\}\\
&=-\left[\sum^N_{\substack{i,j=1, \\ i\neq j}}\frac{d}{2}\log (2\pi)\right]+\sum^N_{\substack{i,j=1, \\ i\neq j}}\left\{\frac{d}{2}\mathbbm{E}_q[\log (\gamma_j)]-\frac{1}{2}\mathbbm{E}_q(\gamma_j)\mathbbm{E}_q(||\bm{v}_{j\leftarrow i}-\bm{u}_j||^2)\right\}\\
&=-\left[\sum^N_{\substack{i,j=1, \\ i\neq j}}\frac{d}{2}\log (2\pi)\right]+\sum^N_{\substack{i,j=1, \\ i\neq j}}\left\{\frac{d}{2}[\bm{\Psi}(\tilde{a}_j)-\log(\tilde{b}_j)]-\frac{1}{2}\frac{\tilde{a}_j}{\tilde{b}_j}[||\bm{\tilde{v}}_{j\leftarrow i}-\bm{\tilde{u}}_j||^2+d(\tilde{\varphi}^2_{ij}+\tilde{\sigma}^2_j)]\right\}\\
&=\texttt{const}+\sum^N_{\substack{i,j=1, \\ i\neq j}}\left\{\frac{d}{2}[\bm{\Psi}(\tilde{a}_j)-\log(\tilde{b}_j)]-\frac{1}{2}\frac{\tilde{a}_j}{\tilde{b}_j}[||\bm{\tilde{v}}_{j\leftarrow i}-\bm{\tilde{u}}_j||^2+d(\tilde{\varphi}^2_{ij}+\tilde{\sigma}^2_j)]\right\}.
\end{split}
\end{equation*}\;\hfill$\square$\\

Since $q_{\bm{\gamma}}(\gamma_j|\tilde{a}_j,\tilde{b}_j)\sim\text{Gamma}(\tilde{a}_j,\tilde{b}_j)$, and $(\bm{v}_{j\leftarrow i}-\bm{u}_j)\sim\text{MVN}_d(\bm{\tilde{v}}_{j\leftarrow i}-\bm{\tilde{u}}_j,(\tilde{\varphi}^2_{ij}+\tilde{\sigma}^2_j)\mathbbm{I}_d)$, the proofs for the above formula is similar to those illustrated in Appendix~\ref{ELBO_pU_Section_Appendix}.


\subsubsection[$z,\Pi$]{Evaluate $\mathbbm{E}_q[\text{\normalfont{log}}\;p(\bm{z}|\bm{\Pi})]$}

\begin{equation*}
\label{ELBO_pz}
\begin{split}
&\mathbbm{E}_q[\log p(\bm{z}|\bm{\Pi})]=
\mathbbm{E}_q\left[\log \prod^N_{i=1}\prod^K_{k=1}f_{\text{multinomial}}(\bm{z}_i|1;\bm{\Pi})\right]=
\mathbbm{E}_q\left(\log \prod^N_{i=1}\prod^K_{k=1}\pi_k^{z_{ik}}\right)\\
&=\mathbbm{E}_q\left(\sum^N_{i=1}\sum^K_{k=1}z_{ik}\log(\pi_k)\right)=
\sum^N_{i=1}\sum^K_{k=1}\mathbbm{E}_q(z_{ik})\mathbbm{E}_q\left[\log(\pi_k)\right]\\
&=\sum^N_{i=1}\sum^K_{k=1}\tilde{\pi}_{ik}\left[\bm{\Psi}\left(\tilde{\delta}_k\right)-\bm{\Psi}\left(\sum^K_{g=1}\tilde{\delta}_g\right)\right].
\end{split}
\end{equation*}\;\hfill$\square$\\

Provided that $q_{\bm{\Pi}}(\pi_1,\dots,\pi_K|\tilde{\delta}_1,\dots,\tilde{\delta}_K)\sim\text{Dirichlet}(\tilde{\delta}_1,\dots,\tilde{\delta}_K)$, we prove the $\mathbbm{E}_q\left[\log(\pi_k)\right]$ as follows.
We first derive the marginal distribution of a Dirichlet distribution, that is, the distribution of the single $\pi_k$.
Without loss of generality, we start from considering that $\sum^K_{g=1}\pi_g=1$, so once we know $(\pi_1,\dots,\pi_{K-1})$ then the variable $\pi_{K}=1-\sum^{K-1}_{g=1}\pi_g$.
This implies that the Probability Density Function (PDF) of the $\text{Dirichlet}(\pi_1,\dots,\pi_K|\tilde{\delta}_1,\dots,\tilde{\delta}_K)$ can be written as:
$$f_{\text{Dirichlet}}(\pi_1,\dots,\pi_{K-1})\propto(1-\pi_1-\dots-\pi_{K-1})^{\tilde{\delta}_K-1}\prod^{K-1}_{g=1}\pi_g^{\tilde{\delta}_g-1}.$$

\textbf{* Determine the distribution of} $(\pi_1,\dots,\pi_{K-2})$: Integrating out $\pi_{K-1}$ gives the PDF of $(\pi_1,\dots,\pi_{K-2})$ as:
\begin{equation*}
\label{Proof_MarginalDirichlet}
\resizebox{1\hsize}{!}{$
\begin{split}
& f(\pi_1,\dots,\pi_{K-2})\propto\int_0^{1-\pi_1-\dots-\pi_{K-2}}(1-\pi_1-\dots-\pi_{K-2}-\pi_{K-1})^{\tilde{\delta}_K-1}\prod^{K-1}_{g=1}\pi_g^{\tilde{\delta}_g-1}\text{d}\pi_{K-1}\\
&=\prod^{K-2}_{g=1}\pi_g^{\tilde{\delta}_g-1}\int_0^{1-\pi_1-\dots-\pi_{K-2}}(1-\pi_1-\dots-\pi_{K-2}-\pi_{K-1})^{\tilde{\delta}_K-1}\pi_{K-1}^{\tilde{\delta}_{K-1}-1}\text{d}\pi_{K-1},\\
&\text{substituting}\;\pi_{K-1}=(1-\pi_1-\dots-\pi_{K-2})x\;\text{giving}\;\text{d}\pi_{K-1}=(1-\pi_1-\dots-\pi_{K-2})\text{d}x,\;\text{then,}\\
&=\prod^{K-2}_{g=1}\pi_g^{\tilde{\delta}_g-1}\int_0^1[(1-\pi_1-\dots-\pi_{K-2})-(1-\pi_1-\dots-\pi_{K-2})x]^{\tilde{\delta}_K-1}[(1-\pi_1-\dots-\pi_{K-2})x]^{\tilde{\delta}_{K-1}-1}\\
&\times(1-\pi_1-\dots-\pi_{K-2})\text{d}x\\
&=\left(\prod^{K-2}_{g=1}\pi_g^{\tilde{\delta}_g-1}\right)(1-\pi_1-\dots-\pi_{K-2})^{\tilde{\delta}_K+\tilde{\delta}_{K-1}-1}\int_0^1(1-x)^{\tilde{\delta}_K-1}x^{\tilde{\delta}_{K-1}-1}\text{d}x\\
&\propto (1-\pi_1-\dots-\pi_{K-2})^{\tilde{\delta}_K+\tilde{\delta}_{K-1}-1}\prod^{K-2}_{g=1}\pi_g^{\tilde{\delta}_g-1},
\end{split}$}
\end{equation*}
which follows the $\text{Dirichlet}(\tilde{\delta}_1,\dots,\tilde{\delta}_{K-2},\tilde{\delta}_{K-1}+\tilde{\delta}_{K})$ distribution.\;\hfill$\square$\\

\textbf{* Determine the distribution of} $\pi_k$: By repeatedly applying the above approach to integrate out all the variables except $\pi_{k}$, we finally end up with $\left(\pi_{k},\sum^K_{\substack{g=1,\\ g\neq k}}\pi_{g}\right)\sim\text{Dirichlet}\left(\tilde{\delta}_{k},\sum^K_{\substack{g=1,\\ g\neq k}}\tilde{\delta}_{g}\right)$, which is equivalent to the fact that $\pi_{k}\sim\text{Beta}\left(\tilde{\delta}_{k},\sum^K_{\substack{g=1,\\ g\neq k}}\tilde{\delta}_{g}\right)$.\;\hfill$\square$\\

The next step is to prove the expectation of log Beta distribution.
We assume the general form:

\textbf{* Find} $\mathbbm{E}\left[\log(x)\right]$, \textbf{where} $x\sim\text{Beta}(a,b)$:
\begin{equation*}
\label{Proof_ElogBeta}
\begin{split}
& \mathbbm{E}\left[\log(x)\right]=\int^1_0\log(x)\frac{x^{a-1}(1-x)^{b-1}}{\text{B}(a,b)}\text{d}x=
\frac{1}{\text{B}(a,b)}\int^1_0\frac{\partial}{\partial a}[x^{a-1}(1-x)^{b-1}]\text{d}x\\
&=\frac{1}{\text{B}(a,b)}\frac{\partial}{\partial a}\int^1_0x^{a-1}(1-x)^{b-1}\text{d}x=
\frac{1}{\text{B}(a,b)}\frac{\partial}{\partial a}\text{B}(a,b)=\frac{\partial}{\partial a}\log\text{B}(a,b)\\
&=\frac{\partial}{\partial a}\log\frac{\Gamma(a)\Gamma(b)}{\Gamma(a+b)}=
\frac{\partial}{\partial a}\log\Gamma(a)-\frac{\partial}{\partial a}\log\Gamma(a+b)=
\bm{\Psi}(a)-\bm{\Psi}(a+b).
\end{split}
\end{equation*}\;\hfill$\square$\\

Combining the above two proofs, we obtain that $$\mathbbm{E}_q\left[\log(\pi_k)\right]=\bm{\Psi}\left(\tilde{\delta}_{k}\right)-\bm{\Psi}\left(\tilde{\delta}_{k}+\sum^K_{\substack{g=1,\\ g\neq k}}\tilde{\delta}_{g}\right)=\bm{\Psi}\left(\tilde{\delta}_{k}\right)-\bm{\Psi}\left(\sum^K_{g=1}\tilde{\delta}_{g}\right).$$\;\hfill$\square$\\


\subsubsection[]{Evaluate $\mathbbm{E}_q[\text{\normalfont{log}}\;\pi(\beta)]$}

\begin{equation*}
\begin{split}
&\mathbbm{E}_q[\log \pi(\beta)]=
\mathbbm{E}_q\left[\log f_{\text{normal}}(\beta|\eta,\rho^2)\right]=
\mathbbm{E}_q\log \left[\frac{1}{\sqrt{2\pi\rho^2}}\exp[-\frac{(\beta-\eta)^2}{2\rho^2}]\right]\\
&=\mathbbm{E}_q\left[-\frac{1}{2}\log2\pi-\frac{1}{2}\log\rho^2-\frac{(\beta-\eta)^2}{2\rho^2}\right]=
-\frac{1}{2}\log2\pi-\frac{1}{2}\log\rho^2-\frac{1}{2\rho^2}\mathbbm{E}_q\left[(\beta-\eta)^2\right]\\
&=-\frac{1}{2}\log2\pi-\frac{1}{2}\log\rho^2-\frac{1}{2\rho^2}\left\{\left[\mathbbm{E}_q(\beta-\eta)\right]^2+\text{Var}_q(\beta-\eta)\right\}\\
&=-\frac{1}{2}\log2\pi-\frac{1}{2}\log\rho^2-\frac{1}{2\rho^2}\left[(\tilde{\eta}-\eta)^2+\tilde{\rho}^2\right]\\
&=\texttt{const}-\frac{1}{2\rho^2}\left[(\tilde{\eta}-\eta)^2+\tilde{\rho}^2\right].
\end{split}
\end{equation*}\;\hfill$\square$\\


\subsubsection[$\mu$]{Evaluate $\mathbbm{E}_q[\text{\normalfont{log}}\;\pi(\bm{\mu})]$}

\begin{equation*}
\resizebox{1\hsize}{!}{$
\begin{split}
&\mathbbm{E}_q[\log \pi(\bm{\mu})]=
\mathbbm{E}_q\left[\log \prod^K_{k=1}f_{\text{MVN}_d}(\bm{\mu}_k|\bm{0},\omega^2\mathbbm{I}_d)\right]=
\mathbbm{E}_q\left[\log \prod^K_{k=1}\frac{\exp(-\frac{\bm{\mu}_k^T\bm{\mu}_k}{2\omega^2})}{\sqrt{(2\pi)^d(\omega^2)^d}}\right]\\
&=\mathbbm{E}_q\left\{\sum^K_{k=1}\left[-\frac{d}{2}\log(2\pi)-\frac{d}{2}\log(\omega^2)-\frac{\bm{\mu}_k^T\bm{\mu}_k}{2\omega^2}\right]\right\}
=\sum^K_{k=1}\left[-\frac{d}{2}\log(2\pi)-\frac{d}{2}\log(\omega^2)-\frac{\mathbbm{E}_q\left(\bm{\mu}_k^T\bm{\mu}_k\right)}{2\omega^2}\right]\\
&=\sum^K_{k=1}\left[-\frac{d}{2}\log(2\pi)-\frac{d}{2}\log(\omega^2)\right]-\sum^K_{k=1}\left[\frac{1}{2\omega^2}\left(\bm{\tilde{\mu}}_k^T\bm{\tilde{\mu}}_k+d\tilde{\omega}^2_k\right)\right]
=\texttt{const}-\sum^K_{k=1}\left[\frac{1}{2\omega^2}\left(\bm{\tilde{\mu}}_k^T\bm{\tilde{\mu}}_k+d\tilde{\omega}^2_k\right)\right].
\end{split}$}
\end{equation*}\;\hfill$\square$\\


\subsubsection[$\tau$]{Evaluate $\mathbbm{E}_q[\text{\normalfont{log}}\;\pi(\bm{\tau})]$}

\begin{equation*}
\begin{split}
&\mathbbm{E}_q[\log \pi(\bm{\tau})]=
\mathbbm{E}_q\left[\log \prod^K_{k=1}f_{\text{gamma}}(\tau_k|\xi,\psi)\right]=
\mathbbm{E}_q\left[\log \prod^K_{k=1}\frac{\psi^\xi}{\Gamma(\xi)}\tau_k^{\xi-1}\exp(-\psi\tau_k)\right]\\
&=\mathbbm{E}_q\left[\sum^K_{k=1}\left[\xi\log\psi-\log\Gamma(\xi)+(\xi-1)\log(\tau_k)-\psi\tau_k\right]\right]\\
&=\sum^K_{k=1}\left[\xi\log\psi-\log\Gamma(\xi)\right]+\sum^K_{k=1}\left\{(\xi-1)\mathbbm{E}_q\left[\log(\tau_k)\right]-\psi\mathbbm{E}_q\left(\tau_k\right)\right\}\\
&=\sum^K_{k=1}\left[\xi\log\psi-\log\Gamma(\xi)\right]+\sum^K_{k=1}\left\{(\xi-1)\left[\bm{\Psi}\left(\tilde{\xi}_k\right)-\log\tilde{\psi}_k\right]-\psi\frac{\tilde{\xi}_k}{\tilde{\psi}_k}\right\}\\
&=\texttt{const}+\sum^K_{k=1}\left\{(\xi-1)\left[\bm{\Psi}\left(\tilde{\xi}_k\right)-\log\tilde{\psi}_k\right]-\psi\frac{\tilde{\xi}_k}{\tilde{\psi}_k}\right\}.
\end{split}
\end{equation*}\;\hfill$\square$\\



\subsubsection[$\gamma$]{Evaluate $\mathbbm{E}_q[\text{\normalfont{log}}\;\pi(\bm{\gamma})]$}

\begin{equation*}
\begin{split}
&\mathbbm{E}_q[\log \pi(\bm{\gamma})]=
\mathbbm{E}_q\left[\log \prod^N_{i=1}f_{\text{gamma}}(\gamma_i|a,b)\right]=
\mathbbm{E}_q\left[\log \prod^N_{i=1}\frac{b^a}{\Gamma(a)}\gamma_i^{a-1}\exp(-b\gamma_i)\right]\\
&=\mathbbm{E}_q\left[\sum^N_{i=1}\left[a\log b-\log\Gamma(a)+(a-1)\log(\gamma_i)-b\gamma_i\right]\right]\\
&=\sum^N_{i=1}\left[a\log b-\log\Gamma(a)\right]+\sum^N_{i=1}\left\{(a-1)\mathbbm{E}_q\left[\log(\gamma_i)\right]-b\mathbbm{E}_q\left(\gamma_i\right)\right\}\\
&=\sum^N_{i=1}\left[a\log b-\log\Gamma(a)\right]+\sum^N_{i=1}\left\{(a-1)\left[\bm{\Psi}\left(\tilde{a}_i\right)-\log\tilde{b}_i\right]-b\frac{\tilde{a}_i}{\tilde{b}_i}\right\}\\
&=\texttt{const}+\sum^N_{i=1}\left\{(a-1)\left[\bm{\Psi}\left(\tilde{a}_i\right)-\log\tilde{b}_i\right]-b\frac{\tilde{a}_i}{\tilde{b}_i}\right\}.
\end{split}
\end{equation*}\;\hfill$\square$\\


\subsubsection[$\Pi$]{Evaluate $\mathbbm{E}_q[\text{\normalfont{log}}\;\pi(\bm{\Pi})]$}

\begin{equation*}
\begin{split}
&\mathbbm{E}_q[\log \pi(\bm{\Pi})]=
\mathbbm{E}_q\left[\log f_{\text{Dirichlet}}(\bm{\Pi}|\bm{\delta})\right]=
\mathbbm{E}_q\log \left[\frac{\Gamma\left(\sum^K_{k=1}\delta_k\right)}{\prod^K_{k=1}\Gamma(\delta_k)}\prod^K_{k=1}\pi_k^{\delta_k-1}\right]\\
&=\mathbbm{E}_q\left[\log \Gamma\left(\sum^K_{k=1}\delta_k\right)-\sum^K_{k=1}\log\Gamma(\delta_k)+\sum^K_{k=1}(\delta_k-1)\log(\pi_k)\right]\\
&=\log \Gamma\left(\sum^K_{k=1}\delta_k\right)-\sum^K_{k=1}\log\Gamma(\delta_k)+\sum^K_{k=1}(\delta_k-1)\mathbbm{E}_q\left[\log(\pi_k)\right]\\
&=\log \Gamma\left(\sum^K_{k=1}\delta_k\right)-\sum^K_{k=1}\log\Gamma(\delta_k)+\sum^K_{k=1}(\delta_k-1)\left[\bm{\Psi}\left(\tilde{\delta}_{k}\right)-\bm{\Psi}\left(\sum^K_{g=1}\tilde{\delta}_{g}\right)\right]\\
&=\texttt{const}+\sum^K_{k=1}(\delta_k-1)\left[\bm{\Psi}\left(\tilde{\delta}_{k}\right)-\bm{\Psi}\left(\sum^K_{g=1}\tilde{\delta}_{g}\right)\right].
\end{split}
\end{equation*}\;\hfill$\square$\\


\subsubsection[$U,\sigma$]{Evaluate $\mathbbm{E}_q[\text{\normalfont{log}}\;q_{\bm{U}}(\bm{U}|\bm{\tilde{U}}, \bm{\tilde{\sigma}^2})]$}

\begin{equation*}
\begin{split}
&\mathbbm{E}_q[\log q_{\bm{U}}(\bm{U}|\bm{\tilde{U}}, \bm{\tilde{\sigma}^2})]=
\mathbbm{E}_q\left[\log \prod^N_{i=1}f_{\text{MVN}_d}(\bm{u}_i|\bm{\tilde{u}}_i, \tilde{\sigma}^2_i\mathbbm{I}_d)\right]\\
&=\mathbbm{E}_q\left[\log \prod^N_{i=1}\frac{1}{(2\pi)^{\frac{d}{2}}|\tilde{\sigma}^2_i\mathbbm{I}_d|^{\frac{1}{2}}}\exp \left[-\frac{1}{2}(\bm{u}_i-\bm{\tilde{u}}_i)^T\frac{1}{\tilde{\sigma}^2_i}\mathbbm{I}_d(\bm{u}_i-\bm{\tilde{u}}_i)\right]\right]\\
&=\mathbbm{E}_q\left\{\sum^N_{i=1}\left[-\frac{d}{2}\log (2\pi)-\frac{d}{2}\log (\tilde{\sigma}^2_i)-\frac{1}{2\tilde{\sigma}^2_i}(\bm{u}_i-\bm{\tilde{u}}_i)^T(\bm{u}_i-\bm{\tilde{u}}_i)\right]\right\}\\
&=\sum^N_{i=1}\left[-\frac{d}{2}\log (2\pi)-\frac{d}{2}\log (\tilde{\sigma}^2_i)-\frac{1}{2\tilde{\sigma}^2_i}\mathbbm{E}_q\left(||\bm{u}_i-\bm{\tilde{u}}_i||^2\right)\right]\\
&=\sum^N_{i=1}\left[-\frac{d}{2}\log (2\pi)-\frac{d}{2}\log (\tilde{\sigma}^2_i)-\frac{1}{2\tilde{\sigma}^2_i}\left(||\bm{\tilde{u}}_i-\bm{\tilde{u}}_i||^2+d\tilde{\sigma}^2_i\right)\right]\\
&=\sum^N_{i=1}\left[-\frac{d}{2}\log (2\pi)-\frac{d}{2}\log (\tilde{\sigma}^2_i)-\frac{d}{2}\right]=
\sum^N_{i=1}\left[-\frac{d}{2}\log (\tilde{\sigma}^2_i)\right]-\sum^N_{i=1}\left[\frac{d}{2}\log (2\pi)+\frac{d}{2}\right]\\
&=\sum^N_{i=1}\left[-\frac{d}{2}\log (\tilde{\sigma}^2_i)\right]-\texttt{const}.
\end{split}
\end{equation*}\;\hfill$\square$\\


\subsubsection[$V,\varphi$]{Evaluate $\mathbbm{E}_q[\text{\normalfont{log}}\;q_{\bm{V}}(\bm{V}|\bm{\tilde{V}}, \bm{\tilde{\varphi}^2})]$}

\begin{equation*}
\begin{split}
&\mathbbm{E}_q[\log q_{\bm{V}}(\bm{V}|\bm{\tilde{V}}, \bm{\tilde{\varphi}^2})]=
\mathbbm{E}_q\left[\log \prod^N_{\substack{i,j=1, \\ i\neq j}}f_{\text{MVN}_d}(\bm{v}_{j\leftarrow i}|\bm{\tilde{v}}_{j\leftarrow i},\tilde{\varphi}^2_{ij}\mathbbm{I}_d)\right]\\
&=\mathbbm{E}_q\left[\log \prod^N_{\substack{i,j=1, \\ i\neq j}}\frac{1}{(2\pi)^{\frac{d}{2}}|\tilde{\varphi}^2_{ij}\mathbbm{I}_d|^{\frac{1}{2}}}\exp \left[-\frac{1}{2\tilde{\varphi}^2_{ij}}(\bm{v}_{j\leftarrow i}-\bm{\tilde{v}}_{j\leftarrow i})^T(\bm{v}_{j\leftarrow i}-\bm{\tilde{v}}_{j\leftarrow i})\right]\right]\\
&=\sum^N_{\substack{i,j=1, \\ i\neq j}}\left[-\frac{d}{2}\log (2\pi)-\frac{d}{2}\log (\tilde{\varphi}^2_{ij})-\frac{1}{2\tilde{\varphi}^2_{ij}}\mathbbm{E}_q\left(||\bm{v}_{j\leftarrow i}-\bm{\tilde{v}}_{j\leftarrow i}||^2\right)\right]\\
&=\sum^N_{\substack{i,j=1, \\ i\neq j}}\left[-\frac{d}{2}\log (2\pi)-\frac{d}{2}\log (\tilde{\varphi}^2_{ij})-\frac{1}{2\tilde{\varphi}^2_{ij}}\left(||\bm{\tilde{v}}_{j\leftarrow i}-\bm{\tilde{v}}_{j\leftarrow i}||^2+d\tilde{\varphi}^2_{ij}\right)\right]\\
&=\sum^N_{\substack{i,j=1, \\ i\neq j}}\left[-\frac{d}{2}\log (2\pi)-\frac{d}{2}\log (\tilde{\varphi}^2_{ij})-\frac{d}{2}\right]=
\sum^N_{i=1}\left[-\frac{d}{2}\log (\tilde{\varphi}^2_{ij})\right]-\sum^N_{i=1}\left[\frac{d}{2}\log (2\pi)+\frac{d}{2}\right]\\
&=\sum^N_{\substack{i,j=1, \\ i\neq j}}\left[-\frac{d}{2}\log (\tilde{\varphi}^2_{ij})\right]-\texttt{const}.
\end{split}
\end{equation*}\;\hfill$\square$\\


\subsubsection[$U,\sigma$]{Evaluate $\mathbbm{E}_q[\text{\normalfont{log}}\;q_{\bm{Z}}(\bm{Z}|\bm{\tilde{\Pi}})]$}

\begin{equation*}
\begin{split}
&\mathbbm{E}_q\left[\log q_{\bm{Z}}(\bm{Z}|\bm{\tilde{\Pi}})\right]=
\mathbbm{E}_q\left[\log \prod^N_{i=1}\prod^K_{k=1}f_{\text{multinomial}}(\bm{z}_i|1;\bm{\tilde{\Pi}})\right]=
\mathbbm{E}_q\left[\log\prod^N_{i=1}\prod^K_{k=1}\tilde{\pi}_{ik}^{z_{ik}}\right]\\
&=\mathbbm{E}_q\left[\sum^N_{i=1}\sum^K_{k=1}z_{ik}\log(\tilde{\pi}_{ik})\right]=
\sum^N_{i=1}\sum^K_{k=1}\mathbbm{E}_q(z_{ik})\log(\tilde{\pi}_{ik})=
\sum^N_{i=1}\sum^K_{k=1}\tilde{\pi}_{ik}\log(\tilde{\pi}_{ik}).
\end{split}
\end{equation*}\;\hfill$\square$\\


\subsubsection[]{Evaluate $\mathbbm{E}_q[\text{\normalfont{log}}\;q_{\beta}(\beta|\tilde{\eta},\tilde{\rho}^2)]$}

\begin{equation*}
\begin{split}
&\mathbbm{E}_q\left[\log q_{\beta}(\beta|\tilde{\eta},\tilde{\rho}^2)\right]=
\mathbbm{E}_q\left[\log f_{\text{normal}}(\beta|\tilde{\eta},\tilde{\rho}^2)\right]=
\mathbbm{E}_q\left\{\log \left[\frac{1}{\sqrt{2\pi\tilde{\rho}^2}}\exp(-\frac{(\beta-\tilde{\eta})^2}{2\tilde{\rho}^2})\right]\right\}\\
&=-\frac{1}{2}\log(2\pi)-\frac{1}{2}\log(\tilde{\rho}^2)-\frac{1}{2\tilde{\rho}^2}\mathbbm{E}_q\left[(\beta-\tilde{\eta})^2\right]\\
&=-\frac{1}{2}\log(2\pi)-\frac{1}{2}\log(\tilde{\rho}^2)-\frac{1}{2\tilde{\rho}^2}\left[\mathbbm{E}^2_q(\beta-\tilde{\eta})+\text{var}_q(\beta-\tilde{\eta})\right]\\
&=-\frac{1}{2}\log(2\pi)-\frac{1}{2}\log(\tilde{\rho}^2)-\frac{1}{2\tilde{\rho}^2}\left(0+\tilde{\rho}^2\right)\\
&=-\frac{1}{2}\log(2\pi)-\frac{1}{2}-\frac{1}{2}\log(\tilde{\rho}^2)=
\texttt{const}-\frac{1}{2}\log(\tilde{\rho}^2).
\end{split}
\end{equation*}\;\hfill$\square$\\


\subsubsection[$\mu$]{Evaluate $\mathbbm{E}_q[\text{\normalfont{log}}\;q_{\bm{\mu}}(\bm{\mu}|\bm{\tilde{\mu}},\bm{\tilde{\omega}^2})]$}

\begin{equation*}
\resizebox{1\hsize}{!}{$
\begin{split}
&\mathbbm{E}_q[\log q_{\bm{\mu}}(\bm{\mu}|\bm{\tilde{\mu}},\bm{\tilde{\omega}^2})]=
\mathbbm{E}_q\left[\log \prod^K_{k=1}f_{\text{MVN}_d}(\bm{\mu}_k|\bm{\tilde{\mu}}_k,\tilde{\omega}^2_k\mathbbm{I}_d)\right]=
\mathbbm{E}_q\left[\log \prod^K_{k=1}\frac{\exp(-\frac{(\bm{\mu}_k-\bm{\tilde{\mu}}_k)^T(\bm{\mu}_k-\bm{\tilde{\mu}}_k)}{2\tilde{\omega}^2_k})}{\sqrt{(2\pi)^d(\tilde{\omega}^2_k)^d}}\right]\\
&=\mathbbm{E}_q\left\{\sum^K_{k=1}\left[-\frac{d}{2}\log(2\pi)-\frac{d}{2}\log(\tilde{\omega}^2_k)-\frac{(\bm{\mu}_k-\bm{\tilde{\mu}}_k)^T(\bm{\mu}_k-\bm{\tilde{\mu}}_k)}{2\tilde{\omega}^2_k}\right]\right\}\\
&=\sum^K_{k=1}\left[-\frac{d}{2}\log(2\pi)-\frac{d}{2}\log(\tilde{\omega}^2_k)-\frac{\mathbbm{E}_q\left[(\bm{\mu}_k-\bm{\tilde{\mu}}_k)^T(\bm{\mu}_k-\bm{\tilde{\mu}}_k)\right]}{2\tilde{\omega}^2_k}\right]\\
&=\sum^K_{k=1}\left[-\frac{d}{2}\log(2\pi)-\frac{d}{2}\log(\tilde{\omega}^2_k)-\frac{1}{2\tilde{\omega}^2_k}\left[(\bm{\tilde{\mu}}_k-\bm{\tilde{\mu}}_k)^T(\bm{\tilde{\mu}}_k-\bm{\tilde{\mu}}_k)+d\tilde{\omega}^2_k\right]\right]\\
&=\sum^K_{k=1}\left[-\frac{d}{2}\log(\tilde{\omega}^2_k)\right]-\sum^K_{k=1}\left[\frac{d}{2}\log(2\pi)+\frac{d}{2}\right]=
\sum^K_{k=1}\left[-\frac{d}{2}\log(\tilde{\omega}^2_k)\right]-\texttt{const}.
\end{split}$}
\end{equation*}\;\hfill$\square$\\


\subsubsection[$\tau,\xi,\psi$]{Evaluate $\mathbbm{E}_q[\text{\normalfont{log}}\;q_{\bm{\tau}}(\bm{\tau}|\bm{\tilde{\xi}},\bm{\tilde{\psi}})]$}

\begin{equation*}
\begin{split}
&\mathbbm{E}_q[\log q_{\bm{\tau}}(\bm{\tau}|\bm{\tilde{\xi}},\bm{\tilde{\psi}})]=
\mathbbm{E}_q\left[\log \prod^K_{k=1}f_{\text{gamma}}(\tau_k|\tilde{\xi}_k,\tilde{\psi}_k)\right]=
\mathbbm{E}_q\left\{\log \prod^K_{k=1}\frac{\tilde{\psi}_k^{\tilde{\xi}_k}}{\Gamma(\tilde{\xi}_k)}\tau_k^{\tilde{\xi}_k-1}\exp(-\tilde{\psi}_k\tau_k)\right]\\
&=\mathbbm{E}_q\left[\sum^K_{k=1}\left[\tilde{\xi}_k\log\tilde{\psi}_k-\log\Gamma(\tilde{\xi}_k)+(\tilde{\xi}_k-1)\log(\tau_k)-\tilde{\psi}_k\tau_k\right]\right\}\\
&=\sum^K_{k=1}\left\{\tilde{\xi}_k\log\tilde{\psi}_k-\log\Gamma(\tilde{\xi}_k)+(\tilde{\xi}_k-1)\mathbbm{E}_q\left[\log(\tau_k)\right]-\tilde{\psi}_k\mathbbm{E}_q\left(\tau_k\right)\right\}\\
&=\sum^K_{k=1}\left\{\tilde{\xi}_k\log\tilde{\psi}_k-\log\Gamma(\tilde{\xi}_k)+(\tilde{\xi}_k-1)\left[\bm{\Psi}\left(\tilde{\xi}_k\right)-\log\tilde{\psi}_k\right]-\tilde{\psi}_k\frac{\tilde{\xi}_k}{\tilde{\psi}_k}\right\}\\
&=\sum^K_{k=1}\left\{-\log\Gamma(\tilde{\xi}_k)+(\tilde{\xi}_k-1)\bm{\Psi}\left(\tilde{\xi}_k\right)+\log\tilde{\psi}_k-\tilde{\xi}_k\right\}.
\end{split}
\end{equation*}\;\hfill$\square$\\


\subsubsection[$\gamma$]{Evaluate $\mathbbm{E}_q[\text{\normalfont{log}}\;q_{\bm{\gamma}}(\bm{\gamma}|\bm{\tilde{a}},\bm{\tilde{b}})]$}

\begin{equation*}
\begin{split}
&\mathbbm{E}_q[\log q_{\bm{\gamma}}(\bm{\gamma}|\bm{\tilde{a}},\bm{\tilde{b}})]=
\mathbbm{E}_q\left[\log \prod^N_{i=1}f_{\text{gamma}}(\gamma_i|\tilde{a}_i,\tilde{b}_i)\right]=
\mathbbm{E}_q\left[\log \prod^N_{i=1}\frac{\tilde{b}_i^{\tilde{a}_i}}{\Gamma(\tilde{a}_i)}\gamma_i^{\tilde{a}_i-1}\exp(-\tilde{b}_i\gamma_i)\right]\\
&=\sum^N_{i=1}\left\{\tilde{a}_i\log \tilde{b}_i-\log\Gamma(\tilde{a}_i)+(\tilde{a}_i-1)\mathbbm{E}_q\left[\log(\gamma_i)\right]-\tilde{b}_i\mathbbm{E}_q\left(\gamma_i\right)\right\}\\
&=\sum^N_{i=1}\left\{\tilde{a}_i\log \tilde{b}_i-\log\Gamma(\tilde{a}_i)+(\tilde{a}_i-1)\left[\bm{\Psi}\left(\tilde{a}_i\right)-\log\tilde{b}_i\right]-\tilde{b}_i\frac{\tilde{a}_i}{\tilde{b}_i}\right\}\\
&=\sum^N_{i=1}\left\{-\log\Gamma(\tilde{a}_i)+(\tilde{a}_i-1)\bm{\Psi}\left(\tilde{a}_i\right)+\log\tilde{b}_i-\tilde{a}_i\right\}.
\end{split}
\end{equation*}\;\hfill$\square$\\


\subsubsection[$\Pi,\delta$]{Evaluate $\mathbbm{E}_q[\text{\normalfont{log}}\;q_{\bm{\Pi}}(\bm{\Pi}|\bm{\tilde{\delta}})]$}

\begin{equation*}
\begin{split}
&\mathbbm{E}_q\left[\log q_{\bm{\Pi}}(\bm{\Pi}|\bm{\tilde{\delta}})\right]=
\mathbbm{E}_q\left[\log f_{\text{Dirichlet}}(\bm{\Pi}|\bm{\tilde{\delta}})\right]=
\mathbbm{E}_q\log \left[\frac{\Gamma\left(\sum^K_{k=1}\tilde{\delta}_k\right)}{\prod^K_{k=1}\Gamma(\tilde{\delta}_k)}\prod^K_{k=1}\pi_k^{\tilde{\delta}_k-1}\right]\\
&=\mathbbm{E}_q\left[\log \Gamma\left(\sum^K_{k=1}\tilde{\delta}_k\right)-\sum^K_{k=1}\log\Gamma(\tilde{\delta}_k)+\sum^K_{k=1}(\tilde{\delta}_k-1)\log(\pi_k)\right]\\
&=\log \Gamma\left(\sum^K_{k=1}\tilde{\delta}_k\right)-\sum^K_{k=1}\log\Gamma(\tilde{\delta}_k)+\sum^K_{k=1}(\tilde{\delta}_k-1)\mathbbm{E}_q\left[\log(\pi_k)\right]\\
&=\log \Gamma\left(\sum^K_{k=1}\tilde{\delta}_k\right)-\sum^K_{k=1}\log\Gamma(\tilde{\delta}_k)+\sum^K_{k=1}(\tilde{\delta}_k-1)\left[\bm{\Psi}\left(\tilde{\delta}_{k}\right)-\bm{\Psi}\left(\sum^K_{g=1}\tilde{\delta}_{g}\right)\right].
\end{split}
\end{equation*}\;\hfill$\square$\\


\subsubsection{Final formula for the ELBO}

Combining the terms above that we have calculated for the ELBO, we finally obtain:
\begin{equation*}
\label{ELBO_Final}
\resizebox{1\hsize}{!}{$
\begin{split}
\mathcal{F}(\bm{\tilde{\Theta}}) & = \mathbbm{E}_q[\log p(\bm{Y}|\bm{U},\bm{V}, \beta)]\\
&+\mathbbm{E}_q[\log p(\bm{U}|\bm{\mu}, \bm{\tau}, \bm{z})]-\mathbbm{E}_q[\log q_{\bm{U}}(\bm{U}|\bm{\tilde{U}},\bm{\tilde{\sigma}^2})]\\
&+\mathbbm{E}_q[\log p(\bm{V}|\bm{U}, \bm{\gamma})]-\mathbbm{E}_q[\log q_{\bm{V}}(\bm{V}|\bm{\tilde{V}},\bm{\tilde{\varphi}^2})]\\
&+\mathbbm{E}_q[\log p(\bm{z}|\bm{\Pi})]-\mathbbm{E}_q[\log q_{\bm{z}}(\bm{z}|\bm{\tilde{\Pi}})]\\
&+\mathbbm{E}_q[\log \pi(\beta)]-\mathbbm{E}_q[\log q_{\beta}(\beta|\tilde{\eta},\tilde{\rho}^2)]\\
&+\mathbbm{E}_q[\log \pi(\bm{\mu})]-\mathbbm{E}_q[\log q_{\bm{\mu}}(\bm{\mu}|\bm{\tilde{\mu}},\bm{\tilde{\omega}^2})]\\
&+\mathbbm{E}_q[\log \pi(\bm{\tau})]-\mathbbm{E}_q[\log q_{\bm{\tau}}(\bm{\tau}|\bm{\tilde{\xi}},\bm{\tilde{\psi}})]\\
&+\mathbbm{E}_q[\log \pi(\bm{\gamma})]-\mathbbm{E}_q[\log q_{\bm{\gamma}}(\bm{\gamma}|\bm{\tilde{a}},\bm{\tilde{b}})]\\
&+\mathbbm{E}_q[\log \pi(\bm{\Pi})]-\mathbbm{E}_q[\log q_{\bm{\Pi}}(\bm{\Pi}|\bm{\tilde{\delta}})]\\
&=\sum^N_{\substack{i,j=1, \\ i\neq j}}\left\{ y_{ij}\left[\tilde{\eta}-||\bm{\tilde{u}}_i-\bm{\tilde{v}}_{j\leftarrow i}||^2-d\left(\tilde{\sigma}^2_i+\tilde{\varphi}^2_{ij}\right)\right]-\frac{\exp \left(\tilde{\eta}+\frac{\tilde{\rho}^2}{2}-\frac{||\bm{\tilde{u}}_i-\bm{\tilde{v}}_{j\leftarrow i}||^2}{1+2\tilde{\sigma}^2_i+2\tilde{\varphi}^2_{ij}}\right)}{(1+2\tilde{\sigma}^2_i+2\tilde{\varphi}^2_{ij})^{d/2}} \right\}\\
&+\sum^N_{\substack{i,j=1, \\ i\neq j}}\left\{\frac{d}{2}\left[\bm{\Psi}(\tilde{a}_j)-\log(\tilde{b}_j)\right]-\frac{1}{2}\frac{\tilde{a}_j}{\tilde{b}_j}[||\bm{\tilde{v}}_{j\leftarrow i}-\bm{\tilde{u}}_j||^2+d(\tilde{\varphi}^2_{ij}+\tilde{\sigma}^2_j)]+\frac{d}{2}\log (\tilde{\varphi}^2_{ij})\right\}\\
&+\sum^N_{i=1}\sum^K_{k=1}\tilde{\pi}_{ik}
\left\{\frac{d}{2}\left[\bm{\Psi}(\tilde{\xi}_k)-\log (\tilde{\psi}_k)\right]-\frac{1}{2}\frac{\tilde{\xi}_k}{\tilde{\psi}_k}\left[||\bm{\tilde{u}}_i-\bm{\tilde{\mu}}_k||^2+d(\tilde{\sigma}^2_i+\tilde{\omega}^2_k)\right]\right\}\\
&+\sum^N_{i=1}\sum^K_{k=1}\tilde{\pi}_{ik}\left[\bm{\Psi}\left(\tilde{\delta}_k\right)-\bm{\Psi}\left(\sum^K_{g=1}\tilde{\delta}_g\right)-\log(\tilde{\pi}_{ik})\right]\\
&+\sum^K_{k=1}\left\{(\delta_k-\tilde{\delta}_k)\left[\bm{\Psi}\left(\tilde{\delta}_{k}\right)-\bm{\Psi}\left(\sum^K_{g=1}\tilde{\delta}_{g}\right)\right]+\log\Gamma(\tilde{\delta}_k)
-\frac{1}{2\omega^2}\left(\bm{\tilde{\mu}}_k^T\bm{\tilde{\mu}}_k+d\tilde{\omega}^2_k\right)+\frac{d}{2}\log(\tilde{\omega}^2_k)\right\}\\
&+\sum^K_{k=1}\left\{(\xi-\tilde{\xi}_k)\bm{\Psi}\left(\tilde{\xi}_k\right)-\xi\log\tilde{\psi}_k-\psi\frac{\tilde{\xi}_k}{\tilde{\psi}_k}+\log\Gamma(\tilde{\xi}_k)+\tilde{\xi}_k\right\}\\
&+\sum^N_{i=1}\left\{(a-\tilde{a}_i)\bm{\Psi}\left(\tilde{a}_i\right)-a\log\tilde{b}_i-b\frac{\tilde{a}_i}{\tilde{b}_i}+\log\Gamma(\tilde{a}_i)+\tilde{a}_i+\frac{d}{2}\log (\tilde{\sigma}^2_i)\right\}\\
&-\frac{1}{2\rho^2}\left[(\tilde{\eta}-\eta)^2+\tilde{\rho}^2\right]+\frac{1}{2}\log(\tilde{\rho}^2)-\log \Gamma\left(\sum^K_{k=1}\tilde{\delta}_k\right)+\texttt{const}.
\end{split}$}
\end{equation*}\;\hfill$\square$\\


\subsection{Maximization of the ELBO for the MLPCM}
\label{MLPCM_ELBO_Maximization_Appendix}


\subsubsection{Proof of the Proposition~\ref{Proposition_Pi_MuOmega2_XiPsi_ab}}
\label{Proof_Proposition_Pi_MuOmega2_XiPsi_ab}
For $j=1,2,\dots,N$ and for $k=1,2,\dots,K$, the 1st-order partial derivative of the ELBO with respect to the $\tilde{\pi}_{jk}$ is obtained as:
\begin{equation*}
\resizebox{1\hsize}{!}{$
\begin{split}
&\frac{\partial\mathcal{F}}{\partial\tilde{\pi}_{jk}}=
\frac{\partial}{\partial\tilde{\pi}_{jk}}\left\{\tilde{\pi}_{jk}
\left[\frac{d}{2}\left[\bm{\Psi}(\tilde{\xi}_k)-\log (\tilde{\psi}_k)\right]-\frac{1}{2}\frac{\tilde{\xi}_k}{\tilde{\psi}_k}\left[||\bm{\tilde{u}}_j-\bm{\tilde{\mu}}_k||^2+d(\tilde{\sigma}^2_j+\tilde{\omega}^2_k)\right]+\bm{\Psi}\left(\tilde{\delta}_k\right)-\bm{\Psi}\left(\sum^K_{g=1}\tilde{\delta}_g\right)\right]\right\}\\
&-\frac{\partial}{\partial\tilde{\pi}_{jk}}\left[\tilde{\pi}_{jk}\log(\tilde{\pi}_{jk})\right]\\
&=\frac{d}{2}\left[\bm{\Psi}(\tilde{\xi}_k)-\log (\tilde{\psi}_k)\right]-\frac{1}{2}\frac{\tilde{\xi}_k}{\tilde{\psi}_k}\left[||\bm{\tilde{u}}_j-\bm{\tilde{\mu}}_k||^2+d(\tilde{\sigma}^2_j+\tilde{\omega}^2_k)\right]+\bm{\Psi}\left(\tilde{\delta}_k\right)-\bm{\Psi}\left(\sum^K_{g=1}\tilde{\delta}_g\right)-\log(\tilde{\pi}_{jk})-1.
\end{split}$}
\end{equation*}
Provided with the fact that $\sum^K_{k=1}\tilde{\pi}_{jk}=1$, we apply the method of Lagrange multipliers to introduce the Lagrange multiplier $\lambda_j$ in order to optimize the ELBO with respect to the constraint of $\tilde{\pi}_{jk}$, that is, letting
$$A_{jk}=\frac{d}{2}\left[\bm{\Psi}(\tilde{\xi}_k)-\log (\tilde{\psi}_k)\right]-\frac{1}{2}\frac{\tilde{\xi}_k}{\tilde{\psi}_k}\left[||\bm{\tilde{u}}_j-\bm{\tilde{\mu}}_k||^2+d(\tilde{\sigma}^2_j+\tilde{\omega}^2_k)\right]+\bm{\Psi}\left(\tilde{\delta}_k\right)-\bm{\Psi}\left(\sum^K_{g=1}\tilde{\delta}_g\right),$$
we solve for the following equations:
\begin{equation*}
\begin{cases} 
\frac{\partial}{\partial\tilde{\pi}_{jk}}\left[\mathcal{F}+\lambda_j\left(\sum^K_{g=1}\tilde{\pi}_{jg}-1\right)\right]=
A_{jk}-\log(\tilde{\pi}_{jk})-1+\lambda_j;\\
\frac{\partial}{\partial\lambda_j}\left[\mathcal{F}+\lambda_i\left(\sum^K_{g=1}\tilde{\pi}_{jg}-1\right)\right]=\sum^K_{g=1}\tilde{\pi}_{jg}-1.
\end{cases}
\end{equation*}
Letting both equations above equal to zero gives:
\begin{equation*}
\begin{cases} 
\tilde{\pi}_{jk}=\exp(A_{jk}-1+\lambda_j);\\
\sum^K_{g=1}\exp(A_{jg}-1+\lambda_j)=1.
\end{cases}\implies
\begin{cases} 
\lambda_j=-\log\left[\sum^K_{g=1}\exp(A_{jg}-1)\right];\\
\tilde{\pi}_{jk}=\frac{\exp(A_{jk}-1)}{\sum^K_{g=1}\exp(A_{jg}-1)}=\frac{\exp(A_{jk})}{\sum^K_{g=1}\exp(A_{jg})}.\\
\end{cases}
\end{equation*}\;\hfill$\square$\\


The 1st-order partial derivative of the ELBO with respect to the $\bm{\tilde{\mu}}_k$ for $k=1,2,\dots, K$ is written as:
\begin{equation*}
\resizebox{1\hsize}{!}{$
\begin{split}
&\frac{\partial\mathcal{F}}{\partial\bm{\tilde{\mu}}_k}=
\frac{\partial}{\partial\bm{\tilde{\mu}}_k}\sum^N_{i=1}
\left(-\frac{\tilde{\pi}_{ik}}{2}\frac{\tilde{\xi}_k}{\tilde{\psi}_k}||\bm{\tilde{u}}_i-\bm{\tilde{\mu}}_k||^2\right)+
\frac{\partial}{\partial\bm{\tilde{\mu}}_k}\left(-\frac{1}{2\omega^2}\bm{\tilde{\mu}}_k^T\bm{\tilde{\mu}}_k\right)
=\left[\sum^N_{i=1}\tilde{\pi}_{ik}\frac{\tilde{\xi}_k}{\tilde{\psi}_k}(\bm{\tilde{u}}_i-\bm{\tilde{\mu}}_k)\right]-\frac{1}{\omega^2}\bm{\tilde{\mu}}_k.
\end{split}$}
\end{equation*}
Letting this 1st-order partial derivative equal to zero gives:
\begin{equation*}
\begin{split}
&\left(\frac{\tilde{\xi}_k}{\tilde{\psi}_k}\sum^N_{i=1}\tilde{\pi}_{ik}\bm{\tilde{u}}_i\right)-
\left(\frac{\tilde{\xi}_k}{\tilde{\psi}_k}\sum^N_{i=1}\tilde{\pi}_{ik}\right)\bm{\tilde{\mu}}_k-
\frac{1}{\omega^2}\bm{\tilde{\mu}}_k=0\\
&\implies\bm{\tilde{\mu}}_k=\frac{\left(\frac{\tilde{\xi}_k}{\tilde{\psi}_k}\sum^N_{i=1}\tilde{\pi}_{ik}\bm{\tilde{u}}_i\right)}{\left(\frac{\tilde{\xi}_k}{\tilde{\psi}_k}\sum^N_{i=1}\tilde{\pi}_{ik}\right)+\frac{1}{\omega^2}}=
\frac{\omega^2\tilde{\xi}_k\left(\sum^N_{i=1}\tilde{\pi}_{ik}\bm{\tilde{u}}_i\right)}{\omega^2\tilde{\xi}_k\left(\sum^N_{i=1}\tilde{\pi}_{ik}\right)+\tilde{\psi}_k}.
\end{split}
\end{equation*}\;\hfill$\square$\\

The 1st-order partial derivative of the ELBO with respect to the $\tilde{\omega}^2_k$ for $k=1,2,\dots, K$ is written as:
\begin{equation*}
\begin{split}
&\frac{\partial\mathcal{F}}{\partial\tilde{\omega}^2_k}=
\frac{\partial}{\partial\tilde{\omega}^2_k}\sum^N_{i=1}\left(-\frac{\tilde{\pi}_{ik}}{2}\frac{\tilde{\xi}_k}{\tilde{\psi}_k}d\cdot\tilde{\omega}^2_k\right)+
\frac{\partial}{\partial\tilde{\omega}^2_k}\left[-\frac{1}{2\omega^2}d\cdot\tilde{\omega}^2_k+\frac{d}{2}\log(\tilde{\omega}^2_k)\right]\\
&=\left[-\frac{d}{2}\sum^N_{i=1}\left(\tilde{\pi}_{ik}\frac{\tilde{\xi}_k}{\tilde{\psi}_k}\right)\right]-\frac{d}{2\omega^2}+\frac{d}{2\tilde{\omega}^2_k}
=\frac{d}{2}\left(\frac{1}{\tilde{\omega}^2_k}-\frac{1}{\omega^2}-\sum^N_{i=1}\tilde{\pi}_{ik}\frac{\tilde{\xi}_k}{\tilde{\psi}_k}\right).
\end{split}
\end{equation*}
Letting this 1st-order partial derivative equal to zero gives:
\begin{equation*}
\begin{split}
&\frac{1}{\tilde{\omega}^2_k}=\frac{1}{\omega^2}+\frac{\tilde{\xi}_k}{\tilde{\psi}_k}\sum^N_{i=1}\tilde{\pi}_{ik}
\implies\tilde{\omega}^2_k=\frac{1}{\frac{1}{\omega^2}+\frac{\tilde{\xi}_k}{\tilde{\psi}_k}\sum^N_{i=1}\tilde{\pi}_{ik}}=
\frac{\omega^2\tilde{\psi}_k}{\tilde{\psi}_k+\omega^2\tilde{\xi}_k\sum^N_{i=1}\tilde{\pi}_{ik}}.
\end{split}
\end{equation*}\;\hfill$\square$\\



The 1st-order partial derivative of the ELBO with respect to the $\tilde{\xi}_k$ for $k=1,2,\dots, K$ is written as:
\begin{equation*}
\begin{split}
&\frac{\partial\mathcal{F}}{\partial\tilde{\xi}_k}=
\frac{\partial}{\partial\tilde{\xi}_k}\sum^N_{i=1}\tilde{\pi}_{ik}
\left\{\frac{d}{2}\bm{\Psi}(\tilde{\xi}_k)-\frac{1}{2}\frac{\tilde{\xi}_k}{\tilde{\psi}_k}\left[||\bm{\tilde{u}}_i-\bm{\tilde{\mu}}_k||^2+d(\tilde{\sigma}^2_i+\tilde{\omega}^2_k)\right]\right\}\\
&+\frac{\partial}{\partial\tilde{\xi}_k}\left[(\xi-\tilde{\xi}_k)\bm{\Psi}\left(\tilde{\xi}_k\right)-\psi\frac{\tilde{\xi}_k}{\tilde{\psi}_k}+\log\Gamma(\tilde{\xi}_k)+\tilde{\xi}_k\right]\\
&=\sum^N_{i=1}\tilde{\pi}_{ik}\left\{\frac{d}{2}\bm{\Psi}'(\tilde{\xi}_k)-\frac{1}{2\tilde{\psi}_k}\left[||\bm{\tilde{u}}_i-\bm{\tilde{\mu}}_k||^2+d(\tilde{\sigma}^2_i+\tilde{\omega}^2_k)\right]\right\}\\
&-\bm{\Psi}\left(\tilde{\xi}_k\right)+(\xi-\tilde{\xi}_k)\bm{\Psi}'\left(\tilde{\xi}_k\right)-\frac{\psi}{\tilde{\psi}_k}+\bm{\Psi}\left(\tilde{\xi}_k\right)+1\\
&=1+(\xi-\tilde{\xi}_k)\bm{\Psi}'\left(\tilde{\xi}_k\right)-\frac{\psi}{\tilde{\psi}_k}+
\sum^N_{i=1}\tilde{\pi}_{ik}\left\{\frac{d}{2}\bm{\Psi}'(\tilde{\xi}_k)-\frac{1}{2\tilde{\psi}_k}\left[||\bm{\tilde{u}}_i-\bm{\tilde{\mu}}_k||^2+d(\tilde{\sigma}^2_i+\tilde{\omega}^2_k)\right]\right\}\\
&=1+\left(\xi-\tilde{\xi}_k+\frac{d}{2}\sum^N_{i=1}\tilde{\pi}_{ik}\right)\bm{\Psi}'\left(\tilde{\xi}_k\right)
-\frac{1}{2\tilde{\psi}_k}\left\{2\psi+\sum^N_{i=1}\tilde{\pi}_{ik}\left[||\bm{\tilde{u}}_i-\bm{\tilde{\mu}}_k||^2+d(\tilde{\sigma}^2_i+\tilde{\omega}^2_k)\right]\right\}.
\end{split}
\end{equation*}

The 1st-order partial derivative of the ELBO with respect to the $\tilde{\psi}_k$ for $k=1,2,\dots, K$ is written as:
\begin{equation*}
\begin{split}
&\frac{\partial\mathcal{F}}{\partial\tilde{\psi}_k}=
\frac{\partial}{\partial\tilde{\psi}_k}\sum^N_{i=1}\tilde{\pi}_{ik}
\left\{-\frac{d}{2}\log(\tilde{\psi}_k)-\frac{1}{2}\frac{\tilde{\xi}_k}{\tilde{\psi}_k}\left[||\bm{\tilde{u}}_i-\bm{\tilde{\mu}}_k||^2+d(\tilde{\sigma}^2_i+\tilde{\omega}^2_k)\right]\right\}\\
&+\frac{\partial}{\partial\tilde{\psi}_k}\left[-\xi\log(\tilde{\psi}_k)-\psi\frac{\tilde{\xi}_k}{\tilde{\psi}_k}\right]\\
&=\left\{\sum^N_{i=1}\tilde{\pi}_{ik}\left[-\frac{d}{2\tilde{\psi}_k}+\frac{1}{2}\frac{\tilde{\xi}_k}{\tilde{\psi}_k^2}\left[||\bm{\tilde{u}}_i-\bm{\tilde{\mu}}_k||^2+d(\tilde{\sigma}^2_i+\tilde{\omega}^2_k)\right]\right]\right\}
-\frac{\xi}{\tilde{\psi}_k}+\psi\frac{\tilde{\xi}_k}{\tilde{\psi}_k^2}\\
&=-\frac{1}{\tilde{\psi}_k}\left(\xi+\frac{d}{2}\sum^N_{i=1}\tilde{\pi}_{ik}\right)+
\frac{\tilde{\xi}_k}{2\tilde{\psi}_k^2}\left\{2\psi+\sum^N_{i=1}\tilde{\pi}_{ik}\left[||\bm{\tilde{u}}_i-\bm{\tilde{\mu}}_k||^2+d(\tilde{\sigma}^2_i+\tilde{\omega}^2_k)\right]\right\}.
\end{split}
\end{equation*}
So provided that
\begin{equation*}
\resizebox{1\hsize}{!}{$
\begin{cases} 
\frac{\partial\mathcal{F}}{\partial\tilde{\psi}_k}=
-\frac{1}{\tilde{\psi}_k}\left(\xi+\frac{d}{2}\sum^N_{i=1}\tilde{\pi}_{ik}\right)+
\frac{\tilde{\xi}_k}{2\tilde{\psi}_k^2}\left\{2\psi+\sum^N_{i=1}\tilde{\pi}_{ik}\left[||\bm{\tilde{u}}_i-\bm{\tilde{\mu}}_k||^2+d(\tilde{\sigma}^2_i+\tilde{\omega}^2_k)\right]\right\};\\
\frac{\partial\mathcal{F}}{\partial\tilde{\xi}_k}=
1+\left(\xi-\tilde{\xi}_k+\frac{d}{2}\sum^N_{i=1}\tilde{\pi}_{ik}\right)\bm{\Psi}'\left(\tilde{\xi}_k\right)
-\frac{1}{2\tilde{\psi}_k}\left\{2\psi+\sum^N_{i=1}\tilde{\pi}_{ik}\left[||\bm{\tilde{u}}_i-\bm{\tilde{\mu}}_k||^2+d(\tilde{\sigma}^2_i+\tilde{\omega}^2_k)\right]\right\},
\end{cases}$}
\end{equation*}
and let both 1st-order partial derivatives above equal to zero gives:
\begin{equation*}
\begin{cases} 
2\tilde{\psi}_k=\frac{\tilde{\xi}_k\left\{2\psi+\sum^N_{i=1}\tilde{\pi}_{ik}\left[||\bm{\tilde{u}}_i-\bm{\tilde{\mu}}_k||^2+d(\tilde{\sigma}^2_i+\tilde{\omega}^2_k)\right]\right\}}{\left(\xi+\frac{d}{2}\sum^N_{i=1}\tilde{\pi}_{ik}\right)};\\
1+\left(\xi-\tilde{\xi}_k+\frac{d}{2}\sum^N_{i=1}\tilde{\pi}_{ik}\right)\bm{\Psi}'\left(\tilde{\xi}_k\right)-
\frac{1}{2\tilde{\psi}_k}\left\{2\psi+\sum^N_{i=1}\tilde{\pi}_{ik}\left[||\bm{\tilde{u}}_i-\bm{\tilde{\mu}}_k||^2+d(\tilde{\sigma}^2_i+\tilde{\omega}^2_k)\right]\right\}=0
\end{cases}
\end{equation*}
Then substituting the 1st equation to the 2nd equation gives:
$$\resizebox{1\hsize}{!}{$1+\left(\xi-\tilde{\xi}_k+\frac{d}{2}\sum^N_{i=1}\tilde{\pi}_{ik}\right)\bm{\Psi}'\left(\tilde{\xi}_k\right)-
\frac{\xi+\frac{d}{2}\sum^N_{i=1}\tilde{\pi}_{ik}}{\tilde{\xi}_k}=
\left(\xi-\tilde{\xi}_k+\frac{d}{2}\sum^N_{i=1}\tilde{\pi}_{ik}\right)\left[\bm{\Psi}'\left(\tilde{\xi}_k\right)-\frac{1}{\tilde{\xi}_k}\right]
=0.$}$$
Here, we can easily check by R code that the $\left[\bm{\Psi}'\left(\tilde{\xi}_k\right)-\frac{1}{\tilde{\xi}_k}\right]$ term is always positive.
So we can finally obtain the solutions:
\begin{equation*}
\begin{cases} 
\tilde{\xi}_k=\xi+\frac{d}{2}\sum^N_{i=1}\tilde{\pi}_{ik};\\
\tilde{\psi}_k=\psi+\frac{1}{2}\sum^N_{i=1}\tilde{\pi}_{ik}\left[||\bm{\tilde{u}}_i-\bm{\tilde{\mu}}_k||^2+d(\tilde{\sigma}^2_i+\tilde{\omega}^2_k)\right].
\end{cases}
\end{equation*}\;\hfill$\square$\\



The 1st-order partial derivative of the ELBO with respect to the $\tilde{a}_j$ for $j=1,2,\dots, N$ is written as:
\begin{equation*}
\begin{split}
&\frac{\partial\mathcal{F}}{\partial\tilde{a}_j}=
\frac{\partial}{\partial\tilde{a}_j}\sum^N_{\substack{i=1, \\ i\neq j}}\left\{\frac{d}{2}\bm{\Psi}(\tilde{a}_j)-\frac{1}{2}\frac{\tilde{a}_j}{\tilde{b}_j}\left[||\bm{\tilde{v}}_{j\leftarrow i}-\bm{\tilde{u}}_j||^2+d(\tilde{\varphi}^2_{ij}+\tilde{\sigma}^2_j)\right]\right\}\\
&+\frac{\partial}{\partial\tilde{a}_j}\left[(a-\tilde{a}_j)\bm{\Psi}\left(\tilde{a}_j\right)-b\frac{\tilde{a}_j}{\tilde{b}_j}+\log\Gamma(\tilde{a}_j)+\tilde{a}_j\right]\\
&=\sum^N_{\substack{i=1, \\ i\neq j}}\left\{\frac{d}{2}\bm{\Psi}'(\tilde{a}_j)-\frac{1}{2\tilde{b}_j}\left[||\bm{\tilde{v}}_{j\leftarrow i}-\bm{\tilde{u}}_j||^2+d(\tilde{\varphi}^2_{ij}+\tilde{\sigma}^2_j)\right]\right\}\\
&-\bm{\Psi}\left(\tilde{a}_j\right)+(a-\tilde{a}_j)\bm{\Psi}'\left(\tilde{a}_j\right)-\frac{b}{\tilde{b}_j}+\bm{\Psi}\left(\tilde{a}_j\right)+1\\
&=1-\frac{b}{\tilde{b}_j}+(a-\tilde{a}_j)\bm{\Psi}'\left(\tilde{a}_j\right)+(N-1)\frac{d}{2}\bm{\Psi}'(\tilde{a}_j)-
\frac{1}{2\tilde{b}_j}\sum^N_{\substack{i=1, \\ i\neq j}}\left[||\bm{\tilde{v}}_{j\leftarrow i}-\bm{\tilde{u}}_j||^2+d(\tilde{\varphi}^2_{ij}+\tilde{\sigma}^2_j)\right]\\
&=1+\left[a-\tilde{a}_j+\frac{d}{2}(N-1)\right]\bm{\Psi}'\left(\tilde{a}_j\right)-
\frac{1}{2\tilde{b}_j}\left\{2b+\sum^N_{\substack{i=1, \\ i\neq j}}\left[||\bm{\tilde{v}}_{j\leftarrow i}-\bm{\tilde{u}}_j||^2+d(\tilde{\varphi}^2_{ij}+\tilde{\sigma}^2_j)\right]\right\}.
\end{split}
\end{equation*}
The 1st-order partial derivative of the ELBO with respect to the $\tilde{b}_j$ for $j=1,2,\dots, N$ is written as:
\begin{equation*}
\begin{split}
&\frac{\partial\mathcal{F}}{\partial\tilde{b}_j}=
\frac{\partial}{\partial\tilde{b}_j}\sum^N_{\substack{i=1, \\ i\neq j}}\left\{-\frac{d}{2}\log(\tilde{b}_j)-\frac{1}{2}\frac{\tilde{a}_j}{\tilde{b}_j}\left[||\bm{\tilde{v}}_{j\leftarrow i}-\bm{\tilde{u}}_j||^2+d(\tilde{\varphi}^2_{ij}+\tilde{\sigma}^2_j)\right]\right\}\\
&+\frac{\partial}{\partial\tilde{a}_j}\left[-a\log(\tilde{b}_j)-b\frac{\tilde{a}_j}{\tilde{b}_j}\right]\\
&=\left\{\sum^N_{\substack{i=1, \\ i\neq j}}\left\{-\frac{d}{2\tilde{b}_j}+\frac{\tilde{a}_j}{2\tilde{b}_j^2}\left[||\bm{\tilde{v}}_{j\leftarrow i}-\bm{\tilde{u}}_j||^2+d(\tilde{\varphi}^2_{ij}+\tilde{\sigma}^2_j)\right]\right\}\right\}
-\frac{a}{\tilde{b}_j}+b\frac{\tilde{a}_j}{\tilde{b}_j^2}\\
&=-\frac{1}{\tilde{b}_j}\left[\frac{d}{2}(N-1)+a\right]+
\frac{\tilde{a}_j}{2\tilde{b}_j^2}\left\{2b+
\sum^N_{\substack{i=1, \\ i\neq j}}\left[||\bm{\tilde{v}}_{j\leftarrow i}-\bm{\tilde{u}}_j||^2+d(\tilde{\varphi}^2_{ij}+\tilde{\sigma}^2_j)\right]\right\}.
\end{split}
\end{equation*}
Provided that
\begin{equation*}
\resizebox{1\hsize}{!}{$
\begin{cases} 
\frac{\partial\mathcal{F}}{\partial\tilde{b}_j}=
-\frac{1}{\tilde{b}_j}\left[\frac{d}{2}(N-1)+a\right]+
\frac{\tilde{a}_j}{2\tilde{b}_j^2}\left\{2b+
\sum^N_{\substack{i=1, \\ i\neq j}}\left[||\bm{\tilde{v}}_{j\leftarrow i}-\bm{\tilde{u}}_j||^2+d(\tilde{\varphi}^2_{ij}+\tilde{\sigma}^2_j)\right]\right\};\\
\frac{\partial\mathcal{F}}{\partial\tilde{a}_j}=
1+\left[a-\tilde{a}_j+\frac{d}{2}(N-1)\right]\bm{\Psi}'\left(\tilde{a}_j\right)-
\frac{1}{2\tilde{b}_j}\left\{2b+\sum^N_{\substack{i=1, \\ i\neq j}}\left[||\bm{\tilde{v}}_{j\leftarrow i}-\bm{\tilde{u}}_j||^2+d(\tilde{\varphi}^2_{ij}+\tilde{\sigma}^2_j)\right]\right\},
\end{cases}$}
\end{equation*}
and let both 1st-order partial derivatives above equal to zero gives:
\begin{equation*}
\begin{split}
&\begin{cases} 
2\tilde{b}_j=\frac{\tilde{a}_j\left\{2b+
\sum^N_{\substack{i=1, \\ i\neq j}}\left[||\bm{\tilde{v}}_{j\leftarrow i}-\bm{\tilde{u}}_j||^2+d(\tilde{\varphi}^2_{ij}+\tilde{\sigma}^2_j)\right]\right\}}{\left[\frac{d}{2}(N-1)+a\right]};\\
1+\left[a-\tilde{a}_j+\frac{d}{2}(N-1)\right]\bm{\Psi}'\left(\tilde{a}_j\right)-\frac{\left[\frac{d}{2}(N-1)+a\right]}{\tilde{a}_j}=0.
\end{cases}\\
&\implies
\begin{cases} 
\left[a-\tilde{a}_j+\frac{d}{2}(N-1)\right]\left[\bm{\Psi}'\left(\tilde{a}_j\right)-\frac{1}{\tilde{a}_j}\right]=0\implies \tilde{a}_j=a+\frac{d}{2}(N-1);\\
\tilde{b}_j=b+\frac{1}{2}\sum^N_{\substack{i=1, \\ i\neq j}}\left[||\bm{\tilde{v}}_{j\leftarrow i}-\bm{\tilde{u}}_j||^2+d(\tilde{\varphi}^2_{ij}+\tilde{\sigma}^2_j)\right].
\end{cases}
\end{split}
\end{equation*}\;\hfill$\square$\\


\subsubsection{Proof of the Proposition~\ref{Proposition_USigma2_VVarphi2_EtaRho2}}
\label{Proof_Proposition_USigma2_VVarphi2_EtaRho2}

The 1st-order partial derivative of the ELBO with respect to the $\bm{\tilde{u}}_i$ for $i=1,2,\dots, N$ is written as:
\begin{equation*}
\resizebox{1\hsize}{!}{$
\begin{split}
&\frac{\partial\mathcal{F}}{\partial\bm{\tilde{u}}_i}=
\frac{\partial}{\partial\bm{\tilde{u}}_i}\sum^N_{\substack{j=1, \\ j\neq i}}\left\{ -y_{ij}||\bm{\tilde{u}}_i-\bm{\tilde{v}}_{j\leftarrow i}||^2-
\frac{\exp(\tilde{\eta}+\frac{\tilde{\rho}^2}{2}-\frac{||\bm{\tilde{u}}_i-\bm{\tilde{v}}_{j\leftarrow i}||^2}{1+2\tilde{\sigma}^2_i+2\tilde{\varphi}^2_{ij}})}{(1+2\tilde{\sigma}^2_i+2\tilde{\varphi}^2_{ij})^{d/2}}-\frac{1}{2}\frac{\tilde{a}_i}{\tilde{b}_i}||\bm{\tilde{v}}_{i\leftarrow j}-\bm{\tilde{u}}_i||^2\right\}\\
&+\frac{\partial}{\partial\bm{\tilde{u}}_i}\sum^K_{k=1}
\left[-\frac{\tilde{\pi}_{ik}}{2}\frac{\tilde{\xi}_k}{\tilde{\psi}_k}||\bm{\tilde{u}}_i-\bm{\tilde{\mu}}_k||^2\right]\\
&=\sum^N_{\substack{j=1, \\ j\neq i}}\left[-2y_{ij}(\bm{\tilde{u}}_i-\bm{\tilde{v}}_{j\leftarrow i})-
\frac{\exp \left(\tilde{\eta}+\frac{\tilde{\rho}^2}{2}-\frac{||\bm{\tilde{u}}_i-\bm{\tilde{v}}_{j\leftarrow i}||^2}{1+2\tilde{\sigma}^2_i+2\tilde{\varphi}^2_{ij}}\right)}{(1+2\tilde{\sigma}^2_i+2\tilde{\varphi}^2_{ij})^{d/2}}\frac{-2(\bm{\tilde{u}}_i-\bm{\tilde{v}}_{j\leftarrow i})}{1+2\tilde{\sigma}^2_i+2\tilde{\varphi}^2_{ij}}+
\frac{\tilde{a}_i}{\tilde{b}_i}(\bm{\tilde{v}}_{i\leftarrow j}-\bm{\tilde{u}}_i)\right]\\
&-\sum^K_{k=1}\left[\tilde{\pi}_{ik}\frac{\tilde{\xi}_k}{\tilde{\psi}_k}(\bm{\tilde{u}}_i-\bm{\tilde{\mu}}_k)\right]\\
&=\sum^N_{\substack{j=1, \\ j\neq i}}\left\{-2\left[y_{ij}-\frac{\exp \left(\tilde{\eta}+\frac{\tilde{\rho}^2}{2}-\frac{||\bm{\tilde{u}}_i-\bm{\tilde{v}}_{j\leftarrow i}||^2}{1+2\tilde{\sigma}^2_i+2\tilde{\varphi}^2_{ij}}\right)}{(1+2\tilde{\sigma}^2_i+2\tilde{\varphi}^2_{ij})^{d/2+1}}\right](\bm{\tilde{u}}_i-\bm{\tilde{v}}_{j\leftarrow i})+\frac{\tilde{a}_i}{\tilde{b}_i}(\bm{\tilde{v}}_{i\leftarrow j}-\bm{\tilde{u}}_i)\right\}-\sum^K_{k=1}\left[\tilde{\pi}_{ik}\frac{\tilde{\xi}_k}{\tilde{\psi}_k}(\bm{\tilde{u}}_i-\bm{\tilde{\mu}}_k)\right].\\
\end{split}$}
\end{equation*}
The 1st-order partial derivative of the ELBO with respect to the $\tilde{\sigma}^2_i$ for $i=1,2,\dots, N$ is written as:
\begin{equation*}
\resizebox{1\hsize}{!}{$
\begin{split}
&\frac{\partial\mathcal{F}}{\partial\tilde{\sigma}^2_i}=
\frac{\partial}{\partial\tilde{\sigma}^2_i}\sum^N_{\substack{j=1, \\ j\neq i}}\left[ -y_{ij}\cdot d\tilde{\sigma}^2_i-\frac{\exp \left(\tilde{\eta}+\frac{\tilde{\rho}^2}{2}-\frac{||\bm{\tilde{u}}_i-\bm{\tilde{v}}_{j\leftarrow i}||^2}{1+2\tilde{\sigma}^2_i+2\tilde{\varphi}^2_{ij}}\right)}{(1+2\tilde{\sigma}^2_i+2\tilde{\varphi}^2_{ij})^{d/2}} -\frac{1}{2}\frac{\tilde{a}_i}{\tilde{b}_i}d\tilde{\sigma}^2_i\right]\\
&-\frac{\partial}{\partial\tilde{\sigma}^2_i}\sum^K_{k=1}
\left(\frac{\tilde{\pi}_{ik}}{2}\frac{\tilde{\xi}_k}{\tilde{\psi}_k}d\tilde{\sigma}^2_i\right)+
\frac{\partial}{\partial\tilde{\sigma}^2_i}\left[\frac{d}{2}\log (\tilde{\sigma}^2_i)\right]\\
&=\sum^N_{\substack{j=1, \\ j\neq i}}\left(-y_{ij}d-\frac{1}{2}\frac{\tilde{a}_i}{\tilde{b}_i}d\right)-
\sum^K_{k=1}\left(\frac{\tilde{\pi}_{ik}}{2}\frac{\tilde{\xi}_k}{\tilde{\psi}_k}d\right)+\frac{d}{2\tilde{\sigma}^2_i}\\
&-\sum^N_{\substack{j=1, \\ j\neq i}}\left[\frac{\exp \left(\tilde{\eta}+\frac{\tilde{\rho}^2}{2}-\frac{||\bm{\tilde{u}}_i-\bm{\tilde{v}}_{j\leftarrow i}||^2}{1+2\tilde{\sigma}^2_i+2\tilde{\varphi}^2_{ij}}\right)\cdot 2||\bm{\tilde{u}}_i-\bm{\tilde{v}}_{j\leftarrow i}||^2(1+2\tilde{\sigma}^2_i+2\tilde{\varphi}^2_{ij})^{-2}}{(1+2\tilde{\sigma}^2_i+2\tilde{\varphi}^2_{ij})^{d/2}}-\frac{d}{2}\frac{\exp \left(\tilde{\eta}+\frac{\tilde{\rho}^2}{2}-\frac{||\bm{\tilde{u}}_i-\bm{\tilde{v}}_{j\leftarrow i}||^2}{1+2\tilde{\sigma}^2_i+2\tilde{\varphi}^2_{ij}}\right)\cdot 2}{(1+2\tilde{\sigma}^2_i+2\tilde{\varphi}^2_{ij})^{d/2+1}}\right]\\
&=\frac{d}{2}\cdot\left[-2\left(\sum^N_{\substack{j=1, \\ j\neq i}}y_{ij}\right)-(N-1)\frac{\tilde{a}_i}{\tilde{b}_i}-\left(\sum^K_{k=1}\tilde{\pi}_{ik}\frac{\tilde{\xi}_k}{\tilde{\psi}_k}\right)+\frac{1}{\tilde{\sigma}^2_i}\right]\\
&-\sum^N_{\substack{j=1, \\ j\neq i}}
\frac{\exp \left(\tilde{\eta}+\frac{\tilde{\rho}^2}{2}-\frac{||\bm{\tilde{u}}_i-\bm{\tilde{v}}_{j\leftarrow i}||^2}{1+2\tilde{\sigma}^2_i+2\tilde{\varphi}^2_{ij}}\right)}{(1+2\tilde{\sigma}^2_i+2\tilde{\varphi}^2_{ij})^{d/2+1}}\left[2\frac{||\bm{\tilde{u}}_i-\bm{\tilde{v}}_{j\leftarrow i}||^2}{1+2\tilde{\sigma}^2_i+2\tilde{\varphi}^2_{ij}}-d\right].
\end{split}$}
\end{equation*}
To obtain the natural gradient ascent updates of $(\bm{\tilde{u}}_i,\tilde{\sigma}^2_i)$, note first that $\tilde{\sigma}^2_i>0$.
So we apply the transformation $\exp(\tilde{\zeta}_i)=\tilde{\sigma}^2_i$ and let 
$$\mathcal{H}\left(\bm{\tilde{u}}_i,\tilde{\zeta}_i,\dots\right)=\mathcal{F}\left(\bm{\tilde{u}}_i,\exp(\tilde{\zeta}_i),\dots\right)=\mathcal{F}\left(\bm{\tilde{u}}_i,\tilde{\sigma}^2_i,\dots\right).$$
That is, we first take the logarithm of $\tilde{\sigma}^2_i$ to obtain the parameter $\tilde{\zeta}_i$ which lies within $(-\infty,\infty)$, and latter we transform it back to $\tilde{\sigma}^2_i$ by taking the exponential of $\tilde{\zeta}_i$.
This transformation implies that
$$\resizebox{1\hsize}{!}{$\frac{\partial}{\partial\tilde{\zeta}_i}\mathcal{H}\left(\bm{\tilde{u}}_i,\tilde{\zeta}_i,\dots\right)=
\frac{\partial}{\partial\tilde{\zeta}_i}\mathcal{F}\left(\bm{\tilde{u}}_i,\exp(\tilde{\zeta}_i),\dots\right)=
\frac{\partial}{\partial\tilde{\sigma}^2_i}\mathcal{F}\left(\bm{\tilde{u}}_i,\tilde{\sigma}^2_i,\dots\right)\frac{\partial\tilde{\sigma}^2_i}{\partial\tilde{\zeta}_i}=
\exp(\tilde{\zeta}_i)\frac{\partial}{\partial\tilde{\sigma}^2_i}\mathcal{F}\left(\bm{\tilde{u}}_i,\tilde{\sigma}^2_i,\dots\right).$}$$
Thus the standard gradient ascent is read as follows:
\begin{equation*}
\begin{cases} 
\bm{\tilde{u}}_i^\star=\bm{\tilde{u}}_i^\dagger+\epsilon\frac{\partial\mathcal{F}}{\partial\bm{\tilde{u}}_i}\left(\bm{\tilde{u}}_i^\dagger,\exp(\tilde{\zeta}_i^\dagger),\dots\right);\\  
\tilde{\zeta}_i^\star=\tilde{\zeta}_i^\dagger+\epsilon\exp(\tilde{\zeta}_i^\dagger)\frac{\partial\mathcal{F}}{\partial\tilde{\sigma}^2_i}\left(\bm{\tilde{u}}_i^\dagger,\tilde{\sigma}^{2^\dagger}_i,\dots\right).
\end{cases}
\end{equation*}
Then we move to evaluate the Fisher information of $(\bm{\tilde{u}}_i,\tilde{\zeta}_i)$ associated to the likelihood $q_{\bm{U}}\left(\bm{u}_i|\bm{\tilde{u}}_i,\exp(\tilde{\zeta}_i)\right)$.
The log-likelihood is written as:
\begin{equation*}
\begin{split}
&\log q_{\bm{U}}\left(\bm{u}_i|\bm{\tilde{u}}_i,\exp(\tilde{\zeta}_i)\right)=
\log f_{\text{MVN}_d}\left(\bm{u}_i|\bm{\tilde{u}}_i,\exp(\tilde{\zeta}_i)\mathbbm{I}_d\right)\\
&=\log\left\{\frac{1}{(2\pi)^{\frac{d}{2}}\left|\exp(\tilde{\zeta}_i)\mathbbm{I}_d\right|^{\frac{1}{2}}}\exp \left[-\frac{1}{2}(\bm{u}_i-\bm{\tilde{u}}_i)^T\frac{1}{\exp(\tilde{\zeta}_i)}\mathbbm{I}_d(\bm{u}_i-\bm{\tilde{u}}_i)\right]\right\}\\
&=-\frac{d}{2}\log(2\pi)-\frac{d}{2}\tilde{\zeta}_i-\frac{1}{2\exp(\tilde{\zeta}_i)}||\bm{u}_i-\bm{\tilde{u}}_i||^2.
\end{split}
\end{equation*}
The 1st-order partial derivatives of the log-likelihood above are thus read as follows:
\begin{equation*}
\label{ZI}
\begin{cases} 
\frac{\partial\log q_{\bm{U}}}{\partial\bm{\tilde{u}}_i}=\frac{1}{\exp(\tilde{\zeta}_i)}\left(\bm{u}_i-\bm{\tilde{u}}_i\right);\\
\frac{\partial\log q_{\bm{U}}}{\partial\tilde{\zeta}_i}=-\frac{d}{2}+\frac{1}{2\exp(\tilde{\zeta}_i)}||\bm{u}_i-\bm{\tilde{u}}_i||^2.
\end{cases}
\end{equation*}
The 2nd-order partial derivatives are thus:
\begin{equation*}
\begin{cases} 
\frac{\partial^2\log q_{\bm{U}}}{\partial\bm{\tilde{u}}_i^2}=-\frac{1}{\exp(\tilde{\zeta}_i)};\\
\frac{\partial^2\log q_{\bm{U}}}{\partial\bm{\tilde{u}}_i\partial\tilde{\zeta}_i}=-\frac{1}{\exp(\tilde{\zeta}_i)}\left(\bm{u}_i-\bm{\tilde{u}}_i\right);\\
\frac{\partial^2\log q_{\bm{U}}}{\partial\tilde{\zeta}_i^2}=-\frac{1}{2\exp(\tilde{\zeta}_i)}||\bm{u}_i-\bm{\tilde{u}}_i||^2.
\end{cases}
\end{equation*}
Taking the negative expectation $-\mathbbm{E}_q$ gives:
\begin{equation*}
\begin{cases} 
-\mathbbm{E}_q\left(\frac{\partial^2\log q_{\bm{U}}}{\partial\bm{\tilde{u}}_i^2}\right)=\frac{1}{\exp(\tilde{\zeta}_i)};\\
-\mathbbm{E}_q\left(\frac{\partial^2\log q_{\bm{U}}}{\partial\bm{\tilde{u}}_i\partial\tilde{\zeta}_i}\right)=\frac{1}{\exp(\tilde{\zeta}_i)}\left(\bm{\tilde{u}}_i-\bm{\tilde{u}}_i\right)=0;\\
-\mathbbm{E}_q\left(\frac{\partial^2\log q_{\bm{U}}}{\partial\tilde{\zeta}_i^2}\right)=\frac{1}{2\exp(\tilde{\zeta}_i)}\left[||\bm{\tilde{u}}_i-\bm{\tilde{u}}_i||^2+d\exp(\tilde{\zeta}_i)\right]=\frac{d}{2}.
\end{cases}
\end{equation*}
These imply that the Fisher information $\mathcal{I}(\bm{\tilde{u}}_i,\tilde{\zeta}_i)$ and the inverse Fisher information $\mathcal{I}^{-1}(\bm{\tilde{u}}_i,\tilde{\zeta}_i)$ are, respectively, written as:
\begin{equation*}
\mathcal{I}(\bm{\tilde{u}}_i,\tilde{\zeta}_i)=\begin{pmatrix}
\frac{1}{\exp(\tilde{\zeta}_i)} & 0 \\ 0 & \frac{d}{2}
\end{pmatrix},\hspace{1em}
\mathcal{I}^{-1}(\bm{\tilde{u}}_i,\tilde{\zeta}_i)=\begin{pmatrix}
\exp(\tilde{\zeta}_i) & 0 \\ 0 & \frac{2}{d}
\end{pmatrix}.
\end{equation*}
Thus the final natural gradient ascent steps read as follows:
\begin{equation*}
\begin{cases} 
\bm{\tilde{u}}_i^\star=\bm{\tilde{u}}_i^\dagger+\epsilon\exp(\tilde{\zeta}^\dagger_i)\frac{\partial\mathcal{F}}{\partial\bm{\tilde{u}}_i}\left(\bm{\tilde{u}}^\dagger_i,\exp(\tilde{\zeta}^\dagger_i),\dots\right);\\  
\tilde{\zeta}_i^\star=\tilde{\zeta}_i^\dagger+\epsilon\frac{2}{d}\exp(\tilde{\zeta}^\dagger_i)\frac{\partial\mathcal{F}}{\partial\tilde{\sigma}^2_i}\left(\bm{\tilde{u}}^\dagger_i,\tilde{\sigma}^{2^\dagger}_i,\dots\right).
\end{cases}
\end{equation*}
Taking the exponential in the 2nd line above and transforming the $\tilde{\zeta}_i$ back to $\tilde{\sigma}^2_i$ gives:
\begin{equation*}
\begin{split}
&\begin{cases} 
\bm{\tilde{u}}_i^\star=\bm{\tilde{u}}_i^\dagger+\epsilon\exp(\tilde{\zeta}^\dagger_i)\frac{\partial\mathcal{F}}{\partial\bm{\tilde{u}}_i}\left(\bm{\tilde{u}}^\dagger_i,\exp(\tilde{\zeta}^\dagger_i),\dots\right);\\  
\exp(\tilde{\zeta}_i^\star)=\exp(\tilde{\zeta}_i^\dagger)\exp[\epsilon\frac{2}{d}\exp(\tilde{\zeta}^\dagger_i)\frac{\partial\mathcal{F}}{\partial\tilde{\sigma}^2_i}\left(\bm{\tilde{u}}^\dagger_i,\tilde{\sigma}^{2^\dagger}_i,\dots\right)].
\end{cases}\\
&=\begin{cases} 
\bm{\tilde{u}}_i^\star=\bm{\tilde{u}}_i^\dagger+\epsilon\tilde{\sigma}^{2^\dagger}_i\frac{\partial\mathcal{F}}{\partial\bm{\tilde{u}}_i}\left(\bm{\tilde{u}}^\dagger_i,\tilde{\sigma}^{2^\dagger}_i,\dots\right);\\  
\tilde{\sigma}^{2^\star}_i=\tilde{\sigma}^{2^\dagger}_i\exp[\epsilon\frac{2}{d}\tilde{\sigma}^{2^\dagger}_i\frac{\partial\mathcal{F}}{\partial\tilde{\sigma}^2_i}\left(\bm{\tilde{u}}^\dagger_i,\tilde{\sigma}^{2^\dagger}_i,\dots\right)].
\end{cases}
\end{split}
\end{equation*}\;\hfill$\square$\\



The 1st-order partial derivative of the ELBO with respect to the $\bm{\tilde{v}}_{j\leftarrow i}$ for $i,j=1,2,\dots, N;\;i\neq j$ is written as:
\begin{equation*}
\begin{split}
&\frac{\partial\mathcal{F}}{\partial\bm{\tilde{v}}_{j\leftarrow i}}=
\frac{\partial}{\partial\bm{\tilde{v}}_{j\leftarrow i}}\left\{ -y_{ij}||\bm{\tilde{u}}_i-\bm{\tilde{v}}_{j\leftarrow i}||^2-\frac{\exp \left(\tilde{\eta}+\frac{\tilde{\rho}^2}{2}-\frac{||\bm{\tilde{u}}_i-\bm{\tilde{v}}_{j\leftarrow i}||^2}{1+2\tilde{\sigma}^2_i+2\tilde{\varphi}^2_{ij}}\right)}{(1+2\tilde{\sigma}^2_i+2\tilde{\varphi}^2_{ij})^{d/2}} -\frac{1}{2}\frac{\tilde{a}_j}{\tilde{b}_j}||\bm{\tilde{v}}_{j\leftarrow i}-\bm{\tilde{u}}_j||^2\right\}\\
&=\left[2y_{ij}(\bm{\tilde{u}}_i-\bm{\tilde{v}}_{j\leftarrow i})-\frac{\tilde{a}_j}{\tilde{b}_j}(\bm{\tilde{v}}_{j\leftarrow i}-\bm{\tilde{u}}_j)\right]
-\frac{\exp \left(\tilde{\eta}+\frac{\tilde{\rho}^2}{2}-\frac{||\bm{\tilde{u}}_i-\bm{\tilde{v}}_{j\leftarrow i}||^2}{1+2\tilde{\sigma}^2_i+2\tilde{\varphi}^2_{ij}}\right)}{(1+2\tilde{\sigma}^2_i+2\tilde{\varphi}^2_{ij})^{d/2}}\frac{2(\bm{\tilde{u}}_i-\bm{\tilde{v}}_{j\leftarrow i})}{1+2\tilde{\sigma}^2_i+2\tilde{\varphi}^2_{ij}}\\
&=\left\{2\left[y_{ij}-\frac{\exp \left(\tilde{\eta}+\frac{\tilde{\rho}^2}{2}-\frac{||\bm{\tilde{u}}_i-\bm{\tilde{v}}_{j\leftarrow i}||^2}{1+2\tilde{\sigma}^2_i+2\tilde{\varphi}^2_{ij}}\right)}{(1+2\tilde{\sigma}^2_i+2\tilde{\varphi}^2_{ij})^{d/2+1}}\right](\bm{\tilde{u}}_i-\bm{\tilde{v}}_{j\leftarrow i})-\frac{\tilde{a}_j}{\tilde{b}_j}(\bm{\tilde{v}}_{j\leftarrow i}-\bm{\tilde{u}}_j)\right\}.
\end{split}
\end{equation*}
The 1st-order partial derivative of the ELBO with respect to the $\tilde{\varphi}^2_{ij}$ for $i,j=1,2,\dots, N;\;i\neq j$ is written as:
\begin{equation*}
\resizebox{1\hsize}{!}{$
\begin{split}
&\frac{\partial\mathcal{F}}{\partial\tilde{\varphi}^2_{ij}}=
\frac{\partial}{\partial\tilde{\varphi}^2_{ij}}\left\{ -y_{ij}\cdot d\tilde{\varphi}^2_{ij}-\frac{\exp \left(\tilde{\eta}+\frac{\tilde{\rho}^2}{2}-\frac{||\bm{\tilde{u}}_i-\bm{\tilde{v}}_{j\leftarrow i}||^2}{1+2\tilde{\sigma}^2_i+2\tilde{\varphi}^2_{ij}}\right)}{(1+2\tilde{\sigma}^2_i+2\tilde{\varphi}^2_{ij})^{d/2}} -\frac{1}{2}\frac{\tilde{a}_j}{\tilde{b}_j}d\tilde{\varphi}^2_{ij}+\frac{d}{2}\log\tilde{\varphi}^2_{ij}\right\}\\
&=-y_{ij}d-\frac{1}{2}\frac{\tilde{a}_j}{\tilde{b}_j}d+\frac{d}{2\tilde{\varphi}^2_{ij}}\\
&-\left[\frac{\exp \left(\tilde{\eta}+\frac{\tilde{\rho}^2}{2}-\frac{||\bm{\tilde{u}}_i-\bm{\tilde{v}}_{j\leftarrow i}||^2}{1+2\tilde{\sigma}^2_i+2\tilde{\varphi}^2_{ij}}\right)\cdot 2||\bm{\tilde{u}}_i-\bm{\tilde{v}}_{j\leftarrow i}||^2(1+2\tilde{\sigma}^2_i+2\tilde{\varphi}^2_{ij})^{-2}}{(1+2\tilde{\sigma}^2_i+2\tilde{\varphi}^2_{ij})^{d/2}}-\frac{d}{2}\frac{\exp \left(\tilde{\eta}+\frac{\tilde{\rho}^2}{2}-\frac{||\bm{\tilde{u}}_i-\bm{\tilde{v}}_{j\leftarrow i}||^2}{1+2\tilde{\sigma}^2_i+2\tilde{\varphi}^2_{ij}}\right)\cdot 2}{(1+2\tilde{\sigma}^2_i+2\tilde{\varphi}^2_{ij})^{d/2+1}}\right]\\
&=\frac{d}{2}\left(-2y_{ij}-\frac{\tilde{a}_j}{\tilde{b}_j}+\frac{1}{\tilde{\varphi}^2_{ij}}\right)
-\frac{
\exp \left(\tilde{\eta}+\frac{\tilde{\rho}^2}{2}-\frac{||\bm{\tilde{u}}_i-\bm{\tilde{v}}_{j\leftarrow i}||^2}{1+2\tilde{\sigma}^2_i+2\tilde{\varphi}^2_{ij}}\right)\left[2\frac{||\bm{\tilde{u}}_i-\bm{\tilde{v}}_{j\leftarrow i}||^2}{1+2\tilde{\sigma}^2_i+2\tilde{\varphi}^2_{ij}}-d\right]
}{(1+2\tilde{\sigma}^2_i+2\tilde{\varphi}^2_{ij})^{d/2+1}}.
\end{split}$}
\end{equation*}
To obtain the natural gradient ascent updates of $(\bm{\tilde{v}}_{j\leftarrow i},\tilde{\varphi}^2_{ij})$, we follow similar procedure as in those of $(\bm{\tilde{u}}_i,\tilde{\sigma}^2_i)$ to let $\exp(\tilde{\nu}_{ij})=\tilde{\varphi}^2_{ij}$ and to let
$$\mathcal{H}\left(\bm{\tilde{u}}_i,\tilde{\nu}_{ij},\dots\right)=\mathcal{F}\left(\bm{\tilde{u}}_i,\exp(\tilde{\nu}_{ij}),\dots\right)=\mathcal{F}\left(\bm{\tilde{u}}_i,\tilde{\varphi}^2_{ij},\dots\right).$$
And thus we have $\frac{\partial}{\partial\tilde{\nu}_{ij}}\mathcal{H}\left(\bm{\tilde{v}}_{j\leftarrow i},\tilde{\nu}_{ij},\dots\right)=
\exp(\tilde{\nu}_{ij})\frac{\partial}{\partial\tilde{\varphi}^2_{ij}}\mathcal{F}\left(\bm{\tilde{v}}_{j\leftarrow i},\tilde{\varphi}^2_{ij},\dots\right).$
The log-likelihood is of the form:
\begin{equation*}
\begin{split}
&\log q_{\bm{V}}\left(\bm{v}_{j\leftarrow i}|\bm{\tilde{v}}_{j\leftarrow i},\exp(\tilde{\nu}_{ij})\right)=
-\frac{d}{2}\log(2\pi)-\frac{d}{2}\tilde{\nu}_{ij}-\frac{1}{2\exp(\tilde{\nu}_{ij})}||\bm{v}_{j\leftarrow i}-\bm{\tilde{v}}_{j\leftarrow i}||^2.
\end{split}
\end{equation*}
The 2nd-order partial derivatives are equal to:
$$
\frac{\partial^2\log q_{\bm{V}}}{\partial\bm{\tilde{v}}_{j\leftarrow i}^2}=-\frac{1}{\exp(\tilde{\nu}_{ij})},\hspace{1em}
\frac{\partial^2\log q_{\bm{V}}}{\partial\bm{\tilde{v}}_{j\leftarrow i}\partial\tilde{\nu}_{ij}}=-\frac{\left(\bm{v}_{j\leftarrow i}-\bm{\tilde{v}}_{j\leftarrow i}\right)}{\exp(\tilde{\nu}_{ij})},\hspace{1em}
\frac{\partial^2\log q_{\bm{V}}}{\partial\tilde{\nu}_{ij}^2}=-\frac{||\bm{v}_{j\leftarrow i}-\bm{\tilde{v}}_{j\leftarrow i}||^2}{2\exp(\tilde{\nu}_{ij})}.
$$
Taking the negative expectation $-\mathbbm{E}_q$ gives:
$$
-\mathbbm{E}_q\left(\frac{\partial^2\log q_{\bm{V}}}{\partial\bm{\tilde{v}}_{j\leftarrow i}^2}\right)=\frac{1}{\exp(\tilde{\nu}_{ij})},\hspace{1em}
-\mathbbm{E}_q\left(\frac{\partial^2\log q_{\bm{V}}}{\partial\bm{\tilde{v}}_{j\leftarrow i}\partial\tilde{\nu}_{ij}}\right)=0,\hspace{1em}
-\mathbbm{E}_q\left(\frac{\partial^2\log q_{\bm{V}}}{\partial\tilde{\nu}_{ij}^2}\right)=\frac{d}{2}.
$$
Thus the Fisher information and its inverse are read as follows:
\begin{equation*}
\mathcal{I}(\bm{\tilde{v}}_{j\leftarrow i},\tilde{\nu}_{ij})=\begin{pmatrix}
\frac{1}{\exp(\tilde{\nu}_{ij})} & 0 \\ 0 & \frac{d}{2}
\end{pmatrix},\hspace{1em}
\mathcal{I}^{-1}(\bm{\tilde{v}}_{j\leftarrow i},\tilde{\nu}_{ij})=\begin{pmatrix}
\exp(\tilde{\nu}_{ij}) & 0 \\ 0 & \frac{2}{d}
\end{pmatrix}.
\end{equation*}
At the end, the final natural gradient ascent steps are of the form:
\begin{equation*}
\begin{split}
&\begin{cases} 
\bm{\tilde{v}}_{j\leftarrow i}^\star=\bm{\tilde{v}}_{j\leftarrow i}^\dagger+\epsilon\exp(\tilde{\nu}_{ij}^\dagger)\frac{\partial\mathcal{F}}{\partial\bm{\tilde{v}}_{j\leftarrow i}}\left(\bm{\tilde{v}}_{j\leftarrow i}^\dagger,\exp(\tilde{\nu}_{ij}^\dagger),\dots\right);\\  
\exp(\tilde{\nu}_{ij}^\star)=\exp(\tilde{\nu}_{ij}^\dagger)\exp[\epsilon\frac{2}{d}\exp(\tilde{\nu}_{ij}^\dagger)\frac{\partial\mathcal{F}}{\partial\tilde{\varphi}^2_{ij}}\left(\bm{\tilde{v}}_{j\leftarrow i}^\dagger,\tilde{\varphi}^{2^\dagger}_{ij},\dots\right)].
\end{cases}\\
&=\begin{cases} 
\bm{\tilde{v}}_{j\leftarrow i}^\star=\bm{\tilde{v}}_{j\leftarrow i}^\dagger+\epsilon\tilde{\varphi}^{2^\dagger}_{ij}\frac{\partial\mathcal{F}}{\partial\bm{\tilde{v}}_{j\leftarrow i}}\left(\bm{\tilde{v}}_{j\leftarrow i}^\dagger,\tilde{\varphi}^{2^\dagger}_{ij},\dots\right);\\  
\tilde{\varphi}^{2^\star}_{ij}=\tilde{\varphi}^{2^\dagger}_{ij}\exp[\epsilon\frac{2}{d}\tilde{\varphi}^{2^\dagger}_{ij}\frac{\partial\mathcal{F}}{\partial\tilde{\varphi}^2_{ij}}\left(\bm{\tilde{v}}_{j\leftarrow i}^\dagger,\tilde{\varphi}^{2^\dagger}_{ij},\dots\right)].
\end{cases}
\end{split}
\end{equation*}\;\hfill$\square$\\



The 1st-order partial derivative of the ELBO with respect to the $\tilde{\eta}$ is written as:
\begin{equation*}
\begin{split}
&\frac{\partial\mathcal{F}}{\partial\tilde{\eta}}=
\frac{\partial}{\partial\tilde{\eta}}\sum^N_{\substack{i,j=1, \\ i\neq j}}\left[ y_{ij}\tilde{\eta}-
\frac{\exp \left(\tilde{\eta}+\frac{\tilde{\rho}^2}{2}-\frac{||\bm{\tilde{u}}_i-\bm{\tilde{v}}_{j\leftarrow i}||^2}{1+2\tilde{\sigma}^2_i+2\tilde{\varphi}^2_{ij}}\right)}{(1+2\tilde{\sigma}^2_i+2\tilde{\varphi}^2_{ij})^{d/2}} \right]-
\frac{\partial}{\partial\tilde{\eta}}\left[\frac{1}{2\rho^2}(\tilde{\eta}-\eta)^2\right]\\
&=\left(\sum^N_{\substack{i,j=1, \\ i\neq j}}y_{ij}\right)-\frac{\tilde{\eta}-\eta}{\rho^2}-
\sum^N_{\substack{i,j=1, \\ i\neq j}}\frac{\exp \left(\tilde{\eta}+\frac{\tilde{\rho}^2}{2}-\frac{||\bm{\tilde{u}}_i-\bm{\tilde{v}}_{j\leftarrow i}||^2}{1+2\tilde{\sigma}^2_i+2\tilde{\varphi}^2_{ij}}\right)}{(1+2\tilde{\sigma}^2_i+2\tilde{\varphi}^2_{ij})^{d/2}}.\\
\end{split}
\end{equation*}
The 1st-order partial derivative of the ELBO with respect to the $\tilde{\rho}^2$ is written as:
\begin{equation*}
\begin{split}
&\frac{\partial\mathcal{F}}{\partial\tilde{\rho}^2}=
\frac{\partial}{\partial\tilde{\rho}^2}\sum^N_{\substack{i,j=1, \\ i\neq j}}\left[ -
\frac{\exp \left(\tilde{\eta}+\frac{\tilde{\rho}^2}{2}-\frac{||\bm{\tilde{u}}_i-\bm{\tilde{v}}_{j\leftarrow i}||^2}{1+2\tilde{\sigma}^2_i+2\tilde{\varphi}^2_{ij}}\right)}{(1+2\tilde{\sigma}^2_i+2\tilde{\varphi}^2_{ij})^{d/2}}\right]+
\frac{\partial}{\partial\tilde{\rho}^2}\left[-\frac{\tilde{\rho}^2}{2\rho^2}+\frac{1}{2}\log\tilde{\rho}^2\right]\\
&=-\frac{1}{2\rho^2}+\frac{1}{2\tilde{\rho}^2}-
\frac{1}{2}\sum^N_{\substack{i,j=1, \\ i\neq j}}\frac{\exp \left(\tilde{\eta}+\frac{\tilde{\rho}^2}{2}-\frac{||\bm{\tilde{u}}_i-\bm{\tilde{v}}_{j\leftarrow i}||^2}{1+2\tilde{\sigma}^2_i+2\tilde{\varphi}^2_{ij}}\right)}{(1+2\tilde{\sigma}^2_i+2\tilde{\varphi}^2_{ij})^{d/2}}.
\end{split}
\end{equation*}
To obtain the natural gradient ascent updates of $(\tilde{\eta},\tilde{\rho}^2)$ where $\tilde{\rho}^2>0$, we let $\exp(\tilde{\iota})=\tilde{\rho}^2$ and let $\mathcal{H}\left(\tilde{\eta},\tilde{\iota},\dots\right)=\mathcal{F}\left(\tilde{\eta},\exp(\tilde{\iota}),\dots\right)=\mathcal{F}\left(\tilde{\eta},\tilde{\rho}^2,\dots\right)$,
then
$$\frac{\partial}{\partial\tilde{\iota}}\mathcal{H}\left(\tilde{\eta},\tilde{\iota},\dots\right)=
\frac{\partial}{\partial\tilde{\iota}}\mathcal{F}\left(\tilde{\eta},\exp(\tilde{\iota}),\dots\right)=
\frac{\partial}{\partial\tilde{\rho}^2}\mathcal{F}\left(\tilde{\eta},\tilde{\rho}^2,\dots\right)\frac{\partial\tilde{\rho}^2}{\partial\tilde{\iota}}=
\exp(\tilde{\iota})\frac{\partial}{\partial\tilde{\rho}^2}\mathcal{F}\left(\tilde{\eta},\tilde{\rho}^2,\dots\right).$$
Recall here that $q_{\beta}(\beta|\tilde{\eta},\tilde{\rho}^2) \sim \text{Normal}(\tilde{\eta},\tilde{\rho}^2)$, so the log-likelihood is of the form:
\begin{equation*}
\begin{split}
&\log q_{\beta}\left(\beta|\tilde{\eta},\exp(\tilde{\iota})\right)=
\log f_{\text{normal}}\left(\beta|\tilde{\eta},\exp(\tilde{\iota})\right)\\
&=
\log\left\{\frac{1}{(2\pi)^{\frac{1}{2}}\exp(\tilde{\iota})^{\frac{1}{2}}}\exp \left[-\frac{1}{2\exp(\tilde{\iota})}(\beta-\tilde{\eta})^2\right]\right\}
=-\frac{1}{2}\log(2\pi)-\frac{1}{2}\tilde{\iota}-\frac{1}{2\exp(\tilde{\iota})}(\beta-\tilde{\eta})^2.
\end{split}
\end{equation*}
The 1st-order partial derivatives are:
$$
\frac{\partial\log q_{\beta}}{\partial\tilde{\eta}}=\frac{\beta-\tilde{\eta}}{\exp(\tilde{\iota})},\hspace{2em}
\frac{\partial\log q_{\beta}}{\partial\tilde{\iota}}=-\frac{1}{2}+\frac{\left(\beta-\tilde{\eta}\right)^2}{2\exp(\tilde{\iota})}.
$$
The 2nd-order partial derivatives are:
$$
\frac{\partial^2\log q_{\beta}}{\partial\tilde{\eta}^2}=-\frac{1}{\exp(\tilde{\iota})},\hspace{1em}
\frac{\partial^2\log q_{\beta}}{\partial\tilde{\eta}\partial\tilde{\iota}}=-\frac{\left(\beta-\tilde{\eta}\right)}{\exp(\tilde{\iota})},\hspace{1em}
\frac{\partial^2\log q_{\beta}}{\partial\tilde{\iota}^2}=-\frac{\left(\beta-\tilde{\eta}\right)^2}{2\exp(\tilde{\iota})}.
$$
Taking the negative expectation $-\mathbbm{E}_q$ gives:
$$
-\mathbbm{E}_q\left(\frac{\partial^2\log q_{\beta}}{\partial\tilde{\eta}^2}\right)=\frac{1}{\exp(\tilde{\iota})},\hspace{1em}
-\mathbbm{E}_q\left(\frac{\partial^2\log q_{\beta}}{\partial\tilde{\eta}\partial\tilde{\iota}}\right)=0,\hspace{1em}
-\mathbbm{E}_q\left(\frac{\partial^2\log q_{\beta}}{\partial\tilde{\iota}^2}\right)=\frac{1}{2}.
$$
Thus the Fisher information and its inverse are written as:
\begin{equation*}
\mathcal{I}(\tilde{\eta},\tilde{\iota})=\begin{pmatrix}
\frac{1}{\exp(\tilde{\iota})} & 0 \\ 0 & \frac{1}{2}
\end{pmatrix},\hspace{1em}
\mathcal{I}^{-1}(\tilde{\eta},\tilde{\iota})=\begin{pmatrix}
\exp(\tilde{\iota}) & 0 \\ 0 & 2
\end{pmatrix}.
\end{equation*}
These give the natural gradient ascent steps of the form:
\begin{equation*}
\begin{split}
&\begin{cases} 
\tilde{\eta}^\star=\tilde{\eta}^\dagger+\epsilon\exp(\tilde{\iota}^\dagger)\frac{\partial\mathcal{F}}{\partial\tilde{\eta}}\left(\tilde{\eta}^\dagger,\exp(\tilde{\iota}^\dagger),\dots\right);\\  
\exp(\tilde{\iota}^\star)=\exp(\tilde{\iota}^\dagger)\exp[\epsilon2\exp(\tilde{\iota}^\dagger)\frac{\partial}{\partial\mathcal{F}\tilde{\rho}^2}\left(\tilde{\eta}^\dagger,\tilde{\rho}^{2^\dagger},\dots\right)].
\end{cases}\\
&=\begin{cases} 
\tilde{\eta}^\star=\tilde{\eta}^\dagger+\epsilon\tilde{\rho}^{2^\dagger}\frac{\partial\mathcal{F}}{\partial\tilde{\eta}}\left(\tilde{\eta}^\dagger,\tilde{\rho}^{2^\dagger},\dots\right);\\  
\tilde{\rho}^{2^\star}=\tilde{\rho}^{2^\dagger}\exp[\epsilon2\tilde{\rho}^{2^\dagger}\frac{\partial\mathcal{F}}{\partial\tilde{\rho}^2}\left(\tilde{\eta}^\dagger,\tilde{\rho}^{2^\dagger},\dots\right)].
\end{cases}
\end{split}
\end{equation*}\;\hfill$\square$\\


\subsubsection{Proof of the Proposition~\ref{Proposition_delta}}
\label{Proof_Proposition_delta}
The 1st-order partial derivative of the ELBO with respect to each $\tilde{\delta}_k$, for $k=1,2,\dots,K$ is written as:
\begin{equation*}
\resizebox{1\hsize}{!}{$
\begin{split}
&\frac{\partial\mathcal{F}}{\partial\tilde{\delta}_k}=
\frac{\partial}{\partial\tilde{\delta}_k}\sum^N_{i=1}\sum^K_{h=1}\tilde{\pi}_{ih}
\left[\bm{\Psi}\left(\tilde{\delta}_h\right)-\bm{\Psi}\left(\sum^K_{g=1}\tilde{\delta}_g\right)\right]\\
&+\frac{\partial}{\partial\tilde{\delta}_k}\sum^K_{h=1}\left\{(\delta_h-\tilde{\delta}_h)\left[\bm{\Psi}\left(\tilde{\delta}_{h}\right)-\bm{\Psi}\left(\sum^K_{g=1}\tilde{\delta}_{g}\right)\right]+\log\Gamma(\tilde{\delta}_h)\right\}
-\frac{\partial}{\partial\tilde{\delta}_k}\log \Gamma\left(\sum^K_{g=1}\tilde{\delta}_g\right)\\
&=\frac{\partial}{\partial\tilde{\delta}_k}\sum^N_{i=1}\sum^K_{h=1}\tilde{\pi}_{ih}\bm{\Psi}\left(\tilde{\delta}_h\right)-
\frac{\partial}{\partial\tilde{\delta}_k}\left[\bm{\Psi}\left(\sum^K_{g=1}\tilde{\delta}_g\right)\cdot\left(\sum^N_{i=1}\sum^K_{h=1}\tilde{\pi}_{ih}\right)\right]+
\frac{\partial}{\partial\tilde{\delta}_k}\sum^K_{h=1}(\delta_h-\tilde{\delta}_h)\bm{\Psi}\left(\tilde{\delta}_{h}\right)\\
&-\frac{\partial}{\partial\tilde{\delta}_k}\left\{\bm{\Psi}\left(\sum^K_{g=1}\tilde{\delta}_{g}\right)\cdot\left[\sum^K_{h=1}(\delta_h-\tilde{\delta}_h)\right]\right\}+
\frac{\partial}{\partial\tilde{\delta}_k}\sum^K_{h=1}\log\Gamma(\tilde{\delta}_h)-
\frac{\partial}{\partial\tilde{\delta}_k}\log \Gamma\left(\sum^K_{g=1}\tilde{\delta}_g\right)\\
&=\sum^N_{i=1}\tilde{\pi}_{ik}\frac{\partial}{\partial\tilde{\delta}_k}\bm{\Psi}\left(\tilde{\delta}_k\right)-
\left[\frac{\partial}{\partial\tilde{\delta}_k}\bm{\Psi}\left(\sum^K_{g=1}\tilde{\delta}_g\right)\right]\cdot\left(\sum^N_{i=1}\sum^K_{h=1}\tilde{\pi}_{ih}\right)-
\bm{\Psi}\left(\tilde{\delta}_{k}\right)+(\delta_k-\tilde{\delta}_k)\frac{\partial}{\partial\tilde{\delta}_k}\bm{\Psi}\left(\tilde{\delta}_{k}\right)\\
&-\left[\frac{\partial}{\partial\tilde{\delta}_k}\bm{\Psi}\left(\sum^K_{g=1}\tilde{\delta}_{g}\right)\right]\cdot\left[\sum^K_{h=1}(\delta_h-\tilde{\delta}_h)\right]+
\bm{\Psi}\left(\sum^K_{g=1}\tilde{\delta}_{g}\right)+\bm{\Psi}\left(\tilde{\delta}_{k}\right)-\bm{\Psi}\left(\sum^K_{g=1}\tilde{\delta}_{g}\right)\\
&=\left[\frac{\partial}{\partial\tilde{\delta}_k}\bm{\Psi}\left(\tilde{\delta}_k\right)\right]\left(\delta_k-\tilde{\delta}_k+\sum^N_{i=1}\tilde{\pi}_{ik}\right)-
\left[\frac{\partial}{\partial\tilde{\delta}_k}\bm{\Psi}\left(\sum^K_{g=1}\tilde{\delta}_g\right)\right]\cdot\left\{\left[\sum^K_{h=1}(\delta_h-\tilde{\delta}_h)\right]+\left(\sum^N_{i=1}\sum^K_{h=1}\tilde{\pi}_{ih}\right)\right\}\\
&=\bm{\Psi}'\left(\tilde{\delta}_k\right)\cdot\left(\delta_k-\tilde{\delta}_k+\sum^N_{i=1}\tilde{\pi}_{ik}\right)-
\bm{\Psi}'\left(\sum^K_{g=1}\tilde{\delta}_g\right)\cdot
\left[\sum^K_{g=1}\left(\delta_g-\tilde{\delta}_g+\sum^N_{i=1}\tilde{\pi}_{ig}\right)\right].
\end{split}$}
\end{equation*}
Let $\exp(\tilde{\alpha}_k)=\tilde{\delta}_k$ and let 
$$\mathcal{H}\left(\tilde{\alpha}_1,\tilde{\alpha}_2,\dots,\tilde{\alpha}_K,\dots\right)=
\mathcal{F}\left(\exp(\tilde{\alpha}_1),\exp(\tilde{\alpha}_2),\dots,\exp(\tilde{\alpha}_K), \dots\right)=
\mathcal{F}\left(\tilde{\delta}_1,\tilde{\delta}_2,\dots,\tilde{\delta}_K,\dots\right),$$
then
\begin{equation*}
\begin{split}
&\frac{\partial}{\partial\tilde{\alpha}_k}\mathcal{H}\left(\tilde{\alpha}_1,\tilde{\alpha}_2,\dots,\tilde{\alpha}_K,\dots\right)=
\frac{\partial}{\partial\tilde{\alpha}_k}\mathcal{F}\left(\exp(\tilde{\alpha}_1),\exp(\tilde{\alpha}_2),\dots,\exp(\tilde{\alpha}_K), \dots\right)\\
&=\frac{\partial}{\partial\tilde{\delta}_k}\mathcal{F}\left(\tilde{\delta}_1,\tilde{\delta}_2,\dots,\tilde{\delta}_K,\dots\right)\frac{\partial\tilde{\delta}_k}{\partial\tilde{\alpha}_k}=
\exp(\tilde{\alpha}_k)\frac{\partial}{\partial\tilde{\delta}_k}\mathcal{F}\left(\tilde{\delta}_1,\tilde{\delta}_2,\dots,\tilde{\delta}_K,\dots\right)
\end{split}
\end{equation*}
Now we first show that the Fisher information associated to the likelihood,
$$f_{\text{Dirichlet}}(\bm{\Pi}|\exp(\tilde{\alpha}_1),\exp(\tilde{\alpha}_2),\dots,\exp(\tilde{\alpha}_K)),$$
is not in the form of diagonal matrix.
Though it is a symmetric matrix, the inverse Fisher information matrix is still computationally demanding.
Thus we finally instead consider the standard gradient ascent for the update of each $\tilde{\delta}_k$ for $k=1,2,\dots,K$.
The logarithm of the likelihood above reads as follows:
\begin{equation*}
\begin{split}
&\log q_{\bm{\Pi}}\left(\bm{\Pi}|\exp(\tilde{\alpha}_1),\exp(\tilde{\alpha}_2),\dots,\exp(\tilde{\alpha}_K)\right)=
\log \frac{\Gamma\left(\sum^K_{k=1}\exp(\tilde{\alpha}_k)\right)\prod^K_{k=1}\pi_k^{\exp(\tilde{\alpha}_k)-1}}{\prod^K_{k=1}\Gamma\left(\exp(\tilde{\alpha}_k)\right)}\\
&=\log\Gamma\left(\sum^K_{k=1}\exp(\tilde{\alpha}_k)\right)-
\sum^K_{k=1}\log\Gamma\left(\exp(\tilde{\alpha}_k)\right)+
\sum^K_{k=1}\left[\exp(\tilde{\alpha}_k)-1\right]\log(\pi_k).
\end{split}
\end{equation*}
The 1st-order partial derivative with respect to $\tilde{\delta}_g$ for $g=1,2,\dots,K$ is of the form:
\begin{equation*}
\begin{split}
&\frac{\partial q_{\bm{\Pi}}}{\partial\tilde{\alpha}_g}=
\Psi\left[\sum^K_{k=1}\exp(\tilde{\alpha}_k)\right]\exp(\tilde{\alpha}_g)-
\Psi\left[\exp(\tilde{\alpha}_g)\right]\exp(\tilde{\alpha}_g)+\exp(\tilde{\alpha}_g)\log(\pi_g)\\
&=\left\{\Psi\left[\sum^K_{k=1}\exp(\tilde{\alpha}_k)\right]-\Psi\left[\exp(\tilde{\alpha}_g)\right]+\log(\pi_g)\right\}\exp(\tilde{\alpha}_g).
\end{split}
\end{equation*}
The 2nd-order partial derivative for $g=1,2,\dots,K$ are of the form:
\begin{equation*}
\begin{split}
&\frac{\partial^2 q_{\bm{\Pi}}}{\partial\tilde{\alpha}_g^2}=
\left\{\Psi'\left[\sum^K_{k=1}\exp(\tilde{\alpha}_k)\right]\exp(\tilde{\alpha}_g)-\Psi'\left[\exp(\tilde{\alpha}_g)\right]\exp(\tilde{\alpha}_g)\right\}\exp(\tilde{\alpha}_g)\\
&+\left\{\Psi\left[\sum^K_{k=1}\exp(\tilde{\alpha}_k)\right]-\Psi\left[\exp(\tilde{\alpha}_g)\right]+\log(\pi_g)\right\}\exp(\tilde{\alpha}_g)\\
&=\left\{\Psi'\left[\sum^K_{k=1}\exp(\tilde{\alpha}_k)\right]-\Psi'\left[\exp(\tilde{\alpha}_g)\right]\right\}\exp(\tilde{\alpha}_g)^2\\
&+\left\{\Psi\left[\sum^K_{k=1}\exp(\tilde{\alpha}_k)\right]-\Psi\left[\exp(\tilde{\alpha}_g)\right]+\log(\pi_g)\right\}\exp(\tilde{\alpha}_g),
\end{split}
\end{equation*}
whereas, for $g,h=1,2,\dots,K;\;g\neq h$, we have
\begin{equation*}
\begin{split}
&\frac{\partial^2 q_{\bm{\Pi}}}{\partial\tilde{\alpha}_g\partial\tilde{\alpha}_h}=
\Psi'\left[\sum^K_{k=1}\exp(\tilde{\alpha}_k)\right]\exp(\tilde{\alpha}_h)\exp(\tilde{\alpha}_g)\\
\end{split}
\end{equation*}
Taking the negative expectation $-\mathbbm{E}_q$ on both equations above gives:
\begin{equation*}
\begin{split}
&-\mathbbm{E}_q\left(\frac{\partial^2 q_{\bm{\Pi}}}{\partial\tilde{\alpha}_g^2}\right)=
-\left\{\Psi'\left[\sum^K_{k=1}\exp(\tilde{\alpha}_k)\right]-\Psi'\left[\exp(\tilde{\alpha}_g)\right]\right\}\exp(\tilde{\alpha}_g)^2\\
&-\left\{\Psi\left[\sum^K_{k=1}\exp(\tilde{\alpha}_k)\right]-\Psi\left[\exp(\tilde{\alpha}_g)\right]+\mathbbm{E}_q\left[\log(\pi_g)\right]\right\}\exp(\tilde{\alpha}_g)\\
&=-\left\{\Psi'\left[\sum^K_{k=1}\exp(\tilde{\alpha}_k)\right]-\Psi'\left[\exp(\tilde{\alpha}_g)\right]\right\}\exp(\tilde{\alpha}_g)^2\\
&-\left\{\Psi\left[\sum^K_{k=1}\exp(\tilde{\alpha}_k)\right]-\Psi\left[\exp(\tilde{\alpha}_g)\right]+\Psi\left[\exp(\tilde{\alpha}_g)\right]-\Psi\left[\sum^K_{k=1}\exp(\tilde{\alpha}_k)\right]\right\}\exp(\tilde{\alpha}_g)\\
&=\left\{\Psi'\left[\exp(\tilde{\alpha}_g)\right]-\Psi'\left[\sum^K_{k=1}\exp(\tilde{\alpha}_k)\right]\right\}\exp(\tilde{\alpha}_g)^2.
\end{split}
\end{equation*}
Similarly,
\begin{equation*}
\begin{split}
&-\mathbbm{E}_q\left(\frac{\partial^2 q_{\bm{\Pi}}}{\partial\tilde{\alpha}_g\partial\tilde{\alpha}_h}\right)=
-\Psi'\left[\sum^K_{k=1}\exp(\tilde{\alpha}_k)\right]\exp(\tilde{\alpha}_h)\exp(\tilde{\alpha}_g).\\
\end{split}
\end{equation*}
Thus the Fisher information $\mathcal{I}\left(\tilde{\alpha}_1,\tilde{\alpha}_2,\dots,\tilde{\alpha}_K\right)$ is a $K \times K$ matrix with each diagonal entry being $-\mathbbm{E}_q\left(\frac{\partial^2 q_{\bm{\Pi}}}{\partial\tilde{\alpha}_g^2}\right)$ for $g=1,2,\dots,K$, and each off-diagonal entry being $-\mathbbm{E}_q\left(\frac{\partial^2 q_{\bm{\Pi}}}{\partial\tilde{\alpha}_g\partial\tilde{\alpha}_h}\right)$ for $g,h=1,2,\dots,K;\;g\neq h$.
It can be expected that the inverse Fisher information is very computational demanding, so we instead consider the standard gradient ascent updates for the inference:
\begin{equation*}
\tilde{\delta}_k^\star=\tilde{\delta}_k^\dagger\exp[\epsilon\tilde{\delta}_k^\dagger\frac{\partial\mathcal{F}}{\partial\tilde{\delta}_k}\left(\tilde{\delta}_1^\dagger,\tilde{\delta}_2^\dagger,\dots,\tilde{\delta}_K^\dagger,\dots\right)],
\end{equation*}
for $k=1,2,\dots,K$.\;\hfill$\square$\\


\subsection{The PICL for the MLPCM (Proof of the Proposition~\ref{Proposition_PICL})}
\label{MLPCM_PICL_Appendix}

Recall here that, provided with the point estimates of the parameters $\bm{\hat{U}}, \bm{\hat{V}}, \bm{\hat{z}}$, the PICL criteria is based on the integrated log-likelihood:
\begin{equation*}
\begin{split}
&\log p(\bm{Y},\bm{\hat{U}}, \bm{\hat{V}}, \bm{\hat{z}}|K)=
\log p(\bm{Y}|\bm{\hat{U}}, \bm{\hat{V}}, \bm{\hat{z}},K)+\log p(\bm{\hat{U}}, \bm{\hat{V}}, \bm{\hat{z}}|K)\\
&=\log p(\bm{Y}|\bm{\hat{U}}, \bm{\hat{V}})+\log p(\bm{\hat{V}}|\bm{\hat{U}}, \bm{\hat{z}},K)+\log p(\bm{\hat{U}}, \bm{\hat{z}}|K)\\
&=\log p(\bm{Y}|\bm{\hat{U}}, \bm{\hat{V}})+\log p(\bm{\hat{V}}|\bm{\hat{U}})+\log p(\bm{\hat{U}}|\bm{\hat{z}})+\log p(\bm{\hat{z}}|K)\\
&=\log \int_{\beta}p(\bm{Y}|\bm{\hat{U}}, \bm{\hat{V}},\beta)\pi(\beta)\text{d}\beta+
\log \int_{\bm{\gamma}}p(\bm{\hat{V}}|\bm{\hat{U}},\bm{\gamma})\pi(\bm{\gamma})\text{d}\bm{\gamma}\\
&+\log \int_{\bm{\tau}}\int_{\bm{\mu}}p(\bm{\hat{U}}|\bm{\hat{z}},\bm{\mu},\bm{\tau})\pi(\bm{\mu})\pi(\bm{\tau})\text{d}\bm{\mu}\text{d}\bm{\tau}+
\log \int_{\bm{\Pi}}p(\bm{\hat{z}}|\bm{\Pi})\pi(\bm{\Pi}|K)\text{d}\bm{\Pi},
\end{split}
\end{equation*}
where we apply the BIC approximation on the 1st $\log p(\bm{Y}|\bm{\hat{U}}, \bm{\hat{V}})$ term, and we apply direct marginalization on the 2nd $\log p(\bm{\hat{V}}|\bm{\hat{U}})$ term and the 4th $\log p(\bm{\hat{z}}|K)$ term.
The combination of the partial integration and BIC approximation is applied on the 3rd $\log p(\bm{\hat{U}}|\bm{\hat{z}})$ term.


\subsubsection[]{BIC approximation for the $\log p(\bm{Y}|\bm{\hat{U}}, \bm{\hat{V}})$}
\label{PICL_logY_Appendix}

\begin{equation*}
\resizebox{1\hsize}{!}{$
\begin{split}
&\log p(\bm{Y}|\bm{\hat{U}}, \bm{\hat{V}})=\log \int_{\beta}p(\bm{Y}|\bm{\hat{U}}, \bm{\hat{V}},\beta)\pi(\beta)\text{d}\beta\overset{\text{\tiny BIC}}{\approx} \max_{\beta} \left[\log p(\bm{Y}|\bm{\hat{U}}, \bm{\hat{V}},\beta)\right] - \frac{1}{2}\log[N(N-1)]\\
&=\max_{\beta} \left[\log\prod^N_{\substack{i,j=1, \\ i\neq j}}f_{\text{Pois}}\left(y_{ij}|\exp(\beta-||\bm{\hat{u}}_i-\bm{\hat{v}}_{j\leftarrow i}||^2)\right)\right]- \frac{1}{2}\log[N(N-1)]\\
&=\max_{\beta}\left[\log\prod^N_{\substack{i,j=1, \\ i\neq j}}\frac{\left[\exp(\beta-||\bm{\hat{u}}_i-\bm{\hat{v}}_{j\leftarrow i}||^2)\right]^{y_{ij}}\exp[-\exp(\beta-||\bm{\hat{u}}_i-\bm{\hat{v}}_{j\leftarrow i}||^2)]}{y_{ij}!}\right]- \frac{1}{2}\log[N(N-1)]\\
&=\max_{\beta}\sum^N_{\substack{i,j=1, \\ i\neq j}}\left[ y_{ij}\left(\beta-||\bm{\hat{u}}_i-\bm{\hat{v}}_{j\leftarrow i}||^2\right)-\exp(\beta-||\bm{\hat{u}}_i-\bm{\hat{v}}_{j\leftarrow i}||^2)-\log(y_{ij}!)\right]- \frac{1}{2}\log[N(N-1)]\\
&=\max_{\beta} \left[\beta\left(\sum^N_{\substack{i,j=1, \\ i\neq j}}y_{ij}\right)-\sum^N_{\substack{i,j=1, \\ i\neq j}}\exp(\beta-||\bm{\hat{u}}_i-\bm{\hat{v}}_{j\leftarrow i}||^2)\right]-\sum^N_{\substack{i,j=1, \\ i\neq j}}\left[y_{ij}||\bm{\hat{u}}_i-\bm{\hat{v}}_{j\leftarrow i}||^2+\log(y_{ij}!)\right]\\
&- \frac{1}{2}\log[N(N-1)].
\end{split}$}
\end{equation*}
Here, defining
\begin{equation*}
\begin{split}
&g(\beta):=\beta\left(\sum^N_{\substack{i,j=1, \\ i\neq j}}y_{ij}\right)-\sum^N_{\substack{i,j=1, \\ i\neq j}}\exp(\beta-||\bm{\hat{u}}_i-\bm{\hat{v}}_{j\leftarrow i}||^2),
\end{split}
\end{equation*}
the maximization of this $g(\beta)$ function has an unique analytical solution, where the 1st derivative with respect to $\beta$ gives:
\begin{equation*}
\begin{split}
& \frac{\partial g(\beta)}{\partial \beta} = \left(\sum^N_{\substack{i,j=1, \\ i\neq j}}y_{ij}\right)-\exp(\beta)\sum^N_{\substack{i,j=1, \\ i\neq j}}\exp(-||\bm{\hat{u}}_i-\bm{\hat{v}}_{j\leftarrow i}||^2):=0\\
&\implies \beta=\log(\sum^N_{\substack{i,j=1, \\ i\neq j}}y_{ij})-\log\left[\sum^N_{\substack{i,j=1, \\ i\neq j}}\exp(-||\bm{\hat{u}}_i-\bm{\hat{v}}_{j\leftarrow i}||^2)\right].
\end{split}
\end{equation*}
It can be checked that the 2nd derivative with respect to $\beta$ is:
\begin{equation*}
\begin{split}
& \frac{\partial^2g(\beta)}{\partial \beta^2} = -\exp(\beta)\sum^N_{\substack{i,j=1, \\ i\neq j}}\exp(-||\bm{\hat{u}}_i-\bm{\hat{v}}_{j\leftarrow i}||^2) <0.
\end{split}
\end{equation*}
This ensures the global maximum at the above obtained $\beta$.
Substituting back to the $\log p(\bm{Y}|\bm{\hat{U}}, \bm{\hat{V}})$ term in PICL gives:
\begin{equation*}
\begin{split}
&\log p(\bm{Y}|\bm{\hat{U}}, \bm{\hat{V}})
\overset{\text{\tiny BIC}}{\approx} \max_{\beta} \left[\beta\left(\sum^N_{\substack{i,j=1, \\ i\neq j}}y_{ij}\right)-\sum^N_{\substack{i,j=1, \\ i\neq j}}\exp(\beta-||\bm{\hat{u}}_i-\bm{\hat{v}}_{j\leftarrow i}||^2)\right]\\
&-\sum^N_{\substack{i,j=1, \\ i\neq j}}\left[y_{ij}||\bm{\hat{u}}_i-\bm{\hat{v}}_{j\leftarrow i}||^2+\log(y_{ij}!)\right]
- \frac{1}{2}\log[N(N-1)]\\
&= \left\{\log(\sum^N_{\substack{i,j=1, \\ i\neq j}}y_{ij})-\log\left[\sum^N_{\substack{i,j=1, \\ i\neq j}}\exp(-||\bm{\hat{u}}_i-\bm{\hat{v}}_{j\leftarrow i}||^2)\right]\right\}\left(\sum^N_{\substack{i,j=1, \\ i\neq j}}y_{ij}\right)\\
&-\frac{\sum^N_{\substack{i,j=1, \\ i\neq j}}y_{ij}}{\sum^N_{\substack{i,j=1, \\ i\neq j}}\exp(-||\bm{\hat{u}}_i-\bm{\hat{v}}_{j\leftarrow i}||^2)}\sum^N_{\substack{i,j=1, \\ i\neq j}}\exp(-||\bm{\hat{u}}_i-\bm{\hat{v}}_{j\leftarrow i}||^2)\\
&-\sum^N_{\substack{i,j=1, \\ i\neq j}}\left[y_{ij}||\bm{\hat{u}}_i-\bm{\hat{v}}_{j\leftarrow i}||^2+\log(y_{ij}!)\right]
- \frac{1}{2}\log[N(N-1)]\\
&=\left\{\log(\sum^N_{\substack{i,j=1, \\ i\neq j}}y_{ij})-\log\left[\sum^N_{\substack{i,j=1, \\ i\neq j}}\exp(-||\bm{\hat{u}}_i-\bm{\hat{v}}_{j\leftarrow i}||^2)\right]-1\right\}\left(\sum^N_{\substack{i,j=1, \\ i\neq j}}y_{ij}\right)\\
&-\sum^N_{\substack{i,j=1, \\ i\neq j}}\left[y_{ij}||\bm{\hat{u}}_i-\bm{\hat{v}}_{j\leftarrow i}||^2+\log(y_{ij}!)\right]
- \frac{1}{2}\log[N(N-1)]\\
&=\texttt{const}-\left(\sum^N_{\substack{i,j=1, \\ i\neq j}}y_{ij}\right)\log\left[\sum^N_{\substack{i,j=1, \\ i\neq j}}\exp(-||\bm{\hat{u}}_i-\bm{\hat{v}}_{j\leftarrow i}||^2)\right]\\
&-\sum^N_{\substack{i,j=1, \\ i\neq j}}\left(y_{ij}||\bm{\hat{u}}_i-\bm{\hat{v}}_{j\leftarrow i}||^2\right).
\end{split}
\end{equation*}\;\hfill$\square$\\


\subsubsection[]{Integrated term of the $\log p(\bm{\hat{V}}|\bm{\hat{U}})$}

\begin{equation*}
\resizebox{1\hsize}{!}{$
\begin{split}
&\log p(\bm{\hat{V}}|\bm{\hat{U}}) = \log \int_{\bm{\gamma}}p(\bm{\hat{V}}|\bm{\hat{U}},\bm{\gamma})\pi(\bm{\gamma})\text{d}\bm{\gamma}\\
&=\log\bigint_{\bm{\gamma}}\prod^N_{j=1}\left\{\left[\prod^N_{\substack{i=1, \\ i\neq j}}f_{\text{MVN}_d}\left(\bm{\hat{v}}_{j\leftarrow i}|\bm{\hat{u}}_j,1/\gamma_j\mathbbm{I}_d\right)\right]f_{\text{Ga}}\left(\gamma_j|a,b\right)\right\}\text{d}\bm{\gamma}\\
&=\log\bigint_{\bm{\gamma}}\prod^N_{j=1}\left\{\left[\prod^N_{\substack{i=1, \\ i\neq j}} \frac{|\gamma_j\mathbbm{I}_d|^{1/2}}{(2\pi)^{d/2}}\exp[-\frac{1}{2}\left(\bm{\hat{v}}_{j\leftarrow i}-\bm{\hat{u}}_j\right)\gamma_j\mathbbm{I}_d\left(\bm{\hat{v}}_{j\leftarrow i}-\bm{\hat{u}}_j\right)^T]
\right]f_{\text{Ga}}\left(\gamma_j|a,b\right)\right\}\text{d}\bm{\gamma}\\
&=\log\bigint_{\bm{\gamma}}\prod^N_{j=1}\left\{
(2\pi)^{-\frac{d}{2}(N-1)}\gamma_j^{\frac{d}{2}(N-1)}\exp[-\frac{1}{2}\gamma_j\sum^N_{\substack{i=1, \\ i\neq j}}||\bm{\hat{v}}_{j\leftarrow i}-\bm{\hat{u}}_j||^2]
\frac{b^a}{\Gamma(a)}\gamma_j^{a-1}\exp(-b\gamma_j)\right\}\text{d}\bm{\gamma}\\
&=\log\prod^N_{j=1}\left\{(2\pi)^{-\frac{d}{2}(N-1)}\frac{b^a}{\Gamma(a)}\bigint_{\gamma_j}
\gamma_j^{a+\frac{d}{2}(N-1)-1}\exp[-\left(b+\frac{1}{2}\sum^N_{\substack{i=1, \\ i\neq j}}||\bm{\hat{v}}_{j\leftarrow i}-\bm{\hat{u}}_j||^2\right)\gamma_j]
\text{d}\gamma_j\right\}\\
&=\log\prod^N_{j=1}\left[(2\pi)^{-\frac{d}{2}(N-1)}\frac{b^a}{\Gamma(a)}\frac{\Gamma\left(a+\frac{d}{2}(N-1)\right)}{\left(b+\frac{1}{2}\sum^N_{\substack{i=1, \\ i\neq j}}||\bm{\hat{v}}_{j\leftarrow i}-\bm{\hat{u}}_j||^2\right)^{a+\frac{d}{2}(N-1)}}\right]\\
&=\sum^N_{j=1}\left[-\frac{d}{2}(N-1)\log(2\pi)+a\log(b)-\log\Gamma(a)+\log\Gamma\left(a+\frac{d}{2}(N-1)\right)\right]\\
&-\sum^N_{j=1}\left[a+\frac{d}{2}(N-1)\right]\log(b+\frac{1}{2}\sum^N_{\substack{i=1, \\ i\neq j}}||\bm{\hat{v}}_{j\leftarrow i}-\bm{\hat{u}}_j||^2)\\
&=\texttt{const}-\left[a+\frac{d}{2}(N-1)\right]\sum^N_{j=1}\log(b+\frac{1}{2}\sum^N_{\substack{i=1, \\ i\neq j}}||\bm{\hat{v}}_{j\leftarrow i}-\bm{\hat{u}}_j||^2).
\end{split}$}
\end{equation*}\;\hfill$\square$\\

\subsubsection[]{Partial integration + BIC approximation for the $\log p(\bm{\hat{U}}|\bm{\hat{z}})$}
\begin{equation*}
\resizebox{1\hsize}{!}{$
\begin{split}
&\log p(\bm{\hat{U}}|\bm{\hat{z}}) = \log \int_{\bm{\tau}}\int_{\bm{\mu}}p(\bm{\hat{U}}|\bm{\hat{z}},\bm{\mu},\bm{\tau})\pi(\bm{\mu})\text{d}\bm{\mu}\;\pi(\bm{\tau})\text{d}\bm{\tau}\\
&=\log \bigint_{\bm{\tau}} \bigint_{\bm{\mu}} \prod^K_{g=1}\left\{ \left[ \prod^N_{\substack{i=1, \\ \hat{z}_i=g}} f_{\text{MVN}_d}\left(\bm{\hat{u}}_i|\bm{\mu}_g,1/\tau_g\mathbbm{I}_d\right) \right] f_{\text{MVN}_d}\left(\bm{\mu}_g|\bm{0},\omega^2\mathbbm{I}_d\right) \right\}\text{d}\bm{\mu}\;\pi(\bm{\tau})\text{d}\bm{\tau}\\
&=\log \bigint_{\bm{\tau}} \left\{ \prod^K_{g=1} \bigint_{\bm{\mu}_g} \left[ \prod^N_{\substack{i=1, \\ \hat{z}_i=g}} 
\frac{|\tau_g\mathbbm{I}_d|^{1/2}}{(2\pi)^{d/2}}\exp(-\frac{\tau_g}{2}||\bm{\hat{u}}_i-\bm{\mu}_g||^2)
\right] f_{\text{MVN}_d}\left(\bm{\mu}_g|\bm{0},\omega^2\mathbbm{I}_d\right)\text{d}\bm{\mu}_g\right\}\pi(\bm{\tau})\text{d}\bm{\tau}\\
&=\log \bigint_{\bm{\tau}} \left\{ \prod^K_{g=1} \bigint_{\bm{\mu}_g} \left[
\left(\frac{\tau_g}{2\pi}\right)^{\frac{d}{2}\hat{n}_g} \exp(- \frac{\tau_g}{2} \sum^N_{\substack{i=1, \\ \hat{z}_i=g}} ||\bm{\hat{u}}_i-\bm{\mu}_g||^2)
\right] \frac{\exp(-\frac{1}{2\omega^2}||\bm{\mu}_g||^2)}{(2\pi\omega^2)^{\frac{d}{2}} }
\text{d}\bm{\mu}_g\right\}\pi(\bm{\tau})\text{d}\bm{\tau}\\
&=\log \bigint_{\bm{\tau}}  \prod^K_{g=1}  
 \frac{ \bigint_{\bm{\mu}_g} \exp[- \frac{\tau_g}{2} \left(\left(\sum_{\substack{i=1, \\ \hat{z}_i=g}}||\bm{\hat{u}}_i||^2\right) - 2\left(\sum^N_{\substack{i=1, \\ \hat{z}_i=g}}\bm{\hat{u}}_i^T\right)\bm{\mu}_g + \hat{n}_g||\bm{\mu}_g||^2 \right) -\frac{1}{2\omega^2}||\bm{\mu}_g||^2 ]\text{d}\bm{\mu}_g }{ \left(\frac{\tau_g}{2\pi}\right)^{-\frac{d}{2}\hat{n}_g} (2\pi\omega^2)^{\frac{d}{2}}}
\pi(\bm{\tau})\text{d}\bm{\tau}\\
&=\log \bigint_{\bm{\tau}}  \prod^K_{g=1}  
 \frac{ \bigint_{\bm{\mu}_g} \exp[
 -\frac{1}{2}\left(\tau_g\hat{n}_g+\frac{1}{\omega^2}\right)||\bm{\mu}_g||^2 + 2\frac{\tau_g}{2}\left(\sum^N_{\substack{i=1, \\ \hat{z}_i=g}}\bm{\hat{u}}_i^T\right)\bm{\mu}_g - \frac{\tau_g}{2} \left(\sum^N_{\substack{i=1, \\ \hat{z}_i=g}}||\bm{\hat{u}}_i||^2\right)
 ]\text{d}\bm{\mu}_g }{ \left(\frac{\tau_g}{2\pi}\right)^{-\frac{d}{2}\hat{n}_g} (2\pi\omega^2)^{\frac{d}{2}}}
\pi(\bm{\tau})\text{d}\bm{\tau}\\
&=\log \bigint_{\bm{\tau}}  \prod^K_{g=1}  
 \frac{ \bigint_{\bm{\mu}_g} \exp[
 -\frac{1}{2}\left(\tau_g\hat{n}_g+\frac{1}{\omega^2}\right)\norm{\bm{\mu}_g-\frac{\tau_g}{\tau_g\hat{n}_g+\frac{1}{\omega^2}}\left(\sum^N_{\substack{i=1, \\ \hat{z}_i=g}}\bm{\hat{u}}_i\right)}^2
 ]\text{d}\bm{\mu}_g }{ \left(\frac{\tau_g}{2\pi}\right)^{-\frac{d}{2}\hat{n}_g} (2\pi\omega^2)^{\frac{d}{2}}}\\
 & \prod^K_{g=1} \exp[
  \frac{1}{2}\left(\tau_g\hat{n}_g+\frac{1}{\omega^2}\right) \left(\frac{\tau_g}{\tau_g\hat{n}_g+\frac{1}{\omega^2}}\right)^2\norm{\sum_{\hat{z}_i=g}\bm{\hat{u}}_i}^2
  - \frac{\tau_g}{2} \left(\sum_{\hat{z}_i=g}||\bm{\hat{u}}_i||^2\right)
 ]\pi(\bm{\tau})\text{d}\bm{\tau}\\
 &=\log \bigint_{\bm{\tau}}  \prod^K_{g=1} \frac{ 
 \exp[
  \frac{1}{2} \frac{\tau_g^2}{\tau_g\hat{n}_g+\frac{1}{\omega^2}}\norm{\sum_{\hat{z}_i=g}\bm{\hat{u}}_i}^2
  - \frac{\tau_g}{2} \left(\sum_{\hat{z}_i=g}||\bm{\hat{u}}_i||^2\right)
 ]
  }{  \left(\frac{\tau_g}{2\pi}\right)^{-\frac{d}{2}\hat{n}_g} \left(\tau_g\hat{n}_g\omega^2+1\right)^{\frac{d}{2}} } \pi(\bm{\tau})\text{d}\bm{\tau}\\
&\overset{\text{\tiny BIC}}{\approx} \max_{\bm{\tau}} \left\{ \log \prod^K_{g=1} \frac{ 
 \exp[
  \frac{1}{2} \frac{\tau_g^2\omega^2}{\tau_g\hat{n}_g\omega^2+1}\norm{\sum_{\hat{z}_i=g}\bm{\hat{u}}_i}^2
  - \frac{\tau_g}{2} \left(\sum_{\hat{z}_i=g}||\bm{\hat{u}}_i||^2\right)
 ]
  }{  \left(\frac{\tau_g}{2\pi}\right)^{-\frac{d}{2}\hat{n}_g} \left(\tau_g\hat{n}_g\omega^2+1\right)^{\frac{d}{2}} } \right\} - \frac{1}{2}K\log(N)\\
  &=  \max_{\bm{\tau}} \sum^K_{g=1} \left[
  \frac{d}{2}\hat{n}_g\log(\tau_g)-\frac{d}{2}\log\left(\tau_g\hat{n}_g\omega^2+1\right)+
  \frac{1}{2} \frac{\tau_g^2\omega^2}{\tau_g\hat{n}_g\omega^2+1}\norm{\sum_{\hat{z}_i=g}\bm{\hat{u}}_i}^2
  - \frac{\tau_g}{2} \left(\sum_{\hat{z}_i=g}||\bm{\hat{u}}_i||^2\right)
\right]\\
&- \frac{d}{2}N\log(2\pi) -\frac{1}{2}K\log(N),
\end{split}$}
\end{equation*}
where $\hat{n}_g:=\sum_{i=1}^N\mathbbm{1}(\hat{z}_i=g)$.
We denote 
\begin{equation*}
\resizebox{1\hsize}{!}{$
\begin{split}
&h(\bm{\tau})=\sum^K_{g=1} \left[
  \frac{d}{2}\hat{n}_g\log(\tau_g)-\frac{d}{2}\log\left(\tau_g\hat{n}_g\omega^2+1\right)+
  \frac{1}{2} \frac{\tau_g^2\omega^2}{\tau_g\hat{n}_g\omega^2+1}\norm{\sum_{\hat{z}_i=g}\bm{\hat{u}}_i}^2
  - \frac{\tau_g}{2} \left(\sum_{\hat{z}_i=g}||\bm{\hat{u}}_i||^2\right)
\right],
\end{split}$}
\end{equation*}
and we leverage the standard gradient ascent for the maximization.
We first obtain the 1st derivative with respect to $\tau_g$:
\begin{equation*}
\resizebox{1\hsize}{!}{$
\begin{split}
&\frac{\partial h(\bm{\tau})}{\partial \tau_g} = \frac{d\hat{n}_g}{2\tau_g}-\frac{d\hat{n}_g\omega^2}{2\left(\tau_g\hat{n}_g\omega^2+1\right)}+
\frac{1}{2}\left[\frac{2\tau_g\omega^2}{\tau_g\hat{n}_g\omega^2+1}-\frac{\tau_g^2\omega^2\hat{n}_g\omega^2}{\left(\tau_g\hat{n}_g\omega^2+1\right)^2}\right]\norm{\sum_{\hat{z}_i=g}\bm{\hat{u}}_i}^2
-\frac{1}{2} \left(\sum_{\hat{z}_i=g}||\bm{\hat{u}}_i||^2\right)\\
&=\frac{d}{2}\hat{n}_g\left(\frac{1}{\tau_g}-\frac{\omega^2}{\tau_g\hat{n}_g\omega^2+1}\right)
+\frac{\tau_g\omega^2\left(\tau_g\hat{n}_g\omega^2+2\right)}{2\left(\tau_g\hat{n}_g\omega^2+1\right)^2}\norm{\sum_{\hat{z}_i=g}\bm{\hat{u}}_i}^2
-\frac{1}{2} \left(\sum_{\hat{z}_i=g}||\bm{\hat{u}}_i||^2\right),
\end{split}$}
\end{equation*}
which is a function of $\tau_g$. Then we follow the standard gradient ascent scheme to find the values of $\{\tau_g:g=1,\dots,K\}$ which achieves the maximum.
However, as $\tau_g>0$ for $g=1,\dots,K$, we let $\exp(r_g)=\tau_g$ and let $l(r_1,\dots,r_K)=h(\exp(r_1),\dots,\exp(r_K))=h(\tau_1,\dots,\tau_K)$, then
\begin{equation*}
\resizebox{1\hsize}{!}{$
\begin{split}
&\frac{\partial}{\partial r_g}l\left(r_1,\dots,r_K\right)=
\frac{\partial}{\partial r_g}h(\exp(r_1),\dots,\exp(r_K))=
\frac{\partial}{\partial \tau_g}h(\tau_1,\dots,\tau_K)\frac{\partial \tau_g}{\partial r_g}=
\exp(r_g)\frac{\partial}{\partial \tau_g}h(\tau_1,\dots,\tau_K).
\end{split}$}
\end{equation*}
This brings the standard gradient ascent scheme as:
\begin{equation*}
\begin{split}
&r_g^\star=r_g^\dagger+\epsilon\exp(r_g^\dagger)\frac{\partial h}{\partial \tau_g}\left(\tau_g^\dagger\right),
\end{split}
\end{equation*}
for $g=1,\dots,K$. Taking the $\exp(\cdot)$ on both sides gives the final standard gradient ascent scheme steps:
\begin{equation*}
\begin{split}
&\exp(r_g^\star)=\exp[r_g^\dagger+\epsilon\exp(r_g^\dagger)\frac{\partial h}{\partial \tau_g}\left(\tau_g^\dagger\right)]\\
&\implies \exp(r_g^\star)=\exp(r_g^\dagger)\exp[\epsilon\exp(r_g^\dagger)\frac{\partial h}{\partial \tau_g}\left(\tau_g^{(t)}\right)]\\
&\implies \tau_g^\star=\tau_g^\dagger\exp[\epsilon\tau_g^\dagger\frac{\partial h}{\partial \tau_g}\left(\tau_g^\dagger\right)].
\end{split}
\end{equation*}
Finally, substituting the obtained $\{\tau_g:g=1,\dots,K\}$, which maximizes $h(\bm{\tau})$, back to the $h(\bm{\tau})$ gives the final PICL criteria value.\;\hfill$\square$\\


\subsubsection[]{Integrated term of the $\log p(\bm{\hat{z}}|K)$}

\begin{equation*}
\resizebox{1\hsize}{!}{$
\begin{split}
&\log p(\bm{\hat{z}}|K) = \log \int_{\bm{\Pi}}p(\bm{\hat{z}}|\bm{\Pi})\pi(\bm{\Pi}|K)\text{d}\bm{\Pi}
=\log \int_{\bm{\Pi}}\left[\prod^N_{i=1}f_{\text{Cat}}(\hat{z}_i|\bm{\Pi})\right]f_{\text{Dirichlet}}(\bm{\Pi}|\delta_1,\dots,\delta_K)\text{d}\bm{\Pi}\\
&= \log \int_{\bm{\Pi}}\left(\prod^K_{k=1}\pi_k^{\hat{n}_k}\right)\cdot \frac{\left(\prod^K_{k=1}\pi_k^{\delta_k-1}\right)}{\text{B}(\delta_1,\dots,\delta_K)}\text{d}\bm{\Pi}
= \log \int_{\bm{\Pi}}\frac{1}{\text{B}(\delta_1,\dots,\delta_K)}\cdot \left(\prod^K_{k=1}\pi_k^{\hat{n}_k+\delta_k-1}\right)\text{d}\bm{\Pi}\\
&=\log\frac{\text{B}(\hat{n}_1+\delta_1,\dots,\hat{n}_K+\delta_K)}{\text{B}(\delta_1,\dots,\delta_K)}
=\log\left[\frac{\prod^K_{k=1}\Gamma(\hat{n}_k+\delta_k)}{\Gamma\left(N+\sum^K_{k=1}\delta_k\right)}\frac{\Gamma\left(\sum^K_{k=1}\delta_k\right)}{\prod^K_{k=1}\Gamma(\delta_k)}\right]\\
&=\left[\sum^K_{k=1}\log\Gamma(\hat{n}_k+\delta_k)\right]-\log\Gamma\left(N+\sum^K_{k=1}\delta_k\right)+\log\Gamma\left(\sum^K_{k=1}\delta_k\right)-\sum^K_{k=1}\log\Gamma(\delta_k).
\end{split}$}
\end{equation*}
When $\delta_k=\delta$ for $k=1,\dots,K$, the above formula reduces to:
\begin{equation*}
\begin{split}
&\log p(\bm{\hat{z}}|K)
=\left[\sum^K_{k=1}\log\Gamma(\hat{n}_k+\delta)\right]-\log\Gamma\left(N+K\delta\right)+\log\Gamma\left(K\delta\right)-K\log\Gamma(\delta).
\end{split}
\end{equation*}\;\hfill$\square$\\


\subsection{Derivation of the ELBO for the PoisLPCM}
\label{PoisLPCM_ELBO_Derivation_Appendix}

The ELBO for the PoisLPCM is of the form:
\begin{equation*}
\begin{split}
&\mathcal{F}(\bm{\tilde{\Theta}})= \mathbbm{E}_q[\log p(\bm{Y}|\bm{U}, \beta)]+
\mathbbm{E}_q[\log p(\bm{U}|\bm{\mu}, \bm{\tau}, \bm{z})]+
\mathbbm{E}_q[\log p(\bm{z}|\bm{\Pi})]\\
&+\mathbbm{E}_q[\log \pi(\beta)]+
\mathbbm{E}_q[\log \pi(\bm{\mu})]+
\mathbbm{E}_q[\log \pi(\bm{\tau})]+
\mathbbm{E}_q[\log \pi(\bm{\Pi})]\\
&-\mathbbm{E}_q[\log q_{\bm{U}}(\bm{U}|\bm{\tilde{U}},\bm{\tilde{\sigma}^2})]-
\mathbbm{E}_q[\log q_{\bm{z}}(\bm{z}|\bm{\tilde{\Pi}})]-
\mathbbm{E}_q[\log q_{\beta}(\beta|\tilde{\eta},\tilde{\rho}^2)]\\
&-\mathbbm{E}_q[\log q_{\bm{\mu}}(\bm{\mu}|\bm{\tilde{\mu}},\bm{\tilde{\omega}^2})]-
\mathbbm{E}_q[\log q_{\bm{\tau}}(\bm{\tau}|\bm{\tilde{\xi}},\bm{\tilde{\psi}})]-
\mathbbm{E}_q[\log q_{\bm{\Pi}}(\bm{\Pi}|\bm{\tilde{\delta}})],
\end{split}
\end{equation*}
where all the terms except the 1st $\mathbbm{E}_q[\log p(\bm{Y}|\bm{U}, \beta)]$ term are already proved in Appendix~\ref{MLPCM_ELBO_Derivation_Appendix}.
This it just remains to show the derivation of such a term here.
\begin{equation*}
\begin{split}
&\mathbbm{E}_q[\log p(\bm{Y}|\bm{U}, \beta)]=
\mathbbm{E}_q\left[\log \prod^N_{\substack{i,j=1, \\ i\neq j}}f_{\text{Pois}}\left(y_{ij}\middle|\exp (\beta-||\bm{u}_i-\bm{u}_j||^2)\right)\right]\\
&=\mathbbm{E}_q\left[\log\prod^N_{\substack{i,j=1, \\ i\neq j}}\frac{\left[\exp(\beta-||\bm{u}_i-\bm{u}_j||^2)\right]^{y_{ij}}\exp[-\exp (\beta-||\bm{u}_i-\bm{u}_j||^2)]}{y_{ij}!}\right]\\
&=\mathbbm{E}_q\left[\sum^N_{\substack{i,j=1, \\ i\neq j}}\left[y_{ij}\left(\beta-||\bm{u}_i-\bm{u}_j||^2\right)-\exp (\beta-||\bm{u}_i-\bm{u}_j||^2)-\log(y_{ij}!)\right]\right]\\
&=\sum^N_{\substack{i,j=1, \\ i\neq j}}\left\{y_{ij}\left[\mathbbm{E}_q(\beta)-\mathbbm{E}_q\left(||\bm{u}_i-\bm{u}_j||^2\right)\right]-\mathbbm{E}_q\left[\exp (\beta-||\bm{u}_i-\bm{u}_j||^2)\right]\right\}-
\sum^N_{\substack{i,j=1, \\ i\neq j}}\log(y_{ij}!)\\
&=\sum^N_{\substack{i,j=1, \\ i\neq j}}\left\{ y_{ij}\left[\tilde{\eta}-||\bm{\tilde{u}}_i-\bm{\tilde{u}}_j||^2-d\left(\tilde{\sigma}^2_i+\tilde{\sigma}^2_j\right)\right]-\frac{\exp \left(\tilde{\eta}+\frac{\tilde{\rho}^2}{2}-\frac{||\bm{\tilde{u}}_i-\bm{\tilde{u}}_j||^2}{1+2\tilde{\sigma}^2_i+2\tilde{\sigma}^2_j}\right)}{(1+2\tilde{\sigma}^2_i+2\tilde{\sigma}^2_j)^{d/2}} \right\}+\texttt{const},
\end{split}
\end{equation*}
where the calculations here are similar to those shown in Appendix~\ref{ELBO_pY_Section_Appendix}. \;\hfill$\square$\\


\subsection{Poisson latent position cluster model}
\label{Poisson_LPCM_Section}

We illustrate the performance of our novel MLPCM by comparing to that of the Poisson Latent Position Cluster Model (PoisLPCM), where we follow \textcite{lu2025zero} to extend the binary LPCM proposed by \textcite{handcock2007model} to non-negative discrete weighted networks: 
for $i,j = 1,\dots,N;\;i\neq j$, we have
\begin{equation*}
\label{ThePoisLPCM1}
\begin{split}
 y_{ij}&\sim\text{Pois}(\lambda_{ij}),\;\;\text{log}(\lambda_{ij})=\beta-||\bm{u}_i-\bm{u}_j||^2,
\end{split}
\end{equation*}
where $\bm{u}_i|z_i=g$ and $z_i$ follow Eq.\eqref{MixedLPCM1} for $i=1,2,\dots,N$.
Thus the interactions $y_{ij}$ and $y_{ji}$, which have different orientations, are assumed to follow the same distribution for $i,j = 1,\dots,N;\;i\neq j$.
Removing step 3 of the Algorithm~\ref{MixedLPCMSimulation} without inputting $\bm{\gamma}$, and replacing step 4 by $y_{ij} \sim \text{Pois}\left[\exp (\beta-||\bm{u}_i-\bm{u}_j||^2)\right]$ therein provides the procedure to simulate from the PoisLPCM.
The inference for the PoisLPCM is also based on the variational Bayes in this paper, where the ELBO for the PoisLPCM shares certain terms with Eq.~\eqref{ELBO_MLPCM} as illustrated in Proposition~\ref{PoisLPCM_ELBO} below.
\begin{proposition}
\label{PoisLPCM_ELBO}
Provided with the variational distributions following Eqs.~\eqref{Variational_distributions} without considering $\{\bm{\tilde{v}}_{j\leftarrow i}\}$ and $\{\gamma_i\}$, and letting $\bm{\tilde{\Phi}}:=\{\bm{\tilde{U}},\bm{\tilde{\sigma}^2},\bm{\tilde{\Pi}},\tilde{\eta},\tilde{\rho}^2,\bm{\tilde{\mu}},\bm{\tilde{\omega}^2},\bm{\tilde{\xi}},\bm{\tilde{\psi}},\bm{\tilde{\delta}}\}$, the ELBO for the PoisLPCM is written as:
\begin{equation}
\label{ELBO_PoisLPCM}
\begin{split}
&\mathcal{F}(\bm{\tilde{\Phi}})=
\sum^N_{\substack{i,j=1, \\ i\neq j}}\left\{ y_{ij}\left[\tilde{\eta}-||\bm{\tilde{u}}_i-\bm{\tilde{u}}_j||^2-d\left(\tilde{\sigma}^2_i+\tilde{\sigma}^2_j\right)\right]-\frac{\exp \left(\tilde{\eta}+\frac{\tilde{\rho}^2}{2}-\frac{||\bm{\tilde{u}}_i-\bm{\tilde{u}}_j||^2}{1+2\tilde{\sigma}^2_i+2\tilde{\sigma}^2_j}\right)}{(1+2\tilde{\sigma}^2_i+2\tilde{\sigma}^2_j)^{d/2}} \right\}\\
&+\sum^N_{i=1}\sum^K_{k=1}\tilde{\pi}_{ik}
\left\{\frac{d}{2}\left[\bm{\Psi}(\tilde{\xi}_k)-\log (\tilde{\psi}_k)\right]-\frac{1}{2}\frac{\tilde{\xi}_k}{\tilde{\psi}_k}\left[||\bm{\tilde{u}}_i-\bm{\tilde{\mu}}_k||^2+d(\tilde{\sigma}^2_i+\tilde{\omega}^2_k)\right]\right\}\\
&+\sum^N_{i=1}\sum^K_{k=1}\tilde{\pi}_{ik}\left[\bm{\Psi}\left(\tilde{\delta}_k\right)-\bm{\Psi}\left(\sum^K_{g=1}\tilde{\delta}_g\right)-\log(\tilde{\pi}_{ik})\right]\\
&+\sum^K_{k=1}\left\{(\delta_k-\tilde{\delta}_k)\left[\bm{\Psi}\left(\tilde{\delta}_{k}\right)-\bm{\Psi}\left(\sum^K_{g=1}\tilde{\delta}_{g}\right)\right]+\log\Gamma(\tilde{\delta}_k)
-\frac{1}{2\omega^2}\left(\bm{\tilde{\mu}}_k^T\bm{\tilde{\mu}}_k+d\tilde{\omega}^2_k\right)+\frac{d}{2}\log(\tilde{\omega}^2_k)\right\}\\
&+\sum^K_{k=1}\left\{(\xi-\tilde{\xi}_k)\bm{\Psi}\left(\tilde{\xi}_k\right)-\xi\log\tilde{\psi}_k-\psi\frac{\tilde{\xi}_k}{\tilde{\psi}_k}+\log\Gamma(\tilde{\xi}_k)+\tilde{\xi}_k\right\}\\
&+\left[\sum^N_{i=1}\frac{d}{2}\log (\tilde{\sigma}^2_i)\right]
-\frac{1}{2\rho^2}\left[(\tilde{\eta}-\eta)^2+\tilde{\rho}^2\right]+\frac{1}{2}\log(\tilde{\rho}^2)-\log \Gamma\left(\sum^K_{k=1}\tilde{\delta}_k\right)+\normalfont{\texttt{const}}.
\end{split}
\end{equation}
\end{proposition}
The proof of the Proposition~\ref{PoisLPCM_ELBO} is provided in Appendix~\ref{PoisLPCM_ELBO_Derivation_Appendix}.
Compared to Eq.~\eqref{ELBO_MLPCM}, most of the terms in Eq.~\eqref{ELBO_PoisLPCM} are in common with the corresponding terms included in Eq.~\eqref{ELBO_MLPCM}, whereas the main differences are raised from the 1st summation in Eq.~\eqref{ELBO_PoisLPCM} that contains multiple variational parameters including $\{\bm{\tilde{U}},\bm{\tilde{\sigma}^2},\tilde{\eta},\tilde{\rho}^2\}$.
These make the PoisLPCM variational Bayes inference procedures, which are relevant to such terms, differ from the corresponding terms of the MLPCM, and other procedures excluding those related to mixed latent positions remain the same.
The details of the variational Bayes inference algorithm as well as the PICL criteria defined for the PoisLPCM are provided in Appendix~\ref{PoisLPCM_VB_Appendix}.


\subsection{Variational Bayes inference for the PoisLPCM}
\label{PoisLPCM_VB_Appendix}

In this appendix, we detail the variational Bayes inference algorithm for the PoisLPCM.
Recall that most of the terms in Eq.~\eqref{ELBO_PoisLPCM} are in common with the corresponding terms included in Eq.~\eqref{ELBO_MLPCM}, whereas the main differences are raised from the 1st summation in Eq.~\eqref{ELBO_PoisLPCM} that contains multiple variational parameters including $\{\bm{\tilde{U}},\bm{\tilde{\sigma}^2},\tilde{\eta},\tilde{\rho}^2\}$.
Thus such differences lead to the following propositions required for the variational Bayes algorithm of the PoisLPCM.

\begin{proposition}
\label{Proposition_USigma2_EtaRho2_PoisLPCM}
For any values of $\bm{\tilde{\Phi}}$ and for a small enough $\epsilon>0$,
\begin{enumerate}
\item the natural gradient ascent updates of $(\bm{\tilde{u}}_i,\tilde{\sigma}^2_i)$, for $i=1,2,\dots,N$, do not decrease the ELBO for the PoisLPCM, and are defined by:
\begin{equation}
\label{USigma2_NGA_PoisLPCM}
\begin{cases} 
\bm{\tilde{u}}_i^\star=\bm{\tilde{u}}_i^\dagger+\epsilon\tilde{\sigma}^{2^\dagger}_i\frac{\partial\mathcal{F}}{\partial\bm{\tilde{u}}_i}\left(\bm{\tilde{u}}_i^\dagger,\tilde{\sigma}^{2^\dagger}_i,\dots\right);\\  
\tilde{\sigma}^{2^\star}_i=\tilde{\sigma}^{2^\dagger}_i\exp[\epsilon\frac{2}{d}\tilde{\sigma}^{2^\dagger}_i\frac{\partial\mathcal{F}}{\partial\tilde{\sigma}^2_i}\left(\bm{\tilde{u}}_i^\dagger,\tilde{\sigma}^{2^\dagger}_i,\dots\right)],
\end{cases}
\end{equation}
where $(\cdot)^\star$ and $(\cdot)^\dagger$ denotes the new and current states of the variational parameters, respectively, and,
\begin{equation}
\label{USigma2_Partial_PoisLPCM}
\resizebox{1\hsize}{!}{$
\begin{split}
&\frac{\partial\mathcal{F}}{\partial\bm{\tilde{u}}_i}
=\sum^N_{\substack{j=1, \\ j\neq i}}\left\{-2\left[y_{ij}+y_{ji}-\frac{2\exp \left(\tilde{\eta}+\frac{\tilde{\rho}^2}{2}-\frac{||\bm{\tilde{u}}_i-\bm{\tilde{u}}_j||^2}{1+2\tilde{\sigma}^2_i+2\tilde{\sigma}^2_j}\right)}{(1+2\tilde{\sigma}^2_i+2\tilde{\sigma}^2_j)^{d/2+1}}\right](\bm{\tilde{u}}_i-\bm{\tilde{u}}_j)\right\}-\sum^K_{k=1}\left[\tilde{\pi}_{ik}\frac{\tilde{\xi}_k}{\tilde{\psi}_k}(\bm{\tilde{u}}_i-\bm{\tilde{\mu}}_k)\right],\\
&\frac{\partial\mathcal{F}}{\partial\tilde{\sigma}^2_i}
=\frac{d}{2}\cdot\left[-2\left(\sum^N_{\substack{j=1, \\ j\neq i}}(y_{ij}+y_{ji})\right)-\left(\sum^K_{k=1}\tilde{\pi}_{ik}\frac{\tilde{\xi}_k}{\tilde{\psi}_k}\right)+\frac{1}{\tilde{\sigma}^2_i}\right]
-\sum^N_{\substack{j=1, \\ j\neq i}}
\frac{2\exp \left(\tilde{\eta}+\frac{\tilde{\rho}^2}{2}-\frac{||\bm{\tilde{u}}_i-\bm{\tilde{u}}_j||^2}{1+2\tilde{\sigma}^2_i+2\tilde{\sigma}^2_j}\right)}{(1+2\tilde{\sigma}^2_i+2\tilde{\sigma}^2_j)^{d/2+1}}\left[2\frac{||\bm{\tilde{u}}_i-\bm{\tilde{u}}_j||^2}{1+2\tilde{\sigma}^2_i+2\tilde{\sigma}^2_j}-d\right].
\end{split}$}
\end{equation}

\item the natural gradient ascent updates of $(\tilde{\eta},\tilde{\rho}^{2})$ do not decrease the ELBO for the PoisLPCM, and are defined by:
\begin{equation}
\label{EtaRho2_NGA_PoisLPCM}
\begin{split}
\begin{cases} 
\tilde{\eta}^\star=\tilde{\eta}^\dagger+\epsilon\tilde{\rho}^{2^\dagger}\frac{\partial\mathcal{F}}{\partial\tilde{\eta}}\left(\tilde{\eta}^\dagger,\tilde{\rho}^{2^\dagger},\dots\right);\\  
\tilde{\rho}^{2^\star}=\tilde{\rho}^{2^\dagger}\exp[\epsilon2\tilde{\rho}^{2^\dagger}\frac{\partial\mathcal{F}}{\partial\tilde{\rho}^2}\left(\tilde{\eta}^\dagger,\tilde{\rho}^{2^\dagger},\dots\right)],
\end{cases}
\end{split}
\end{equation}
where $(\cdot)^\star$ and $(\cdot)^\dagger$ denotes the new and current states of the variational parameters, respectively, and,
\begin{equation*}
\begin{split}
&\frac{\partial\mathcal{F}}{\partial\tilde{\eta}}
=\left(\sum^N_{\substack{i,j=1, \\ i\neq j}}y_{ij}\right)-\frac{\tilde{\eta}-\eta}{\rho^2}-
\sum^N_{\substack{i,j=1, \\ i\neq j}}\frac{\exp \left(\tilde{\eta}+\frac{\tilde{\rho}^2}{2}-\frac{||\bm{\tilde{u}}_i-\bm{\tilde{u}}_j||^2}{1+2\tilde{\sigma}^2_i+2\tilde{\sigma}^2_j}\right)}{(1+2\tilde{\sigma}^2_i+2\tilde{\sigma}^2_j)^{d/2}},\\
&\frac{\partial\mathcal{F}}{\partial\tilde{\rho}^2}
=-\frac{1}{2\rho^2}+\frac{1}{2\tilde{\rho}^2}-
\frac{1}{2}\sum^N_{\substack{i,j=1, \\ i\neq j}}\frac{\exp \left(\tilde{\eta}+\frac{\tilde{\rho}^2}{2}-\frac{||\bm{\tilde{u}}_i-\bm{\tilde{u}}_j||^2}{1+2\tilde{\sigma}^2_i+2\tilde{\sigma}^2_j}\right)}{(1+2\tilde{\sigma}^2_i+2\tilde{\sigma}^2_j)^{d/2}}.
\end{split}
\end{equation*}

\end{enumerate}

\end{proposition}
The natural gradient ascent schemes in Proposition~\ref{Proposition_USigma2_EtaRho2_PoisLPCM} are generally identical to the corresponding schemes illustrated in Proposition~\ref{Proposition_USigma2_VVarphi2_EtaRho2}, while the key differences are the 1st-order partial derivatives, of which the proofs are provided in Appendix~\ref{Proof_Proposition_USigma2_EtaRho2_PoisLPCM}.
Note that the 1st-order partial derivatives in Eq.~\eqref{USigma2_Partial_PoisLPCM} for the Pois-LPCM depend on the whole set of the variational parameters $\bm{\tilde{U}}$ and $\bm{\tilde{\sigma}^2}$, and this characteristic is different from that of the 1st-order partial derivatives illustrated in Eq.~\eqref{USigma2_partial}, which only depend on the pair of variational parameters $(\bm{\tilde{u}}_i,\tilde{\sigma}^2_i)$, for the MLPCM.
Thus the natural gradient ascent updates shown as Eqs.~\eqref{USigma2_NGA_PoisLPCM} and \eqref{USigma2_Partial_PoisLPCM} further require conditioning on the newest state of each element of $\bm{\tilde{U}}$ and $\bm{\tilde{\sigma}^2}$ as illustrated in Algorithm~\ref{VB_algorithm_PoisLPCM}.
\begin{algorithm}[htbp!]
\caption{The variational Bayes inference algorithm for the PoisLPCM.}
\label{VB_algorithm_PoisLPCM}
\begin{algorithmic} 
\State \textbf{Input}: $\bm{Y},d,K,\eta,\rho^2,\omega^2,\xi,\psi,\bm{\delta}$,
the tolerance level \texttt{tol}, the gradient ascent step sizes $\{\epsilon_{\bm{u}_i}\}^N_{i=1},\epsilon_{\beta},\epsilon_{\bm{\Pi}}$,
and the initial state:
 $$\bm{\tilde{\Phi}}^{(0)}=\{\bm{\tilde{U}}^{(0)},\bm{\tilde{\sigma}^{2^{\textnormal{(0)}}}},\bm{\tilde{\Pi}}^{(0)},\tilde{\eta}^{(0)},\tilde{\rho}^{2^{\textnormal{(0)}}},\bm{\tilde{\mu}}^{(0)},\bm{\tilde{\omega}^{2^{\textnormal{(0)}}}},\bm{\tilde{\xi}}^{(0)},\bm{\tilde{\psi}}^{(0)},\bm{\tilde{\delta}}^{(0)}\}.$$
Set \texttt{stop}=\texttt{FALSE} and $t=0$. Calculate ELBO at initial state: $\mathcal{F}(\bm{\tilde{\Phi}}^{(0)})$.

\While {\texttt{stop} $\neq$ \texttt{TRUE}}

\State Initialize $\bm{\tilde{\Phi}}^{(t+1)}=\bm{\tilde{\Phi}}^{(t)}$.

\For {$i = 1,\dots,N$}
\State \textcolor{red}{1.} Conditional on the $\{\bm{\tilde{u}}_i^{(t)},\bm{\tilde{U}}^{(t+1)}/\bm{\tilde{u}}_i^{(t+1)}\}$ and $\{\tilde{\sigma}_i^{2^{(t)}},\bm{\tilde{\sigma}^{2^{\textnormal{(t+1)}}}}/\tilde{\sigma}_i^{2^{(t+1)}}\}$, update $(\bm{\tilde{u}}_i^{(t+1)},\tilde{\sigma}_i^{2^{(t+1)}})$ following Eq.~\eqref{USigma2_NGA_PoisLPCM} with $\epsilon=\epsilon_{\bm{u}_i}$.
\If {$\mathcal{F}(\bm{\tilde{u}}_i^{(t+1)},\tilde{\sigma}^{2^{(t+1)}}_i,\dots)<\mathcal{F}(\bm{\tilde{u}}_i^{(t)},\tilde{\sigma}^{2^{(t)}}_i,\dots)$} 
$\epsilon_{\bm{u}_i}=\epsilon_{\bm{u}_i}/2$; rerun Step~\textcolor{red}{1}.
\EndIf
\EndFor

\For {$i = 1,\dots,N$}
\For {$k = 1,\dots,K$}
\State \textcolor{red}{2.} Update $\tilde{\pi}_{ik}^{(t+1)}$ following Eq.~\eqref{Pi_Analytical_Sol}.
\EndFor
\EndFor

\State \textcolor{red}{3.} Update $(\tilde{\eta}^{(t+1)},\tilde{\rho}^{2^{(t+1)}})$ following Eq.~\eqref{EtaRho2_NGA_PoisLPCM} with $\epsilon=\epsilon_{\beta}$.
\If {$\mathcal{F}(\tilde{\eta}^{(t+1)},\tilde{\rho}^{2^{(t+1)}},\dots)<\mathcal{F}(\tilde{\eta}^{(t)},\tilde{\rho}^{2^{(t)}},\dots)$} 
$\epsilon_{\beta}=\epsilon_{\beta}/2$; rerun Step~\textcolor{red}{3}.
\EndIf

\For {$k = 1,\dots,K$}
\State \textcolor{red}{4.} Update $\bm{\tilde{\mu}}_k^{(t+1)}$ and $\tilde{\omega}_k^{2^{(t+1)}}$ following Eq.~\eqref{MuOmega2_Analytical_Sol}.
\State \textcolor{red}{5.} Update $(\tilde{\xi}_k^{(t+1)},\tilde{\psi}_k^{(t+1)})$ following Eq.~\eqref{XiPsi_Analytical_Sol}.
\State \textcolor{red}{6.} Update $\tilde{\delta}_k^{(t+1)}$ following Eq.~\eqref{delta_SGA} with $\epsilon=\epsilon_{\bm{\Pi}}$.
\EndFor
\If {$\mathcal{F}(\bm{\tilde{\delta}}^{(t+1)},\dots)<\mathcal{F}(\bm{\tilde{\delta}}^{(t)},\dots)$} 
$\epsilon_{\bm{\Pi}}=\epsilon_{\bm{\Pi}}/2$; rerun Step~\textcolor{red}{7} for $k = 1,\dots,K$.
\EndIf

\If {$\mathcal{F}(\bm{\tilde{\Phi}}^{(t+1)})-\mathcal{F}(\bm{\tilde{\Phi}}^{(t)})<\texttt{tol}$} 
\texttt{stop}=\texttt{TRUE}
\EndIf

\hspace{-0.8em} Set $t=t+1$.
\EndWhile

\State \textbf{Output}: $\{\bm{\tilde{\Phi}}^{(s)},\mathcal{F}(\bm{\tilde{\Phi}}^{(s)}):s = 1,2,\dots,t\}$.
\end{algorithmic}
\end{algorithm}
As detailed in Proposition~\ref{Proposition_PICL_PoisLPCM} below, we follow the same idea behind Proposition~\ref{Proposition_PICL} to leverage the PICL model selection criteria for the best choice of the number of clusters.
The proof of Proposition~\ref{Proposition_PICL_PoisLPCM} is similar to that of Proposition~\ref{Proposition_PICL} shown in Appendix~\ref{MLPCM_PICL_Appendix}, and can be easily obtained by first removing the $\log p(\bm{\hat{V}}|\bm{\hat{U}})$ term in the integrated log-likelihood and then replacing all the $\bm{\hat{v}}_{j\leftarrow i}$ with $\bm{\hat{u}}_j$ in Appendix~\ref{PICL_logY_Appendix}.

\begin{proposition}
\label{Proposition_PICL_PoisLPCM}
Provided with the prior distributions following Eq.~\eqref{Priors} without considering the $\bm{\gamma}$ prior, along with the point estimates of the model parameters $\bm{\hat{U}}, \bm{\hat{z}}$ for the PoisLPCM, the PICL criteria of the PoisLPCM is written as:
\begin{equation*}
\begin{split}
&PICL=\normalfont{\texttt{const}}-\left(\sum^N_{\substack{i,j=1, \\ i\neq j}}y_{ij}\right)\log\left[\sum^N_{\substack{i,j=1, \\ i\neq j}}\exp(-||\bm{\hat{u}}_i-\bm{\hat{u}}_j||^2)\right]
-\sum^N_{\substack{i,j=1, \\ i\neq j}}\left(y_{ij}||\bm{\hat{u}}_i-\bm{\hat{u}}_j||^2\right)\\
&+\max_{\bm{\tau}} \sum^K_{g=1} \left[
  \frac{d}{2}\hat{n}_g\log(\tau_g)-\frac{d}{2}\log\left(\tau_g\hat{n}_g\omega^2+1\right)+
  \frac{1}{2} \frac{\tau_g^2\omega^2}{\tau_g\hat{n}_g\omega^2+1}\norm{\sum_{\hat{z}_i=g}\bm{\hat{u}}_i}^2
  - \frac{\tau_g}{2} \left(\sum_{\hat{z}_i=g}||\bm{\hat{u}}_i||^2\right)
\right]\\
& -\frac{1}{2}K\log(N)+\left[\sum^K_{k=1}\log\Gamma(\hat{n}_k+\delta)\right]-\log\Gamma\left(N+K\delta\right)+\log\Gamma\left(K\delta\right)-K\log\Gamma(\delta),
\end{split}
\end{equation*}
where $\hat{n}_g:=\sum_{i=1}^N\mathbbm{1}(\hat{z}_i=g)$ and,
\begin{equation*}
\resizebox{1\hsize}{!}{$
\begin{split}
&\normalfont{\texttt{const}}=\left(\sum^N_{\substack{i,j=1, \\ i\neq j}}y_{ij}\right) 
\left[\log(\sum^N_{\substack{i,j=1, \\ i\neq j}}y_{ij})-1\right]
-\left[\sum^N_{\substack{i,j=1, \\ i\neq j}}\log(y_{ij}!)\right] - \frac{1}{2}\log[N(N-1)] - \frac{d}{2}N\log(2\pi),
\end{split}$}
\end{equation*}
and the $\max_{\bm{\tau}}(\cdot)$ term is obtained by the same standard gradient ascent updates shown in Proposition~\ref{Proposition_PICL}.
\end{proposition}


\subsection{Maximization of the ELBO for the PoisLPCM}
\label{PoisLPCM_ELBO_Maximization_Appendix}

Since the closed form optimization steps and the standard gradient step as well as the natural gradient ascent schemes of the ELBO for the PoisLPCM are in common with the corresponding steps for the MLPCM and are already proved in Appendix~\ref{MLPCM_ELBO_Maximization_Appendix}, it remains to prove the 1st-order partial derivatives included in Proposition~\ref{Proposition_USigma2_EtaRho2_PoisLPCM}.


\subsubsection{Proof of the Proposition~\ref{Proposition_USigma2_EtaRho2_PoisLPCM}}
\label{Proof_Proposition_USigma2_EtaRho2_PoisLPCM}

The natural gradient ascent scheme of the $(\bm{\tilde{u}}_i,\tilde{\sigma}^2_i)$, for $i=1,2,\dots, N$, for the PoisLPCM is in line with the corresponding scheme proved in Appendix~\ref{Proof_Proposition_USigma2_VVarphi2_EtaRho2} for the MLPCM.
The key differences are the 1st-order partial derivatives of the ELBO as calculated below.

The 1st-order partial derivative of the ELBO with respect to the $\bm{\tilde{u}}_i$ for $i=1,2,\dots, N$ is written as:
\begin{equation*}
\resizebox{1\hsize}{!}{$
\begin{split}
&\frac{\partial\mathcal{F}}{\partial\bm{\tilde{u}}_i}=
\frac{\partial}{\partial\bm{\tilde{u}}_i}\sum^N_{\substack{j=1, \\ j\neq i}}\left\{ -y_{ij}||\bm{\tilde{u}}_i-\bm{\tilde{u}}_j||^2-
\frac{\exp(\tilde{\eta}+\frac{\tilde{\rho}^2}{2}-\frac{||\bm{\tilde{u}}_i-\bm{\tilde{u}}_j||^2}{1+2\tilde{\sigma}^2_i+2\tilde{\sigma}^2_j})}{(1+2\tilde{\sigma}^2_i+2\tilde{\sigma}^2_j)^{d/2}}\right\}+\frac{\partial}{\partial\bm{\tilde{u}}_i}\sum^K_{k=1}
\left[-\frac{\tilde{\pi}_{ik}}{2}\frac{\tilde{\xi}_k}{\tilde{\psi}_k}||\bm{\tilde{u}}_i-\bm{\tilde{\mu}}_k||^2\right]+\\
&+\frac{\partial}{\partial\bm{\tilde{u}}_i}\sum^N_{\substack{j=1, \\ j\neq i}}\left\{ -y_{ji}||\bm{\tilde{u}}_j-\bm{\tilde{u}}_i||^2-
\frac{\exp(\tilde{\eta}+\frac{\tilde{\rho}^2}{2}-\frac{||\bm{\tilde{u}}_j-\bm{\tilde{u}}_i||^2}{1+2\tilde{\sigma}^2_j+2\tilde{\sigma}^2_i})}{(1+2\tilde{\sigma}^2_j+2\tilde{\sigma}^2_i)^{d/2}}\right\}\\
&=\sum^N_{\substack{j=1, \\ j\neq i}}\left[-2(y_{ij}+y_{ji})(\bm{\tilde{u}}_i-\bm{\tilde{u}}_j)-
\frac{2\exp \left(\tilde{\eta}+\frac{\tilde{\rho}^2}{2}-\frac{||\bm{\tilde{u}}_i-\bm{\tilde{u}}_j||^2}{1+2\tilde{\sigma}^2_i+2\tilde{\sigma}^2_j}\right)}{(1+2\tilde{\sigma}^2_i+2\tilde{\sigma}^2_j)^{d/2}}\frac{-2(\bm{\tilde{u}}_i-\bm{\tilde{u}}_j)}{1+2\tilde{\sigma}^2_i+2\tilde{\sigma}^2_j}\right]-\sum^K_{k=1}\left[\tilde{\pi}_{ik}\frac{\tilde{\xi}_k}{\tilde{\psi}_k}(\bm{\tilde{u}}_i-\bm{\tilde{\mu}}_k)\right]\\
&=\sum^N_{\substack{j=1, \\ j\neq i}}\left\{-2\left[y_{ij}+y_{ji}-\frac{2\exp \left(\tilde{\eta}+\frac{\tilde{\rho}^2}{2}-\frac{||\bm{\tilde{u}}_i-\bm{\tilde{u}}_j||^2}{1+2\tilde{\sigma}^2_i+2\tilde{\sigma}^2_j}\right)}{(1+2\tilde{\sigma}^2_i+2\tilde{\sigma}^2_j)^{d/2+1}}\right](\bm{\tilde{u}}_i-\bm{\tilde{u}}_j)\right\}-\sum^K_{k=1}\left[\tilde{\pi}_{ik}\frac{\tilde{\xi}_k}{\tilde{\psi}_k}(\bm{\tilde{u}}_i-\bm{\tilde{\mu}}_k)\right].\\
\end{split}$}
\end{equation*}
The 1st-order partial derivative of the ELBO with respect to the $\tilde{\sigma}^2_i$ for $i=1,2,\dots, N$ is written as:
\begin{equation*}
\resizebox{1\hsize}{!}{$
\begin{split}
&\frac{\partial\mathcal{F}}{\partial\tilde{\sigma}^2_i}=
\frac{\partial}{\partial\tilde{\sigma}^2_i}\sum^N_{\substack{j=1, \\ j\neq i}}\left[ -y_{ij}\cdot d\tilde{\sigma}^2_i-\frac{\exp \left(\tilde{\eta}+\frac{\tilde{\rho}^2}{2}-\frac{||\bm{\tilde{u}}_i-\bm{\tilde{u}}_j||^2}{1+2\tilde{\sigma}^2_i+2\tilde{\sigma}^2_j}\right)}{(1+2\tilde{\sigma}^2_i+2\tilde{\sigma}^2_j)^{d/2}} \right]+\\
&+\frac{\partial}{\partial\tilde{\sigma}^2_i}\sum^N_{\substack{j=1, \\ j\neq i}}\left[ -y_{ji}\cdot d\tilde{\sigma}^2_i-\frac{\exp \left(\tilde{\eta}+\frac{\tilde{\rho}^2}{2}-\frac{||\bm{\tilde{u}}_j-\bm{\tilde{u}}_i||^2}{1+2\tilde{\sigma}^2_j+2\tilde{\sigma}^2_i}\right)}{(1+2\tilde{\sigma}^2_j+2\tilde{\sigma}^2_i)^{d/2}} \right]
-\frac{\partial}{\partial\tilde{\sigma}^2_i}\sum^K_{k=1}
\left(\frac{\tilde{\pi}_{ik}}{2}\frac{\tilde{\xi}_k}{\tilde{\psi}_k}d\tilde{\sigma}^2_i\right)+
\frac{\partial}{\partial\tilde{\sigma}^2_i}\left[\frac{d}{2}\log (\tilde{\sigma}^2_i)\right]\\
&=\sum^N_{\substack{j=1, \\ j\neq i}}\left[-(y_{ij}+y_{ji})d\right]-
\sum^K_{k=1}\left(\frac{\tilde{\pi}_{ik}}{2}\frac{\tilde{\xi}_k}{\tilde{\psi}_k}d\right)+\frac{d}{2\tilde{\sigma}^2_i}\\
&-\sum^N_{\substack{j=1, \\ j\neq i}}2\left[\frac{\exp \left(\tilde{\eta}+\frac{\tilde{\rho}^2}{2}-\frac{||\bm{\tilde{u}}_i-\bm{\tilde{u}}_j||^2}{1+2\tilde{\sigma}^2_i+2\tilde{\sigma}^2_j}\right)\cdot 2||\bm{\tilde{u}}_i-\bm{\tilde{u}}_j||^2(1+2\tilde{\sigma}^2_i+2\tilde{\sigma}^2_j)^{-2}}{(1+2\tilde{\sigma}^2_i+2\tilde{\sigma}^2_j)^{d/2}}-\frac{d}{2}\frac{\exp \left(\tilde{\eta}+\frac{\tilde{\rho}^2}{2}-\frac{||\bm{\tilde{u}}_i-\bm{\tilde{u}}_j||^2}{1+2\tilde{\sigma}^2_i+2\tilde{\sigma}^2_j}\right)\cdot 2}{(1+2\tilde{\sigma}^2_i+2\tilde{\sigma}^2_j)^{d/2+1}}\right]\\
&=\frac{d}{2}\cdot\left[-2\left(\sum^N_{\substack{j=1, \\ j\neq i}}(y_{ij}+y_{ji})\right)-\left(\sum^K_{k=1}\tilde{\pi}_{ik}\frac{\tilde{\xi}_k}{\tilde{\psi}_k}\right)+\frac{1}{\tilde{\sigma}^2_i}\right]
-\sum^N_{\substack{j=1, \\ j\neq i}}
\frac{2\exp \left(\tilde{\eta}+\frac{\tilde{\rho}^2}{2}-\frac{||\bm{\tilde{u}}_i-\bm{\tilde{u}}_j||^2}{1+2\tilde{\sigma}^2_i+2\tilde{\sigma}^2_j}\right)}{(1+2\tilde{\sigma}^2_i+2\tilde{\sigma}^2_j)^{d/2+1}}\left[2\frac{||\bm{\tilde{u}}_i-\bm{\tilde{u}}_j||^2}{1+2\tilde{\sigma}^2_i+2\tilde{\sigma}^2_j}-d\right].
\end{split}$}
\end{equation*}\;\hfill$\square$\\



The natural gradient ascent scheme of the $(\tilde{\eta},\tilde{\rho}^2)$ for the PoisLPCM is in line with the corresponding scheme proved in Appendix~\ref{Proof_Proposition_USigma2_VVarphi2_EtaRho2} for the MLPCM.
The key differences are the 1st-order partial derivatives of the ELBO as calculated below.

The 1st-order partial derivative of the ELBO with respect to the $\tilde{\eta}$ is written as:
\begin{equation*}
\begin{split}
&\frac{\partial\mathcal{F}}{\partial\tilde{\eta}}=
\frac{\partial}{\partial\tilde{\eta}}\sum^N_{\substack{i,j=1, \\ i\neq j}}\left[ y_{ij}\tilde{\eta}-
\frac{\exp \left(\tilde{\eta}+\frac{\tilde{\rho}^2}{2}-\frac{||\bm{\tilde{u}}_i-\bm{\tilde{u}}_j||^2}{1+2\tilde{\sigma}^2_i+2\tilde{\sigma}^2_j}\right)}{(1+2\tilde{\sigma}^2_i+2\tilde{\sigma}^2_j)^{d/2}} \right]-
\frac{\partial}{\partial\tilde{\eta}}\left[\frac{1}{2\rho^2}(\tilde{\eta}-\eta)^2\right]\\
&=\left(\sum^N_{\substack{i,j=1, \\ i\neq j}}y_{ij}\right)-\frac{\tilde{\eta}-\eta}{\rho^2}-
\sum^N_{\substack{i,j=1, \\ i\neq j}}\frac{\exp \left(\tilde{\eta}+\frac{\tilde{\rho}^2}{2}-\frac{||\bm{\tilde{u}}_i-\bm{\tilde{u}}_j||^2}{1+2\tilde{\sigma}^2_i+2\tilde{\sigma}^2_j}\right)}{(1+2\tilde{\sigma}^2_i+2\tilde{\sigma}^2_j)^{d/2}}.
\end{split}
\end{equation*}
The 1st-order partial derivative of the ELBO with respect to the $\tilde{\rho}^2$ is written as:
\begin{equation*}
\begin{split}
&\frac{\partial\mathcal{F}}{\partial\tilde{\rho}^2}=
\frac{\partial}{\partial\tilde{\rho}^2}\sum^N_{\substack{i,j=1, \\ i\neq j}}\left[ -
\frac{\exp \left(\tilde{\eta}+\frac{\tilde{\rho}^2}{2}-\frac{||\bm{\tilde{u}}_i-\bm{\tilde{u}}_j||^2}{1+2\tilde{\sigma}^2_i+2\tilde{\sigma}^2_j}\right)}{(1+2\tilde{\sigma}^2_i+2\tilde{\sigma}^2_j)^{d/2}}\right]+
\frac{\partial}{\partial\tilde{\rho}^2}\left[-\frac{\tilde{\rho}^2}{2\rho^2}+\frac{1}{2}\log\tilde{\rho}^2\right]\\
&=-\frac{1}{2\rho^2}+\frac{1}{2\tilde{\rho}^2}-
\frac{1}{2}\sum^N_{\substack{i,j=1, \\ i\neq j}}\frac{\exp \left(\tilde{\eta}+\frac{\tilde{\rho}^2}{2}-\frac{||\bm{\tilde{u}}_i-\bm{\tilde{u}}_j||^2}{1+2\tilde{\sigma}^2_i+2\tilde{\sigma}^2_j}\right)}{(1+2\tilde{\sigma}^2_i+2\tilde{\sigma}^2_j)^{d/2}}.
\end{split}
\end{equation*}\;\hfill$\square$\\


\subsection{2024 Egocentric ArmsTransfers K=5 latent representations}
\label{RDA_K5LatentRepresentations_Appendix}

\begin{figure}[htbp!]
\centering
\includegraphics[scale=0.45]{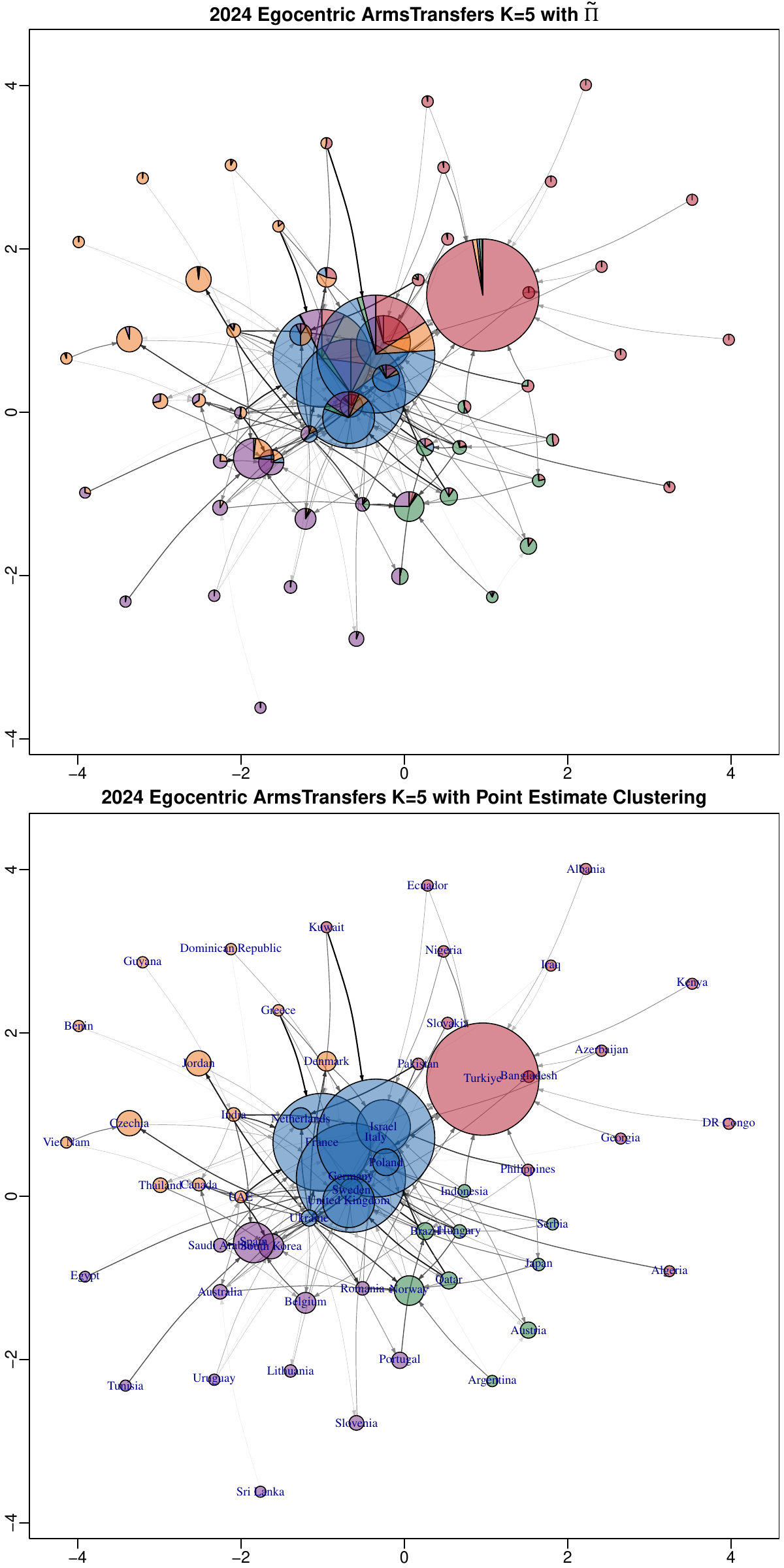}
\caption{2024 Egocentric ArmsTransfers real data application: inferred latent space for $K=5$. For both plots, node positions correspond to $\bm{\hat{U}}$ and node sizes are proportional to the inferred $\{1/\hat{\gamma}_j=\tilde{b}_j/\tilde{a}_j\}$, while edge widths and colors are proportional to edge weights.
Top: each node pie chart illustrates the corresponding inferred variational probability of group assignment. Bottom: the node colors correspond to the maximimum a posteriori clustering; each node is labeled by the corresponding country name.}
\label{RDA_K5LatentRepresentations}
\end{figure}

\end{document}